\documentclass[12pt]{article}
\def\baselinestretch{1.2}

\usepackage{jheppub, bm, wrapfig,float,array}
\usepackage[utf8]{inputenc}
\numberwithin{equation}{section}
\renewcommand\afterTocRuleSpace{\bigskip}

\usepackage{amsfonts}
\usepackage{amsmath}
\usepackage{color}
\usepackage{graphicx}
\usepackage{float}
\usepackage[T1]{fontenc}
\usepackage[utf8]{inputenc}
\usepackage{alphabeta}
\usepackage{fancyvrb}
\usepackage[dvipsnames]{xcolor}
\usepackage{bold-extra}
\usepackage{lmodern}

\usepackage{tikz}
\usetikzlibrary{arrows}
\usetikzlibrary{shapes.geometric,calc,arrows, positioning,shapes.misc,decorations.markings}
\tikzset{
  big arrow/.style={
    decoration={markings,mark=at position 1 with {\arrow[scale=2,#1]{>}}},
    postaction={decorate},
    shorten >=0.4pt},
  big arrow/.default=black}

\newcommand{\bea}{\begin{eqnarray}}
\newcommand{\eea}{\end{eqnarray}}
\newcommand{\be}{\begin{equation}}
\newcommand{\ee}{\end{equation}}
\newcommand{\bit}{\begin{itemize}}
\newcommand{\eit}{\end{itemize}}
\newcommand{\ben}{\begin{enumerate}}
\newcommand{\een}{\end{enumerate}}
\newcommand{\nn}{\nonumber}
\renewcommand{\ni}{\noindent}

\newcommand{\wt}{\widetilde}
\newcommand{\wh}{\widehat}
\newcommand{\ot}{\otimes}
\newcommand{\half}{\frac{1}{2}}

\newcommand{\mbb}{\mathbb}
\newcommand{\Z}{{\mathbb Z}}
\newcommand{\R}{{\mathbb R}}
\newcommand{\C}{{\mathbb C}}
\newcommand{\Q}{{\mathbb Q}}
\renewcommand{\P}{{\mathbb P}}

\newcommand{\cA}{\mathcal{A}}
\newcommand{\cB}{\mathcal{B}}
\newcommand{\cC}{\mathcal{C}}
\newcommand{\cD}{\mathcal{D}}
\newcommand{\cE}{\mathcal{E}}
\newcommand{\cF}{\mathcal{F}}
\newcommand{\cG}{\mathcal{G}}
\newcommand{\cH}{\mathcal{H}}
\newcommand{\cI}{\mathcal{I}}
\newcommand{\cJ}{\mathcal{J}}
\newcommand{\cK}{\mathcal{K}}
\newcommand{\cL}{\mathcal{L}}
\newcommand{\cM}{\mathcal{M}}
\newcommand{\cN}{\mathcal{N}}
\newcommand{\cO}{\mathcal{O}}
\newcommand{\cP}{\mathcal{P}}
\newcommand{\cQ}{\mathcal{Q}}
\newcommand{\cR}{\mathcal{R}}
\newcommand{\cS}{\mathcal{S}}
\newcommand{\cT}{\mathcal{T}}
\newcommand{\cU}{\mathcal{U}}
\newcommand{\cV}{\mathcal{V}}
\newcommand{\cW}{\mathcal{W}}
\newcommand{\cX}{\mathcal{X}}
\newcommand{\cY}{\mathcal{Y}}
\newcommand{\cZ}{\mathcal{Z}}

\newcommand{\F}{\mathsf{F}}
\renewcommand{\S}{\mathsf{S}}
\newcommand{\Asym}{\mathsf{\Lambda}^2}
\newcommand{\bAsym}{\bar{\mathsf{\Lambda}}^2}
\newcommand{\tAsym}{\mathsf{\Lambda}^3}
\newcommand{\btAsym}{\bar{\mathsf{\Lambda}}^3}
\newcommand{\nAsym}{\mathsf{\Lambda}^n}
\newcommand{\bnAsym}{\bar{\mathsf{\Lambda}}^n}
\newcommand{\mAsym}{\mathsf{\Lambda}^m}
\newcommand{\Sym}{\mathsf{S}^2}
\newcommand{\bSym}{\bar{\mathsf{S}}^2}
\renewcommand{\C}{\mathsf{C}}
\newcommand{\A}{\mathsf{A}}
\newcommand{\B}{\mathsf{B}}
\renewcommand{\C}{\mathsf{C}}
\newcommand{\D}{\mathsf{D}}
\renewcommand{\L}{\mathsf{\Lambda}}

\newcommand{\mf}{\mathfrak}
\newcommand{\fT}{\mathfrak{T}}
\newcommand{\fB}{\mathfrak{B}}
\newcommand{\fe}{\mathfrak{e}}
\newcommand{\ff}{\mathfrak{f}}
\newcommand{\fg}{\mathfrak{g}}
\newcommand{\fh}{\mathfrak{h}}
\newcommand{\su}{\mathfrak{su}}
\renewcommand{\sp}{\mathfrak{sp}}
\newcommand{\so}{\mathfrak{so}}
\renewcommand{\u}{\mathfrak{u}}

\newcommand{\PP}{Pin$^+$}
\newcommand{\PM}{Pin$^-$}

\newcommand{\ubf}[1]{\underline{\bf #1}}
\newcommand{\res}[1]{\left(#1\right)|_\Sigma}
\def\tr{\mathop{\mathrm{tr}}\nolimits}

\newcommand{\lra}{\leftrightarrow}
\newcommand{\llra}{\longleftrightarrow}
\newcommand{\Lra}{\longrightarrow}

\newcommand{\bF}{{\mathbb F}}
\newcommand{\dP}{\mathbf{dP}}

\title{Classification of $5d$ $\cN=1$ gauge theories}

\author{Lakshya Bhardwaj$\,^a$, Gabi Zafrir$\,^{b, c}$}

\affiliation{$^a$ Department of Physics, Harvard University\\ \hspace*{8pt}17 Oxford St, Cambridge, MA 02138, USA\\
$^b$ Kavli IPMU (WPI), UTIAS, the University of Tokyo, Kashiwa, Chiba 277-8583, Japan\\ $^c$ Dipartimento di Fisica, Universit\`a di Milano-Bicocca \& INFN, Sezione di Milano-Bicocca,\\ I-20126 Milano, Italy}

\abstract{We classify $5d$ $\cN=1$ gauge theories carrying a simple gauge group that can arise by mass-deforming $5d$ SCFTs and $6d$ SCFTs (compactified on a circle, possibly with a twist). For theories having a $6d$ UV completion, we determine the tensor branch data of the $6d$ SCFT and capture the twist in terms of the tensor branch data. We also determine the dualities between these $5d$ gauge theories, thus determining the sets of gauge theories having a common UV completion.
}

\begin{document}

\maketitle

\section{Introduction} \label{intro}

The study of five and six dimensional supersymmetric gauge theories provides an interesting window to the study of the strong coupling behavior of quantum field theory. This comes about as these theories are perturbatively non-renormalizable, yet appear to exist at low energies when interacting fixed points in these dimensions are mass-deformed. As a result, the underlying microscopic theories in these cases are intrinsically strongly coupled conformal quantum field theories, and it is hoped that a better understanding of this relation can teach us much about the strong coupling behavior of quantum field theory. Additionally, by compactifying these fixed point theories on various manifolds, many interesting theories in lower dimensions can be generated, and much of their surprising behavior elucidated, as was originally advocated in \cite{Gaiotto:2009we}. Thus, the study of higher dimensional theories also has the potential to teach us much about the behavior of lower dimensional ones.

While what was said so far may also be relevant for non-supersymmetric gauge theories, most of the study on higher dimensional gauge theories has been devoted to the supersymmetric cases, as the added supersymmetry provides us with tools that greatly facilitates this study from either the field theory or string theory directions\footnote{For a recent attempt to study a non-supersymmetric $5d$ fixed point see \cite{Genolini:2020htk}, and \cite{Peskin:1980ay,Dienes:2004rt,Morris:2004mg} for some less recent ones.}. In the case of supersymmetric five dimensional gauge theories, these were initially studied in the past from field theory \cite{Seiberg:1996bd,Morrison:1996xf,Intriligator:1997pq}, using brane systems \cite{Aharony:1997ju,Aharony:1997bh,DeWolfe:1999hik}, and from geometry using compactifications of M-theory on Calabi-Yau three folds \cite{Douglas:1996xp}. Recently the interest in this field of study has been rekindled, and much work has been done to further the study on all fronts notably from field theory \cite{Kim:2012gu,Bergman:2013rgz,Bergman:2013aca,Zafrir:2014ywa,Tachikawa:2015mha,Zafrir:2015uaa,Yonekura:2015ksa,Cremonesi:2015lsa,Jefferson:2017ahm,Ferlito:2017xdq}, using brane systems \cite{Bao:2011pty,Bergman:2014kza,Kim:2015jba,Hayashi:2015fsa,Gaiotto:2017hck,Bergman:2015dpa,Zafrir:2015rga,Hayashi:2015zka,Ohmori:2016shy,Zafrir:2015ftn,Hayashi:2015vhy,Zafrir:2016csd,Hayashi:2016abm,Hayashi:2017skf,Hayashi:2018bkd,Hayashi:2018lyv,Cabrera:2019hya,Hayashi:2019yxj,Hayashi:2020sly}, and even more recently from geometry \cite{DelZotto:2017pti,Xie:2017pfl,EJeK,EJaK1,EKY,Jefferson:2018irk,EK1,EK2,Bhardwaj:2018yhy,Bhardwaj:2018vuu,Apruzzi:2018nre,Closset:2018bjz,EJaK2,EJ,Apruzzi:2019vpe,Apruzzi:2019opn,Apruzzi:2019enx,Bhardwaj:2019jtr,Bhardwaj:2019fzv,Bhardwaj:2019ngx,Saxena:2019wuy,Bhardwaj:2019xeg,Apruzzi:2019syw,Closset:2019mdz}. 

These recent series of works addressed many questions of interest in the study of higher dimensional theories. One notable such question is the classification of five dimensional gauge theories and five dimensional SCFTs. The latter refers to the task of enumerating all $5d$ SCFTs, while the former refers to the question of which $5d$ gauge theories can be generated by a mass deformation of a $5d$ SCFT\footnote{Some clarifications appropriate for the five dimensional case are in order. For $5d$ supersymmetric gauge theories the relation is generically that there is an underlying $5d$ SCFT that can be made to flow to the gauge theory via a mass deformation. The mass deformations used are then manifested in the low-energy gauge theory as the gauge coupling constants. The interesting aspect of this relation is that it appears that many of the states in the $5d$ SCFT that were made massive by the deformation can still be recovered in the gauge theory where they appear as instantonic states. This is most notable in the study of supersymmetric partition function of these theories, notably the superconformal index, which generically forms characters of the global symmetry of the $5d$ SCFT rather than just the global symmetry of the $5d$ gauge theory, see for instance \cite{Kim:2012gu,Bergman:2013aca,Zafrir:2014ywa}. It is not clear at this moment whether or not this extends also to the non-BPS spectrum. Regardless, in this paper, when talking about the relation between $5d$ gauge theories and SCFTs, we shall mean it in this context.}, that is what are all the $5d$ gauge theories that exist as microscopic $5d$ theories. It is convenient in this context to extend the definition slightly and also allow $5d$ gauge theories that can arise via a mass-deformation of a $6d$ SCFT compactified on a circle\footnote{Here, similar clarifications as mentioned in the purely $5d$ case, also apply.}, that is to consider the full space of $5d$ gauge theories that have a UV completion as a quantum field theory. It should be noted that these two classification programs are, while related, distinct. This comes about as, first, a single $5d$ SCFT can be deformed in multiple ways so as to lead to different $5d$ gauge theories, a phenomena referred to as continuation past infinite coupling, fiber-base duality or simply as duality, see for instance \cite{Aharony:1997ju,Bao:2011pty,Bergman:2013aca,Zafrir:2014ywa}. Alternatively, there are $5d$ SCFTs that cannot be mass deformed to a $5d$ gauge theory, the first known example of which is probably the so called $E_0$ SCFT discovered in \cite{Morrison:1996xf}. Therefore, given a classification of $5d$ SCFTs, one would also need to understand their possible mass deformations in order to also get a classification of $5d$ gauge theories. Similarly, given a classification of $5d$ gauge theories, one would need to supplement this with the list of all $5d$ SCFTs without a gauge theory deformation, as well as understanding the various dualities between then in order to also get a classification of $5d$ SCFTs.

The purpose of this article is to begin an exploration of the classification of $5d$ supersymmetric gauge theories using the geometric approach. In any classification attempt some sort of strategy, or a set of simplifying assumptions is required. Unlike the case in $4d$ or $6d$ for gauge theories with the same amount of supersymmetries \cite{Bhardwaj:2013wy,Bhardwaj:2015xxa}, there is no obvious field theoretic criteria for when a $5d$ supersymmetric gauge theory possesses an SCFT UV completion. So far, the most promising criteria appear to be the ones proposed in \cite{Jefferson:2017ahm}, which are a set of constraints on the prepotential of the gauge theory. For the most part we will not have need of the explicit conditions in this article, and so would not review them here, rather reverting to mentioning several points of note. 

Depending on how a given gauge theory meets the criteria, the theory is deemed as either ruled in, ruled out or marginal. A ruled in gauge theory should have a $5d$ SCFT UV completion, a marginal theory should have a $6d$ SCFT UV completion and a ruled out should have no SCFT UV completion. It should be noted though that these criterea are thought to be be necessary, but are known to be insufficient, that is a $5d$ supersymmetric gauge theory obeying these criteria may still not have an SCFT UV completion\footnote{More accurately, it is expected that the behavior of the theory be no better than that expected from the criteria in the following order: $5d$ SCFT UV completion, $6d$ SCFT UV completion, no UV completion. In other words, a theory not obeying the conditions has no UV completion, marginal theories should have either a $6d$ SCFT UV completion or none, and ruled in theories may behave in either one of the three ways.}. Here, we shall assume that these criteria are indeed necessary and try to verify which of the gauge theories obeying these criteria indeed exist. The latter is to be accomplished using geometrical methods. As there are many possible gauge theories, we shall here concentrate on the simpler cases of gauge theories containing only a single simple gauge group. We leave open the analysis of quiver theories to future works. 

As the list of all such gauge theories obeying the criteria of \cite{Jefferson:2017ahm} were already determined in that work, all that remains for us here is to go over the list of theories and check whether these indeed have an SCFT UV completion. To do this, we analyze a local geometric setup in M-theory constructing each marginal theory appearing in \cite{Jefferson:2017ahm}. The rules for translating $5d$ gauge theories into local portions of Calabi-Yau threefolds, and vice-versa, are discussed in Section 2 of \cite{Bhardwaj:2019ngx} and in Section \ref{5dgauge} of the present paper. Performing flops and isomorphisms on this local geometric setup, it is often possible to represent the local geometry for the marginal theory in a form from which it is manifest that it describes a $6d$ SCFT compactified on a circle with a twist\footnote{As discussed in Sections \ref{6dg} and \ref{5dkk}, this form of the geometry satisfies conditions proposed in \cite{Jefferson:2018irk} which should guarantee that this local geometric piece can be shrunk and the physics associated to it be decoupled from the rest of M-theory.}. The information about the corresponding $6d$ SCFT and the twist can be read from the details of the geometry when it is represented in this form. See Sections \ref{6dg} and \ref{5dkk} for more details. Once we find that a shrinkable geometry exists for a marginal theory, we are guaranteed that the geometries for theories obtained by integrating out matter from the marginal theory will be shrinkable as well. Not only that, these geometries are guaranteed to satisfy conditions proposed in \cite{Jefferson:2018irk} which should guarantee that the corresponding geometries give rise to $5d$ SCFTs. Our results also include some $5d$ gauge theories which UV complete into $5d$ SCFTs but cannot be obtained by integrating out matter from a $5d$ KK theory. The geometries corresponding to these theories were shown to satisfy the shrinkability criteria of \cite{Jefferson:2018irk} in the recent work \cite{Bhardwaj:2019xeg}. As discussed in \cite{Bhardwaj:2019xeg}, these $5d$ SCFTs can still be obtain from $5d$ KK theories if one allows more complicated processes as compared to simple integration out of matter. Integrating out matter can be thought as integrating out BPS particles from the extended Coulomb branch of the $5d$ KK theory. A more general process involves integrating out both BPS strings and BPS particles from the extended Coulomb branch of the $5d$ KK theory. See \cite{Bhardwaj:2019xeg} for more details.

In this paper, we also uncover all the dualities between $5d$ gauge theories with a simple gauge group, having UV completions as $5d$ SCFTs and $5d$ KK theories\footnote{We use the term ``$5d$ KK theory'' to mean a $6d$ SCFT compactified on a circle possibly with some twist.}. To identify these, we use the results discussed in last paragraph and collect all the $5d$ gauge theories having UV completion into the same $5d$ KK theory. These gauge theories must be dual to each other. Identifying dualities between $5d$ gauge theories UV completing into $5d$ SCFTs requires some more work but these dualities can be obtained from dualities of $5d$ gauge theories having UV completion as KK theories\footnote{This exhausts the list of all possible dualities between $5d$ gauge theories whose UV completion is a $5d$ SCFT that can be obtained by integrating out matter from a $5d$ KK theory. This is because as pointed out in \cite{Bhardwaj:2019ngx}, once a duality between two $5d$ gauge theories is found, one can add matter to both sides of the duality and the resulting gauge theories remain dual, until we reach gauge theories having a $6d$ SCFT UV completion. For $5d$ gauge theories whose UV completion is a $5d$ SCFT that cannot be obtained by integrating out matter from a $5d$ KK theory, we find the corresponding dualities by performing operations discussed in \cite{Bhardwaj:2019ngx}.}. A duality between two gauge theories means that geometries corresponding to the two $5d$ gauge theories should be the same upto flops and isomorphisms. Since we already know all dualities between $5d$ gauge theories having a $6d$ UV completion, we find a sequence of geometric manipulations (i.e. flops and isomorphisms) taking the geometry associated to gauge theory on one side of each such duality to the geometry associated to the gauge theory on the other side of the duality. Then we integrate out matter from both sides of the duality which corresponds to blowing down the two geometries. If the sequence of geometric manipulations implementing duality is left undisturbed after the blowdown, the resulting $5d$ gauge theories are dual to each other. If the sequence of geometric manipulations is obstructed by the blowdowns, the resulting $5d$ gauge theories are not dual to each other. In this way, we find all the possible dualities between the $5d$ gauge theories having $5d$ SCFT UV completion that we consider here.

The structure of this article is as follows. In Section \ref{summary}, we collect all the results obtained in this paper in one place for the ease and convenience of the reader. Section \ref{kkr} collects all the $5d$ gauge theories having a UV completion as a $5d$ KK theory, organized according to the gauge algebras. Section \ref{scftr} collects all the $5d$ gauge theories having a UV completion as a $5d$ SCFT, organized according to the gauge algebras. Section \ref{it} collects all the $5d$ gauge theories which are allowed by the criteria of \cite{Jefferson:2017ahm} but which we can rule out based using our geometric analysis. Section \ref{ur} collects all the $5d$ gauge theories allowed by the criteria of \cite{Jefferson:2017ahm} but which we cannot rule out or rule in using our geometric analysis. Section \ref{dr} collects all the dualities between $5d$ gauge theories having UV completion either as a $5d$ KK theory or a $5d$ SCFT. Section \ref{rr} discusses the connection of our work with the classification program for $5d$ SCFTs. Section \ref{gd} describes the general features of our geometric methods in detail. Section \ref{gf} discusses general consistency conditions that all local geometries need to satisfy. Section \ref{5dgauge} discusses the structure of a geometry corresponding to a $5d$ gauge theory. Section \ref{6dg} discusses the structure of a geometry corresponding to a twisted circle compactification of a $6d$ gauge theory. Section \ref{5dkk} discusses the structure of a geometry corresponding to a $5d$ KK theory and how to read the data of the $6d$ SCFT and twist from the geometry. Section \ref{da} provides detailed arguments for the results presented in this paper, organized according to rank.

\section{Summary of results}\label{summary}

In this section we shall summarize our results for the theories, where there is evidence from geometry that they have a $5d$ or $6d$ UV completion. The generic structure is that these cases can be grouped into families, where at the top we have a gauge theory with a $6d$ SCFT UV completion, and the rest of the gauge theories in the family can be generated by integrating out matter, and have a $5d$ SCFT UV completion. It should be noted though that there are a few exceptions to this behavior \cite{Bhardwaj:2019xeg}. We shall next write down our results for the cases with a $6d$ SCFT UV completion, which we refer to as $5d$ KK theories. Cases with a $5d$ SCFT UV completion are then obtained by integrating out matter from these cases, in addition to the handful of cases that don't descend from integrating matter out of a KK gauge theory. These cases will be discussed afterward. Finally, there are a handful of cases where we were not able to determine whether the theory has an SCFT UV completion or not, and these cases will be reported at the end. We also collect theories satisfying the criteria of \cite{Jefferson:2017ahm} but which are ruled to be inconsistent by our methods.

Many of the theories we find from geometry were previously found using other methods, notably brane systems. The latter usually fall to one of two types. One is the type I$'$ string theory configuration involving a system of D$4$-branes and D$8$-branes probing an $O8^-$ background. This type of systems was originally used in \cite{Seiberg:1996bd} to realize $5d$ SCFTs, and can be generalized by the addition of an orbifold singularity \cite{Bergman:2012rgz}. The second type is brane webs \cite{Aharony:1997ju,Aharony:1997bh,DeWolfe:1999hik}, which involve a type IIB configuration of D$5$-branes, NS$5$-branes and D$7$-branes. These can be generalized by the addition of orientifold planes \cite{Brunner:1997ndk,Bergman:2015dpa,Zafrir:2015ftn,Hayashi:2015vhy}. One other method to study $5d$ SCFTs is using holography through a gravity dual. This method, however, is related to the previous one as all known holographic duals of $5d$ SCFTs are thought to be near horizon limits of one of the two types of brane systems. Notably, there is the older gravity dual of \cite{Brandhuber:1999yo}, and its orbifold generalization \cite{Bergman:2012rgz}, that are based on the type I$'$ brane system. More recently, gravity duals believed to describe the near horizon limit of $5$-brane web systems were found \cite{DHoker:2016wak,DHoker:2017prl,DHoker:2017muf}. These have since been extended to also cover brane web systems involving mutually local $7$-branes \cite{DHoker:2017gcu} and orientifold $7$-planes \cite{Uhlemann:2019ors}. Thus, in many cases having a brane realization also implies the existence of a holographic dual, though there are still types of brane systems with no known holographic dual, like ones involving mutually non-local $7$-branes or orientifold $5$-planes, at least at this point in time. We shall try here to give reference to known brane constructions when they exist, though there are also many cases with no known brane construction, or other previous realizations, and so are new. 

To enumerate the gauge theories, we shall mostly adopt the notation of \cite{Bhardwaj:2019ngx}. The gauge theories contain a single gauge group of type $G$ and a collection of $n_i$ hypermultiplets in the representation $\bold{R}_i$, where $n_i$ can be half-integers in representations where half-hypers are possible. To denote representations, we shall use the shorthand notations: $\F$ for the fundamental representation, $\L^k$ for the rank $k$ antisymmetric representation, $\S^k$ for the rank $k$ symmetric representation, and $\A$ for the adjoint representation. For $Spin$ groups, we shall also use $\S$ for the spinor representation and also $\C$ for the other spinor representation if it exists.

For KK theories, we also write down the $6d$ SCFT lift expected from the geometry. To write these we use the F-theory notation of \cite{Heckman:2013pva,Heckman:2015bfa}. Additionally, some of the reductions are done with a twist in a discrete symmetry and we use the notation introduced in \cite{Bhardwaj:2019fzv} to denote that. We turn now to a short review of this notation. The twists are denoted by how they act on the basic matter multiplets: tensors, vectors and hypers. The twist may act on the vectors as the outer automorphism of the associated gauge symmetry. To denote that, we shall use a superscript above the gauge algebra, where $(1)$ signifies a compactification without such a twist and $(2)$ or $(3)$ signify that the compactification is done with a $\Z_2$ or $\Z_3$ outer automorphism twist. The superscript $(3)$ is only used for $\so(8)$ to denote its $\Z_3$ outer automorphism, while $(2)$ denotes its $\Z_2$ outer automorphism. Additionally, the twist may act by permuting the tensor multiplets. This permutation is captured by folding the graph associated to the $6d$ SCFT according to this permutation.

For instance:
\be \nonumber
\begin{tikzpicture} [scale=1.9]
\node (v1) at (-2.7,0.8) {2};
\node at (-2.7,1.1) {$\su(m)^{(1)}$};
\node (v2) at (-3.9,0.8) {2};
\node at (-3.9,1.1) {$\su(m)^{(1)}$};
\draw  (v2) edge (v1);
\draw (v2) .. controls (-4.4,0.3) and (-3.4,0.3) .. (v2);
\end{tikzpicture}
\ee
stands for the twisted compactification of the $6d$ SCFT with tensor branch description as a linear quiver of four $SU(m)$ groups, with $m$ flavor hypermultiplet for both edge groups, where the twist acts on the quiver via a reflection. In other words, the quiver is shaped like an $A_4$ Dynkin diagram and the discrete symmetry act on it in the same way charge conjugation acts on the $A_4$ Dynkin diagram. 

As another example, consider:
\be \nonumber
\begin{tikzpicture} [scale=1.9]
\node (v1) at (-1.5,0.8) {2};
\node at (-1.5,1.1) {$\su(1)^{(1)}$};
\node (v3) at (-2.9,0.8) {2};
\node at (-2.9,1.1) {$\su(1)^{(1)}$};
\node (v2) at (-2.2,0.8) {$\cdots$};
\draw  (v2) edge (v3);
\draw  (v2) edge (v1);
\node (v4) at (-3.8,0.8) {2};
\node at (-3.8,1.1) {$\su(1)^{(1)}$};
\begin{scope}[shift={(0,0.45)}]
\node at (-2.2,-0.15) {$m-1$};
\draw (-3.1,0.15) .. controls (-3.1,0.1) and (-3.1,0.05) .. (-3,0.05);
\draw (-3,0.05) -- (-2.3,0.05);
\draw (-2.2,0) .. controls (-2.2,0.05) and (-2.25,0.05) .. (-2.3,0.05);
\draw (-2.2,0) .. controls (-2.2,0.05) and (-2.15,0.05) .. (-2.1,0.05);
\draw (-2.1,0.05) -- (-1.4,0.05);
\draw (-1.3,0.15) .. controls (-1.3,0.1) and (-1.3,0.05) .. (-1.4,0.05);
\end{scope}
\node (v5) at (-3.35,0.8) {\tiny{2}};
\draw [<-] (v4) edge (v5);
\draw  (v5) edge (v3);
\end{tikzpicture}
\ee
which denotes the twisted compactification of $D_{m+1}$ $(2,0)$ theory by its outer automorphism discrete symmetry, while
\be \nonumber
\begin{tikzpicture} [scale=1.9]
\node (v1) at (-1.5,0.8) {2};
\node at (-1.5,1.1) {$\su(1)^{(1)}$};
\node (v3) at (-2.9,0.8) {2};
\node at (-2.9,1.1) {$\su(1)^{(1)}$};
\node (v2) at (-2.2,0.8) {$\cdots$};
\draw  (v2) edge (v3);
\draw  (v2) edge (v1);
\node (v4) at (-3.8,0.8) {2};
\node at (-3.8,1.1) {$\su(1)^{(1)}$};
\begin{scope}[shift={(0,0.45)}]
\node at (-2.2,-0.15) {$m-1$};
\draw (-3.1,0.15) .. controls (-3.1,0.1) and (-3.1,0.05) .. (-3,0.05);
\draw (-3,0.05) -- (-2.3,0.05);
\draw (-2.2,0) .. controls (-2.2,0.05) and (-2.25,0.05) .. (-2.3,0.05);
\draw (-2.2,0) .. controls (-2.2,0.05) and (-2.15,0.05) .. (-2.1,0.05);
\draw (-2.1,0.05) -- (-1.4,0.05);
\draw (-1.3,0.15) .. controls (-1.3,0.1) and (-1.3,0.05) .. (-1.4,0.05);
\end{scope}
\node (v5) at (-3.35,0.8) {\tiny{2}};
\draw  (v4) edge (v5);
\draw [->] (v5) edge (v3);
\end{tikzpicture}
\ee
denotes the twisted compactification of $A_{2m-1}$ $(2,0)$ by its outer automorphism discrete symmetry. Additionally, there are cases where the twist acts as a combination of quiver reflections and outer automorphism transformations on some of the gauge groups.

\subsection{KK theories}\label{kkr}

Here we shall enumerate the cases of $5d$ gauge theories with a simple gauge group that have a $6d$ SCFT UV completion. It is convenient to break this to two cases. One are cases that exist for arbitrary rank, while the other are special cases that occur only for low rank. We shall first deal with the general cases and then move on to discuss the special cases.

\subsubsection{General Rank}

We begin with the cases that exist for generic rank. These cases include the maximally supersymmetric classical groups, as well as several $\mathcal{N}=1$ only cases. The $6d$ lifts for the maximally supersymmetric Yang-Mills cases are well known, see for instance \cite{Tachikawa:2011ch}, and the results from geometry are consistent with that. For the $\mathcal{N}=1$ only cases, the $6d$ lifts for most cases is well known, see \cite{Jefferson:2017ahm} and references within, and our geometric results are consistent with these. There are, however, a few cases that were undetermined, and the geometric methods allow us to determine them as well. We shall next list our findings for these cases based on the gauge group. 

\subsubsection*{\ubf{$\su(m)$}:}
\label{SSGRSUm}

\be \label{SUmMSSYM}
\begin{tikzpicture} [scale=1.9]
\node at (-3.6,0.9) {$=$};
\node (v1) at (-1.5,0.8) {2};
\node at (-1.5,1.1) {$\su(1)^{(1)}$};
\node (v3) at (-2.9,0.8) {2};
\node at (-2.9,1.1) {$\su(1)^{(1)}$};
\node (v2) at (-2.2,0.8) {$\cdots$};
\draw  (v2) edge (v3);
\draw  (v2) edge (v1);
\begin{scope}[shift={(0,0.45)}]
\node at (-2.2,-0.15) {$m-1$};
\draw (-3.1,0.15) .. controls (-3.1,0.1) and (-3.1,0.05) .. (-3,0.05);
\draw (-3,0.05) -- (-2.3,0.05);
\draw (-2.2,0) .. controls (-2.2,0.05) and (-2.25,0.05) .. (-2.3,0.05);
\draw (-2.2,0) .. controls (-2.2,0.05) and (-2.15,0.05) .. (-2.1,0.05);
\draw (-2.1,0.05) -- (-1.4,0.05);
\draw (-1.3,0.15) .. controls (-1.3,0.1) and (-1.3,0.05) .. (-1.4,0.05);
\end{scope}
\node at (-4.6,0.9) {$\su(m)_0+\A$};
\end{tikzpicture}
\ee

\be \label{SUmwFO}
\begin{tikzpicture} [scale=1.9]
\node at (-6.0,1.1) {$\sp(m-2)^{(1)}$};
\node (v2) at (-6.0,0.8) {1};
\node at (-8.3,0.9) {$\su(m)_0+(2m+4)\F$};
\node at (-6.9,0.9) {$=$};
\end{tikzpicture}
\ee

\be \label{SUmwFa1ASnoCS}
\begin{tikzpicture} [scale=1.9]
\node at (-5.9,0.9) {$\su(m)_0+\L^2+(m+6)\F$};
\node at (-4.3,0.9) {$=$};
\node (v2) at (-3.4,0.8) {1};
\node at (-3.4,1.1) {$\su(m-1)^{(1)}$};
\end{tikzpicture}
\ee

\be \label{SUmwFa1ASaCS}
\begin{tikzpicture} [scale=1.9]
\node (v1) at (-3.9,0.9) {2};
\node at (-3.9,1.2) {$\su(1)^{(1)}$};
\node at (-8.3,0.9) {$\su(m)_{\frac m2}+\L^2+8\F$};
\node at (-6.9,0.9) {$=$};
\node (v3) at (-5.3,0.9) {2};
\node at (-5.3,1.2) {$\su(1)^{(1)}$};
\node (v2) at (-4.6,0.9) {$\cdots$};
\draw  (v2) edge (v3);
\draw  (v2) edge (v1);
\node (v4) at (-6.2,0.9) {1};
\node at (-6.2,1.2) {$\sp(0)^{(1)}$};
\draw  (v4) edge (v3);
\begin{scope}[shift={(-2.4,-0.25)}]
\node at (-2.2,0.55) {$m-2$};
\draw (-3.1,0.85) .. controls (-3.1,0.9) and (-3.1,0.85) .. (-3,0.85);
\draw (-3,0.85) -- (-2.3,0.85);
\draw (-2.2,0.8) .. controls (-2.2,0.85) and (-2.25,0.85) .. (-2.3,0.85);
\draw (-2.2,0.8) .. controls (-2.2,0.85) and (-2.15,0.85) .. (-2.1,0.85);
\draw (-2.1,0.85) -- (-1.4,0.85);
\draw (-1.3,0.95) .. controls (-1.3,0.9) and (-1.3,0.85) .. (-1.4,0.85);
\end{scope}
\end{tikzpicture}
\ee

\be \label{SUmevenw8Fa2AS}
\begin{tikzpicture} [scale=1.9]
\node at (-5.9,0.9) {$\su(2m)_0+2\L^2+8\F$};
\node at (-4.5,0.9) {$=$};
\node (v1) at (-1.5,0.8) {2};
\node at (-1.5,1.1) {$\su(2)^{(1)}$};
\node (v3) at (-2.9,0.8) {2};
\node at (-2.9,1.1) {$\su(2)^{(1)}$};
\node (v2) at (-2.2,0.8) {$\cdots$};
\draw  (v2) edge (v3);
\draw  (v2) edge (v1);
\node (v4) at (-3.8,0.8) {1};
\node at (-3.8,1.1) {$\sp(0)^{(1)}$};
\draw  (v4) edge (v3);
\begin{scope}[shift={(0,0.45)}]
\node at (-2.2,-0.15) {$m-1$};
\draw (-3.1,0.15) .. controls (-3.1,0.1) and (-3.1,0.05) .. (-3,0.05);
\draw (-3,0.05) -- (-2.3,0.05);
\draw (-2.2,0) .. controls (-2.2,0.05) and (-2.25,0.05) .. (-2.3,0.05);
\draw (-2.2,0) .. controls (-2.2,0.05) and (-2.15,0.05) .. (-2.1,0.05);
\draw (-2.1,0.05) -- (-1.4,0.05);
\draw (-1.3,0.15) .. controls (-1.3,0.1) and (-1.3,0.05) .. (-1.4,0.05);
\end{scope}
\end{tikzpicture}
\ee

\be \label{SUmoddw8Fa2AS}
\begin{tikzpicture} [scale=1.9]
\node at (-5.9,0.9) {$\su(2m+1)_0+2\L^2+8\F$};
\node at (-4.5,0.9) {$=$};
\node (v1) at (-1.5,0.8) {2};
\node at (-1.5,1.1) {$\su(2)^{(1)}$};
\node (v3) at (-2.9,0.8) {2};
\node at (-2.9,1.1) {$\su(2)^{(1)}$};
\node (v2) at (-2.2,0.8) {$\cdots$};
\draw  (v2) edge (v3);
\draw  (v2) edge (v1);
\node (v4) at (-3.8,0.8) {1};
\node at (-3.8,1.1) {$\sp(1)^{(1)}$};
\draw  (v4) edge (v3);
\begin{scope}[shift={(0,0.45)}]
\node at (-2.2,-0.15) {$m-1$};
\draw (-3.1,0.15) .. controls (-3.1,0.1) and (-3.1,0.05) .. (-3,0.05);
\draw (-3,0.05) -- (-2.3,0.05);
\draw (-2.2,0) .. controls (-2.2,0.05) and (-2.25,0.05) .. (-2.3,0.05);
\draw (-2.2,0) .. controls (-2.2,0.05) and (-2.15,0.05) .. (-2.1,0.05);
\draw (-2.1,0.05) -- (-1.4,0.05);
\draw (-1.3,0.15) .. controls (-1.3,0.1) and (-1.3,0.05) .. (-1.4,0.05);
\end{scope}
\end{tikzpicture}
\ee

\be
\begin{tikzpicture} [scale=1.9] \label{SUmoddw7Fa2AS}
\node at (-5.9,0.9) {$\su(2m+1)_\frac32+2\L^2+7\F$};
\node at (-4.5,0.9) {$=$};
\node (v1) at (-1.5,0.8) {2};
\node at (-1.5,1.1) {$\su(2)^{(1)}$};
\node (v3) at (-2.9,0.8) {2};
\node at (-2.9,1.1) {$\su(2)^{(1)}$};
\node (v2) at (-2.2,0.8) {$\cdots$};
\draw  (v2) edge (v3);
\draw  (v2) edge (v1);
\node (v4) at (-3.8,0.8) {1};
\node at (-3.8,1.1) {$\sp(0)^{(1)}$};
\draw  (v4) edge (v3);
\begin{scope}[shift={(0,0.45)}]
\node at (-2.2,-0.15) {$m-1$};
\draw (-3.1,0.15) .. controls (-3.1,0.1) and (-3.1,0.05) .. (-3,0.05);
\draw (-3,0.05) -- (-2.3,0.05);
\draw (-2.2,0) .. controls (-2.2,0.05) and (-2.25,0.05) .. (-2.3,0.05);
\draw (-2.2,0) .. controls (-2.2,0.05) and (-2.15,0.05) .. (-2.1,0.05);
\draw (-2.1,0.05) -- (-1.4,0.05);
\draw (-1.3,0.15) .. controls (-1.3,0.1) and (-1.3,0.05) .. (-1.4,0.05);
\end{scope}
\node (v0) at (-0.6,0.8) {2};
\node at (-0.6,1.1) {$\su(1)^{(1)}$};
\draw  (v1) edge (v0);
\end{tikzpicture}
\ee

\be \label{SUmevenw7Fa2AS}
\begin{tikzpicture} [scale=1.9]
\node at (-5.9,0.9) {$\su(2m)_\frac32+2\L^2+7\F$};
\node at (-4.5,0.9) {$=$};
\node (v1) at (-1.5,0.8) {2};
\node at (-1.5,1.1) {$\su(2)^{(1)}$};
\node (v3) at (-2.9,0.8) {2};
\node at (-2.9,1.1) {$\su(2)^{(1)}$};
\node (v2) at (-2.2,0.8) {$\cdots$};
\draw  (v2) edge (v3);
\draw  (v2) edge (v1);
\node (v4) at (-3.8,0.8) {1};
\node at (-3.8,1.1) {$\sp(1)^{(1)}$};
\draw  (v4) edge (v3);
\begin{scope}[shift={(0,0.45)}]
\node at (-2.2,-0.15) {$m-2$};
\draw (-3.1,0.15) .. controls (-3.1,0.1) and (-3.1,0.05) .. (-3,0.05);
\draw (-3,0.05) -- (-2.3,0.05);
\draw (-2.2,0) .. controls (-2.2,0.05) and (-2.25,0.05) .. (-2.3,0.05);
\draw (-2.2,0) .. controls (-2.2,0.05) and (-2.15,0.05) .. (-2.1,0.05);
\draw (-2.1,0.05) -- (-1.4,0.05);
\draw (-1.3,0.15) .. controls (-1.3,0.1) and (-1.3,0.05) .. (-1.4,0.05);
\end{scope}
\node (v0) at (-0.6,0.8) {2};
\node at (-0.6,1.1) {$\su(1)^{(1)}$};
\draw  (v1) edge (v0);
\end{tikzpicture}
\ee

\be \label{SUmwFaS}
\begin{tikzpicture} [scale=1.9]
\node at (-5.9,0.9) {$\su(m)_0+\S^2+(m-2)\F$};
\node at (-4.3,0.9) {$=$};
\node (v2) at (-3.5,0.8) {2};
\node at (-3.5,1.1) {$\su(m-1)^{(1)}$};
\draw (v2) .. controls (-4,0.3) and (-3,0.3) .. (v2);
\end{tikzpicture}
\ee

\be \label{SUmevenwSaCS}
\begin{tikzpicture} [scale=1.9]
\node at (-7.7,0.9) {$\su(2m)_{m}+\S^2$};
\node (v1) at (-3.5,0.8) {2};
\node at (-3.5,1.1) {$\su(1)^{(1)}$};
\node (v3) at (-4.9,0.8) {2};
\node at (-4.9,1.1) {$\su(1)^{(1)}$};
\node (v2) at (-4.2,0.8) {$\cdots$};
\draw  (v2) edge (v3);
\draw  (v2) edge (v1);
\node (v4) at (-5.8,0.8) {2};
\node at (-5.8,1.1) {$\su(1)^{(1)}$};
\begin{scope}[shift={(0,0.45)}]
\node at (-4.2,-0.15) {$2m-2$};
\draw (-5.1,0.15) .. controls (-5.1,0.1) and (-5.1,0.05) .. (-5,0.05);
\draw (-5,0.05) -- (-4.3,0.05);
\draw (-4.2,0) .. controls (-4.2,0.05) and (-4.25,0.05) .. (-4.3,0.05);
\draw (-4.2,0) .. controls (-4.2,0.05) and (-4.15,0.05) .. (-4.1,0.05);
\draw (-4.1,0.05) -- (-3.4,0.05);
\draw (-3.3,0.15) .. controls (-3.3,0.1) and (-3.3,0.05) .. (-3.4,0.05);
\end{scope}
\node (v5) at (-5.35,0.8) {\tiny{2}};
\draw [<-] (v4) edge (v5);
\draw  (v5) edge (v3);
\node at (-6.7,0.9) {$=$};
\end{tikzpicture}
\ee

\be \label{SUmoddwFaS}
\begin{tikzpicture} [scale=1.9]
\node at (-7.5,0.9) {$\su(2m+1)_{m+\half}+\S^2$};
\node (v1) at (-3,0.8) {2};
\node at (-3,1.1) {$\su(1)^{(1)}$};
\node (v3) at (-4.4,0.8) {2};
\node at (-4.4,1.1) {$\su(1)^{(1)}$};
\node (v2) at (-3.7,0.8) {$\cdots$};
\draw  (v2) edge (v3);
\draw  (v2) edge (v1);
\node (v4) at (-5.3,0.8) {2};
\node at (-5.3,1.1) {$\su(1)^{(1)}$};
\begin{scope}[shift={(0,0.45)}]
\node at (-3.7,-0.15) {$2m-1$};
\draw (-4.6,0.15) .. controls (-4.6,0.1) and (-4.6,0.05) .. (-4.5,0.05);
\draw (-4.5,0.05) -- (-3.8,0.05);
\draw (-3.7,0) .. controls (-3.7,0.05) and (-3.75,0.05) .. (-3.8,0.05);
\draw (-3.7,0) .. controls (-3.7,0.05) and (-3.65,0.05) .. (-3.6,0.05);
\draw (-3.6,0.05) -- (-2.9,0.05);
\draw (-2.8,0.15) .. controls (-2.8,0.1) and (-2.8,0.05) .. (-2.9,0.05);
\end{scope}
\node at (-6.3,0.9) {$=$};
\draw  (v4) edge (v3);
\draw (v4) .. controls (-5.8,0.3) and (-4.8,0.3) .. (v4);
\end{tikzpicture}
\ee

\be \label{SUmevenwASaS}
\begin{tikzpicture} [scale=1.9]
\node at (-5.8,0.9) {$\su(2m)_0+\S^2+\L^2$};
\node at (-4.5,0.9) {$=$};
\node (v1) at (-1.5,0.8) {2};
\node at (-1.5,1.1) {$\su(2)^{(1)}$};
\node (v3) at (-3.1,0.8) {$\cdots$};
\node at (-2.4,1.1) {$\su(2)^{(1)}$};
\node (v2) at (-2.5,0.8) {2};
\draw  (v2) edge (v3);
\draw  (v2) edge (v1);
\node (v4) at (-3.8,0.8) {2};
\node at (-3.8,1.1) {$\su(2)^{(1)}$};
\begin{scope}[shift={(0,0.45)}]
\node at (-2.7,-0.15) {$m-1$};
\draw (-4,0.15) .. controls (-4,0.1) and (-4,0.05) .. (-3.9,0.05);
\draw (-3.9,0.05) -- (-2.8,0.05);
\draw (-2.7,0) .. controls (-2.7,0.05) and (-2.75,0.05) .. (-2.8,0.05);
\draw (-2.7,0) .. controls (-2.7,0.05) and (-2.65,0.05) .. (-2.6,0.05);
\draw (-2.6,0.05) -- (-1.4,0.05);
\draw (-1.3,0.15) .. controls (-1.3,0.1) and (-1.3,0.05) .. (-1.4,0.05);
\end{scope}
\node (v0) at (-0.6,0.8) {2};
\node at (-0.6,1.1) {$\su(1)^{(1)}$};
\draw  (v4) edge (v3);
\node (v5) at (-1.05,0.8) {\tiny{2}};
\draw  (v1) edge (v5);
\draw [->] (v5) edge (v0);
\end{tikzpicture}
\ee

\be \label{SUmoddwASaS}
\begin{tikzpicture} [scale=1.9]
\node at (-5.8,0.9) {$\su(2m+1)_0+\S^2+\L^2$};
\node at (-4.5,0.9) {$=$};
\node (v1) at (-1.5,0.8) {2};
\node at (-1.5,1.1) {$\su(2)^{(1)}$};
\node (v3) at (-2.9,0.8) {2};
\node at (-2.9,1.1) {$\su(2)^{(1)}$};
\node (v2) at (-2.2,0.8) {$\cdots$};
\draw  (v2) edge (v3);
\draw  (v2) edge (v1);
\node (v4) at (-3.8,0.8) {2};
\node at (-3.8,1.1) {$\su(2)^{(1)}$};
\begin{scope}[shift={(0,0.45)}]
\node at (-2.2,-0.15) {$m-1$};
\draw (-3.1,0.15) .. controls (-3.1,0.1) and (-3.1,0.05) .. (-3,0.05);
\draw (-3,0.05) -- (-2.3,0.05);
\draw (-2.2,0) .. controls (-2.2,0.05) and (-2.25,0.05) .. (-2.3,0.05);
\draw (-2.2,0) .. controls (-2.2,0.05) and (-2.15,0.05) .. (-2.1,0.05);
\draw (-2.1,0.05) -- (-1.4,0.05);
\draw (-1.3,0.15) .. controls (-1.3,0.1) and (-1.3,0.05) .. (-1.4,0.05);
\end{scope}
\draw  (v4) edge (v3);
\draw (v4) .. controls (-4.3,0.3) and (-3.3,0.3) .. (v4);
\end{tikzpicture}
\ee

Our results from geometry are consistent with many of the existing proposals in the literature. Specifically, case \eqref{SUmMSSYM} is just the well known relation between the $6d$ $(2,0)$ theory and $5d$ maximally supersymmetric Yang-Mills theory \cite{Douglas:2010iu,Lambert:2010iw}. Case \eqref{SUmwFO} matches the original proposal of \cite{Hayashi:2015fsa,Yonekura:2015ksa}. Likewise, case \eqref{SUmwFa1ASnoCS} matches the original proposal of \cite{Zafrir:2015rga,Hayashi:2015zka}. In cases \eqref{SUmwFa1ASaCS} and \eqref{SUmoddwASaS}, our results are consistent with the conjectures in \cite{Jefferson:2017ahm}. Cases \eqref{SUmevenw8Fa2AS}, \eqref{SUmoddw8Fa2AS}, \eqref{SUmoddw7Fa2AS} and \eqref{SUmevenw7Fa2AS} match the $6d$ lifts proposed for these theories in \cite{Zafrir:2015rga}. Finally, our results for case \eqref{SUmwFaS} matches the lift proposed for this case in \cite{Hayashi:2015vhy}. 

There are a few cases where the geometrical results improve upon the results already known in the literature. Notably, the $6d$ lift of \eqref{SUmevenwASaS} was to our knowledge not previously discussed. Our results for cases \eqref{SUmevenwSaCS} and \eqref{SUmoddwFaS} are consistent with the results found in \cite{Jefferson:2017ahm} for the case of $\su(3)$. However, it was conjectured there, based on this case, that the $6d$ lift for higher $m$ is also a twisted compactification of an $A$ type $(2,0)$ theory, while the geometric methods reveal that this is only true for odd $m$, and the even $m$ cases lift to a twisted compactification of a $D$ type $(2,0)$ theory instead. See Section \ref{duality}.

\subsubsection*{\ubf{$\sp(m)$}:}

\be \label{SpmMSSYM0}
\begin{tikzpicture} [scale=1.9]
\node at (-4.5,0.9) {$=$};
\node (v1) at (-1.5,0.8) {2};
\node at (-1.5,1.1) {$\su(1)^{(1)}$};
\node (v3) at (-2.9,0.8) {2};
\node at (-2.9,1.1) {$\su(1)^{(1)}$};
\node (v2) at (-2.2,0.8) {$\cdots$};
\draw  (v2) edge (v3);
\draw  (v2) edge (v1);
\node (v4) at (-3.8,0.8) {2};
\node at (-3.8,1.1) {$\su(1)^{(1)}$};
\begin{scope}[shift={(0,0.45)}]
\node at (-2.2,-0.15) {$m-1$};
\draw (-3.1,0.15) .. controls (-3.1,0.1) and (-3.1,0.05) .. (-3,0.05);
\draw (-3,0.05) -- (-2.3,0.05);
\draw (-2.2,0) .. controls (-2.2,0.05) and (-2.25,0.05) .. (-2.3,0.05);
\draw (-2.2,0) .. controls (-2.2,0.05) and (-2.15,0.05) .. (-2.1,0.05);
\draw (-2.1,0.05) -- (-1.4,0.05);
\draw (-1.3,0.15) .. controls (-1.3,0.1) and (-1.3,0.05) .. (-1.4,0.05);
\end{scope}
\node (v5) at (-3.35,0.8) {\tiny{2}};
\draw [<-] (v4) edge (v5);
\draw  (v5) edge (v3);
\node at (-5.4,0.9) {$\sp(m)_0+\A$};
\end{tikzpicture}
\ee

\be \label{SpmMSSYMpi}
\begin{tikzpicture} [scale=1.9]
\node at (-4.5,0.9) {$=$};
\node (v1) at (-1.5,0.8) {2};
\node at (-1.5,1.1) {$\su(1)^{(1)}$};
\node (v3) at (-2.9,0.8) {2};
\node at (-2.9,1.1) {$\su(1)^{(1)}$};
\node (v2) at (-2.2,0.8) {$\cdots$};
\draw  (v2) edge (v3);
\draw  (v2) edge (v1);
\node (v4) at (-3.8,0.8) {2};
\node at (-3.8,1.1) {$\su(1)^{(1)}$};
\begin{scope}[shift={(0,0.45)}]
\node at (-2.2,-0.15) {$m-1$};
\draw (-3.1,0.15) .. controls (-3.1,0.1) and (-3.1,0.05) .. (-3,0.05);
\draw (-3,0.05) -- (-2.3,0.05);
\draw (-2.2,0) .. controls (-2.2,0.05) and (-2.25,0.05) .. (-2.3,0.05);
\draw (-2.2,0) .. controls (-2.2,0.05) and (-2.15,0.05) .. (-2.1,0.05);
\draw (-2.1,0.05) -- (-1.4,0.05);
\draw (-1.3,0.15) .. controls (-1.3,0.1) and (-1.3,0.05) .. (-1.4,0.05);
\end{scope}
\node at (-5.6,0.9) {$\sp(m)_\pi+\A$};
\draw  (v4) edge (v3);
\draw (v4) .. controls (-4.3,0.3) and (-3.3,0.3) .. (v4);
\end{tikzpicture}
\ee

\be \label{SpmwF} 
\begin{tikzpicture} [scale=1.9]
\node at (-5.6,0.9) {$\sp(m)+(2m+6)\F$};
\node at (-4.3,0.9) {$=$};
\node (v2) at (-3.5,0.8) {1};
\node at (-3.5,1.1) {$\sp(m-1)^{(1)}$};
\end{tikzpicture}
\ee

\be \label{SpmwFaAS}
\begin{tikzpicture} [scale=1.9]
\node (v1) at (-3.9,0.9) {2};
\node at (-3.9,1.2) {$\su(1)^{(1)}$};
\node at (-8.3,0.9) {$\sp(m)+\L^2+8\F$};
\node at (-7.1,0.9) {$=$};
\node (v3) at (-5.3,0.9) {2};
\node at (-5.3,1.2) {$\su(1)^{(1)}$};
\node (v2) at (-4.6,0.9) {$\cdots$};
\draw  (v2) edge (v3);
\draw  (v2) edge (v1);
\node (v4) at (-6.2,0.9) {1};
\node at (-6.2,1.2) {$\sp(0)^{(1)}$};
\draw  (v4) edge (v3);
\begin{scope}[shift={(-2.4,-0.25)}]
\node at (-2.2,0.55) {$m-1$};
\draw (-3.1,0.85) .. controls (-3.1,0.9) and (-3.1,0.85) .. (-3,0.85);
\draw (-3,0.85) -- (-2.3,0.85);
\draw (-2.2,0.8) .. controls (-2.2,0.85) and (-2.25,0.85) .. (-2.3,0.85);
\draw (-2.2,0.8) .. controls (-2.2,0.85) and (-2.15,0.85) .. (-2.1,0.85);
\draw (-2.1,0.85) -- (-1.4,0.85);
\draw (-1.3,0.95) .. controls (-1.3,0.9) and (-1.3,0.85) .. (-1.4,0.85);
\end{scope}
\end{tikzpicture}
\ee

For $\sp$ groups the $6d$ lifts were all previously known in the literature, and our results are consistent with this. Specifically, cases \eqref{SpmMSSYM0} and \eqref{SpmMSSYMpi} are the $5d$ maximally supersymmetric $\sp$ Yang-Mills theories \cite{Tachikawa:2011ch}. Case \eqref{SpmwF} matches the known lift in \cite{Hayashi:2015zka}. Finally, that case \eqref{SpmwFaAS} lifts to the rank $m$ E-string theory is well known from the work of \cite{Ganor:1996pc}. We also note that all the cases here are dual to cases in the $\su$ part. 

\subsubsection*{\ubf{$\so(m)$}:}

\be \label{SOmoddMSSYM}
\begin{tikzpicture} [scale=1.9]
\node at (-4.5,0.9) {$=$};
\node (v1) at (-1.5,0.8) {2};
\node at (-1.5,1.1) {$\su(1)^{(1)}$};
\node (v3) at (-2.9,0.8) {2};
\node at (-2.9,1.1) {$\su(1)^{(1)}$};
\node (v2) at (-2.2,0.8) {$\cdots$};
\draw  (v2) edge (v3);
\draw  (v2) edge (v1);
\node (v4) at (-3.8,0.8) {2};
\node at (-3.8,1.1) {$\su(1)^{(1)}$};
\begin{scope}[shift={(0,0.45)}]
\node at (-2.2,-0.15) {$m-1$};
\draw (-3.1,0.15) .. controls (-3.1,0.1) and (-3.1,0.05) .. (-3,0.05);
\draw (-3,0.05) -- (-2.3,0.05);
\draw (-2.2,0) .. controls (-2.2,0.05) and (-2.25,0.05) .. (-2.3,0.05);
\draw (-2.2,0) .. controls (-2.2,0.05) and (-2.15,0.05) .. (-2.1,0.05);
\draw (-2.1,0.05) -- (-1.4,0.05);
\draw (-1.3,0.15) .. controls (-1.3,0.1) and (-1.3,0.05) .. (-1.4,0.05);
\end{scope}
\node (v5) at (-3.35,0.8) {\tiny{2}};
\draw  (v4) edge (v5);
\draw [->] (v5) edge (v3);
\node at (-5.5,0.9) {$\so(2m+1)+\A$};
\end{tikzpicture}
\ee

\be \label{SOmevenMSSYM}
\begin{tikzpicture} [scale=1.9]
\node at (-4.5,0.9) {$=$};
\node (v1) at (-1.5,0.8) {2};
\node at (-1.5,1.1) {$\su(1)^{(1)}$};
\node (v3) at (-2.9,0.8) {2};
\node (v6) at (-2.9,1.1) {$\su(1)^{(1)}$};
\node (v2) at (-2.2,0.8) {$\cdots$};
\draw  (v2) edge (v3);
\draw  (v2) edge (v1);
\node (v4) at (-3.8,0.8) {2};
\node at (-3.8,1.1) {$\su(1)^{(1)}$};
\begin{scope}[shift={(0,0.45)}]
\node at (-2.2,-0.15) {$m-2$};
\draw (-3.1,0.15) .. controls (-3.1,0.1) and (-3.1,0.05) .. (-3,0.05);
\draw (-3,0.05) -- (-2.3,0.05);
\draw (-2.2,0) .. controls (-2.2,0.05) and (-2.25,0.05) .. (-2.3,0.05);
\draw (-2.2,0) .. controls (-2.2,0.05) and (-2.15,0.05) .. (-2.1,0.05);
\draw (-2.1,0.05) -- (-1.4,0.05);
\draw (-1.3,0.15) .. controls (-1.3,0.1) and (-1.3,0.05) .. (-1.4,0.05);
\end{scope}
\node at (-5.5,0.9) {$\so(2m)+\A$};
\draw  (v4) edge (v3);
\node (v5) at (-2.9,1.8) {2};
\node at (-2.9,2.1) {$\su(1)^{(1)}$};
\draw  (v5) edge (v6);
\end{tikzpicture}
\ee

\be \label{SOmwV}
\begin{tikzpicture} [scale=1.9]
\node at (-5.4,0.9) {$\so(m)+(m-2)\F$};
\node at (-4.3,0.9) {$=$};
\node (v2) at (-3.5,0.8) {2};
\node at (-3.5,1.1) {$\su(m-2)^{(2)}$};
\end{tikzpicture}
\ee

The lifts for generic $\so$ cases were, also, all previously known in the literature, and our results are consistent with this. Specifically, cases \eqref{SOmoddMSSYM} and \eqref{SOmevenMSSYM} are the $5d$ maximally supersymmetric $\so$ Yang-Mills theories \cite{Tachikawa:2011ch}, and case \eqref{SOmwV} matches the $6d$ lift proposed for this case in \cite{Hayashi:2015vhy}. 

Next we are going to consider the cases that only exist for small rank.

\subsubsection{Rank 2}

\subsubsection*{\ubf{$\su(3)$}:}

\be \label{SU3CS7L}
\begin{tikzpicture} [scale=1.9]
\node (v1) at (-5.9,0.8) {1};
\node at (-5.9,1.1) {$\su(3)^{(2)}$};
\node at (-7.7,0.9) {$\su(3)_4+6\F$};
\node at (-6.6,0.9) {$=$};
\end{tikzpicture}
\ee

\be \label{SU3CS8L}
\begin{tikzpicture} [scale=1.9]
\node at (-5.7,0.9) {$=$};
\node (v1) at (-5,0.8) {2};
\node at (-5,1.1) {$\su(1)^{(1)}$};
\node at (-6.4,0.9) {$\su(3)_\frac{15}2+\F$};
\node (v2) at (-4,0.8) {2};
\node at (-4,1.1) {$\su(1)^{(1)}$};
\node (v3) at (-4.5,0.8) {\tiny{3}};
\draw  (v1) edge (v3);
\draw[->]  (v3) edge (v2);
\end{tikzpicture}
\ee

\be \label{SU3CS9L}
\begin{tikzpicture} [scale=1.9]
\node at (-6,0.9) {$\su(3)_9$};
\node at (-5.3,0.9) {$=$};
\node at (-4.6,0.8) {3};
\node at (-4.6,1.1) {$\su(3)^{(2)}$};
\end{tikzpicture}
\ee

All these cases have already appeared in the literature. Specifically, cases \eqref{SU3CS7L} and \eqref{SU3CS9L} were originally discovered from geometry in \cite{Jefferson:2018irk}, with brane realizations following in \cite{Hayashi:2018lyv}. Case \eqref{SU3CS8L} was found more recently in \cite{Bhardwaj:2019jtr}, also from geometry.

\subsubsection*{\ubf{$\sp(2)$}:}

\be \label{SP2w2AS}
\begin{tikzpicture} [scale=1.9]
\node at (-7.7,0.9) {$\sp(2)+2\L^2+4\F$};
\node (v1) at (-5.9,0.8) {1};
\node at (-5.9,1.1) {$\su(3)^{(2)}$};
\node at (-6.6,0.9) {$=$};
\end{tikzpicture}
\ee

\be \label{SP2w3AS}
\begin{tikzpicture} [scale=1.9]
\node at (-6.2,0.9) {$\sp(2)_0+3\L^2$};
\node at (-5.3,0.9) {$=$};
\node at (-4.6,0.8) {2};
\node at (-4.6,1.1) {$\su(3)^{(2)}$};
\end{tikzpicture}
\ee

Here as well all cases have already appeared in the literature. Specifically, case \eqref{SP2w2AS} is dual to \eqref{SU3CS7L}, a result originally found in \cite{Jefferson:2018irk}. Since $\sp(2)=\so(5)$, case \eqref{SP2w3AS} is in fact just the $m=5$ case of \eqref{SOmwV}. However, we have here separated it as for this case there is also the possibility to turn on a theta angle for the $\sp(2)$. The $m=5$ case of \eqref{SOmwV} is then the one with $\theta=0$, while the $\theta=\pi$ case appears to have no SCFT UV completion. This was first noted in the geometrical work of \cite{Jefferson:2018irk}, and our results further support this as they suggest that its associated geometry is dual to $\fg_2+\A+2\F$ which is not a UV complete theory as it has more matter than the KK theory $\fg_2+\A$.

\subsubsection*{\ubf{$\fg_2$}:}

\be \label{G2MSSYM}
\begin{tikzpicture} [scale=1.9]
\node at (-7.4,0.9) {$\fg_2+\A$};
\node at (-6.7,0.9) {$=$};
\node (v1) at (-6,0.8) {2};
\node at (-6,1.1) {$\su(1)^{(1)}$};
\node (v2) at (-5,0.8) {2};
\node at (-5,1.1) {$\su(1)^{(1)}$};
\node (v3) at (-5.5,0.8) {\tiny{3}};
\draw  (v1) edge (v3);
\draw[->]  (v3) edge (v2);
\end{tikzpicture}
\ee

\be \label{G2wF}
\begin{tikzpicture} [scale=1.9]
\node at (-7.2,0.9) {$=$};
\node (v1) at (-6.5,0.8) {1};
\node at (-6.5,1.1) {$\su(3)^{(2)}$};
\node at (-7.9,0.9) {$\fg_2+6\F$};
\end{tikzpicture}
\ee

Case \eqref{G2MSSYM} is again one of the $5d$ maximally supersymmetric Yang-Mills theories, whose $6d$ lift was worked out previously \cite{Tachikawa:2011ch}. Case \eqref{G2wF} is also dual to \eqref{SU3CS7L} and \eqref{SP2w2AS}, a result originally found in \cite{Jefferson:2018irk}.

\subsubsection{Rank 3}

\subsubsection*{\ubf{$\su(4)$}:}

\be \label{SU4CS7L}
\begin{tikzpicture} [scale=1.9]
\node at (-6.1,0.9) {$=$};
\node (v1) at (-5.4,0.8) {3};
\node at (-5.4,1.1) {$\su(3)^{(2)}$};
\node at (-7,0.9) {$\su(4)_4+6\F$};
\node (v2) at (-4.4,0.8) {1};
\node at (-4.4,1.1) {$\sp(0)^{(1)}$};
\node (v3) at (-4.9,0.8) {\tiny{2}};
\draw  (v1) edge (v3);
\draw[->]  (v3) edge (v2);
\end{tikzpicture}
\ee

\be \label{SU4CS8L}
\begin{tikzpicture} [scale=1.9]
\node at (-5.7,0.9) {$\su(4)_8$};
\node at (-5,0.9) {$=$};
\node (v1) at (-4.3,0.8) {4};
\node at (-4.3,1.1) {$\so(8)^{(3)}$};
\end{tikzpicture}
\ee

\be \label{SU4w2ASaCS}
\begin{tikzpicture} [scale=1.9]
\node at (-5.5,0.8) {2};
\node at (-5.5,1.1) {$\so(8)^{(3)}$};
\node at (-7.2,0.9) {$\su(4)_6+2\L^2$};
\node at (-6.2,0.9) {$=$};
\end{tikzpicture}
\ee

\be \label{SU4w3ASaFa0CS}
\begin{tikzpicture} [scale=1.9]
\node at (-6.3,0.9) {$=$};
\node at (-5.5,0.8) {1};
\node at (-5.5,1.1) {$\su(4)^{(2)}$};
\node at (-7.5,0.9) {$\su(4)_0+3\L^2+4\F$};
\end{tikzpicture}
\ee

\be \label{SU4w3ASaFa1CS}
\begin{tikzpicture} [scale=1.9]
\node at (-6.3,0.9) {$=$};
\node at (-5.5,0.8) {1};
\node at (-5.5,1.1) {$\fg_2^{(1)}$};
\node at (-7.5,0.9) {$\su(4)_1+3\L^2+4\F$};
\end{tikzpicture}
\ee

\be \label{SU4w3ASaFa2CS}
\begin{tikzpicture} [scale=1.9]
\node at (-7.5,0.9) {$\su(4)_2+3\L^2+4\F$};
\node at (-6.3,0.9) {$=$};
\node (v1) at (-5.5,0.8) {1};
\node at (-5.5,1.1) {$\sp(0)^{(1)}$};
\node (v2) at (-4.5,0.8) {2};
\node at (-4.5,1.1) {$\su(3)^{(2)}$};
\draw  (v1) edge (v2);
\end{tikzpicture}
\ee

\be \label{SU4w3ASaCS}
\begin{tikzpicture} [scale=1.9]
\node at (-7.4,0.9) {$\su(4)_5+3\L^2$};
\node at (-6.3,0.9) {$=$};
\node (v1) at (-5.5,0.8) {1};
\node at (-5.5,1.1) {$\so(8)^{(3)}$};
\end{tikzpicture}
\ee

\be \label{SU4w4ASaCS0}
\begin{tikzpicture} [scale=1.9]
\node at (-5.8,0.9) {$\su(4)_0+4\L^2$};
\node at (-4.9,0.9) {$=$};
\node (v1) at (-4.3,0.8) {2};
\node at (-4.3,1.1) {$\su(4)^{(2)}$};
\end{tikzpicture}
\ee

\be \label{SU4w4ASaCS2}
\begin{tikzpicture} [scale=1.9]
\node at (-5.8,0.9) {$\su(4)_2+4\L^2$};
\node at (-4.9,0.9) {$=$};
\node (v1) at (-4.3,0.8) {2};
\node at (-4.3,1.1) {$\fg_2^{(1)}$};
\end{tikzpicture}
\ee

\be \label{SU4w4ASaCS4}
\begin{tikzpicture} [scale=1.9]
\node at (-5.8,0.9) {$\su(4)_4+4\L^2$};
\node at (-4.9,0.9) {$=$};
\node (v1) at (-4.2,0.8) {1};
\node at (-4.2,1.1) {$\sp(0)^{(1)}$};
\node (v2) at (-3.2,0.8) {3};
\node at (-3.2,1.1) {$\su(3)^{(2)}$};
\draw  (v1) edge (v2);
\end{tikzpicture}
\ee

Most of the cases here are, to our knowledge, new. The notable exceptions are cases \eqref{SU4CS8L} and \eqref{SU4w4ASaCS0}. Case \eqref{SU4w4ASaCS0} is just the $m=6$ case of \eqref{SOmwV}, though we have singled it out here since for this case there is also the possibility of a Chern-Simons level, and the case fitting in \eqref{SOmwV} is the one with Chern-Simons level zero. The lift for case \eqref{SU4CS8L} was conjectured in \cite{Razamat:2018gro}, and our results are consistent with the conjecture there. Note also that while all the reductions in this section were discussed in \cite{Bhardwaj:2019jtr}, they were not given a gauge theory interpretation there. 

\subsubsection*{\ubf{$\sp(3)$}:}

\be \label{SP3w2AS}
\begin{tikzpicture} [scale=1.9]
\node at (-7.2,0.9) {$\sp(3)_0+2\L^2$};
\node at (-6.3,0.9) {$=$};
\node at (-5.6,0.8) {2};
\node at (-5.6,1.1) {$\so(8)^{(3)}$};
\end{tikzpicture}
\ee

\be \label{SP3wHTASaF}
\begin{tikzpicture} [scale=1.9]
\node at (-5.4,0.9) {$=$};
\node (v1) at (-4.7,0.8) {1};
\node at (-4.7,1.1) {$\sp(1)^{(1)}$};
\node (v2) at (-3.7,0.8) {2};
\node at (-3.7,1.1) {$\su(1)^{(1)}$};
\draw  (v1) edge (v2);
\node at (-6.5,0.9) {$\sp(3)+\frac12\L^3+\frac{19}2\F$};
\end{tikzpicture}
\ee

\be \label{SP3wHTASaAS}
\begin{tikzpicture} [scale=1.9]
\node at (-5.5,0.9) {$=$};
\node (v1) at (-4.8,0.8) {1};
\node at (-4.8,1.1) {$\so(8)^{(3)}$};
\node at (-6.8,0.9) {$\sp(3)+\frac12\L^3+\L^2+\frac{5}2\F$};
\end{tikzpicture}
\ee

\be \label{SP3wTASaF}
\begin{tikzpicture} [scale=1.9]
\node at (-6.6,0.9) {$\sp(3)+\L^3+5\F$};
\node at (-5.6,0.9) {$=$};
\node (v1) at (-4.9,0.8) {3};
\node at (-4.9,1.1) {$\su(3)^{(2)}$};
\node (v2) at (-3.9,0.8) {1};
\node at (-3.9,1.1) {$\sp(0)^{(1)}$};
\node (v3) at (-4.4,0.8) {\tiny{2}};
\draw  (v1) edge (v3);
\draw[->]  (v3) edge (v2);
\end{tikzpicture}
\ee

Here, cases \eqref{SP3w2AS} and \eqref{SP3wHTASaAS} are new. In cases \eqref{SP3wHTASaF} and \eqref{SP3wTASaF}, a brane web was found in \cite{Hayashi:2019yxj} from which it was suggested that these cases lift to $6d$, though the exact $6d$ SCFTs they lift to were not determined. We also note that all cases here are dual to various $\su(4)$ cases. See Section \ref{duality}.

\subsubsection*{\ubf{$\so(7)$}:}

\be \label{SO7w6Sa1V}
\begin{tikzpicture} [scale=1.9]
\node at (-7.3,0.9) {$\so(7)+6\S+\F$};
\node at (-6.3,0.9) {$=$};
\node at (-5.6,0.8) {1};
\node at (-5.6,1.1) {$\su(4)^{(2)}$};
\end{tikzpicture}
\ee

\be \label{SO7w5Sa2V}
\begin{tikzpicture} [scale=1.9]
\node at (-7.9,0.9) {$\so(7)+5\S+2\F$};
\node at (-6.1,0.8) {1};
\node at (-6.1,1.1) {$\fg_2^{(1)}$};
\node at (-6.9,0.9) {$=$};
\end{tikzpicture}
\ee

\be \label{SO7w4Sa3V}
\begin{tikzpicture} [scale=1.9]
\node at (-5.4,0.9) {$=$};
\node (v1) at (-4.7,0.8) {1};
\node at (-4.7,1.1) {$\sp(0)^{(1)}$};
\node (v2) at (-3.7,0.8) {2};
\node at (-3.7,1.1) {$\su(3)^{(2)}$};
\draw  (v1) edge (v2);
\node at (-6.4,0.9) {$\so(7)+4\S+3\F$};
\end{tikzpicture}
\ee

\be \label{SO7w2Sa4V}
\begin{tikzpicture} [scale=1.9]
\node at (-5.9,0.9) {$\so(7)+2\S+4\F$};
\node at (-4.9,0.9) {$=$};
\node (v1) at (-4.2,0.8) {1};
\node at (-4.2,1.1) {$\su(5)^{(2)}$};
\end{tikzpicture}
\ee

\be \label{SO7w7S}
\begin{tikzpicture} [scale=1.9]
\node at (-6.3,0.9) {$=$};
\node at (-5.7,0.8) {1};
\node at (-5.7,1.1) {$\fg_2^{(1)}$};
\node at (-7.1,0.9) {$\so(7)+7\S$};
\end{tikzpicture}
\ee

The $6d$ lifts for some of the cases here were already considered in the literature, but there are new cases as well. Notably, the $6d$ lifts for cases \eqref{SO7w6Sa1V} and \eqref{SO7w4Sa3V} were conjectured by \cite{Razamat:2018gbu}, and our findings from geometry support these conjectures. Cases \eqref{SO7w5Sa2V}, \eqref{SO7w2Sa4V} and \eqref{SO7w7S} are, to our knowledge, new. We also note that the cases \eqref{SO7w5Sa2V} and \eqref{SO7w7S} are dual to each-other. See Section \ref{duality} for other dualities.

\subsubsection{Rank 4}

\subsubsection*{\ubf{$\su(5)$}:}

\be \label{SU5w3ASa3F}
\begin{tikzpicture} [scale=1.9]
\node at (-8.3,0.9) {$\su(5)_0+3\L^2+3\F$};
\node (v1) at (-6.2,0.8) {1};
\node at (-6.2,1.1) {$\so(8)^{(2)}$};
\node at (-7.2,0.9) {$=$};
\end{tikzpicture}
\ee

\be \label{SU5w3ASa2F}
\begin{tikzpicture} [scale=1.9]
\node (v1) at (-5.9,0.8) {2};
\node at (-5.9,1.1) {$\su(1)^{(1)}$};
\node at (-8.1,0.9) {$\su(5)_{\frac{3}2}+3\L^2+2\F$};
\node at (-6.9,0.9) {$=$};
\node (v2) at (-4.9,0.8) {1};
\node at (-4.9,1.1) {$\sp(0)^{(1)}$};
\draw  (v1) edge (v2);
\node (v3) at (-3.9,0.8) {3};
\node at (-3.9,1.1) {$\su(3)^{(2)}$};
\draw  (v2) edge (v3);
\end{tikzpicture}
\ee

Both cases are, to our knowledge, new.

\subsubsection*{\ubf{$\sp(4)$}:}

\be \label{SP4wTAS}
\begin{tikzpicture} [scale=1.9]
\node at (-6.2,0.9) {$\sp(4)+\half\L^3+4\F$};
\node at (-5.1,0.9) {$=$};
\node (v1) at (-4.4,0.8) {4};
\node at (-4.4,1.1) {$\so(8)^{(3)}$};
\node (v2) at (-3.4,0.8) {1};
\node at (-3.4,1.1) {$\sp(0)^{(1)}$};
\node (v3) at (-3.9,0.8) {\tiny{3}};
\draw  (v1) edge (v3);
\draw[->]  (v3) edge (v2);
\end{tikzpicture}
\ee

This case is, to our knowledge, new.

\subsubsection*{\ubf{$\so(8)$}:}

\be \label{SO8w4Sa4F}
\begin{tikzpicture} [scale=1.9]
\node at (-5.9,0.9) {$\so(8)+4\S+4\F$};
\node at (-4.9,0.9) {$=$};
\node (v1) at (-4.2,0.8) {1};
\node at (-4.2,1.1) {$\sp(0)^{(1)}$};
\node (v2) at (-3.2,0.8) {3};
\node at (-3.2,1.1) {$\su(3)^{(2)}$};
\draw  (v1) edge (v2);
\node (v3) at (-2.2,0.8) {1};
\node at (-2.2,1.1) {$\sp(0)^{(1)}$};
\draw  (v2) edge (v3);
\end{tikzpicture}
\ee

\be \label{SO8w2Sa5F}
\begin{tikzpicture} [scale=1.9]
\node at (-5.2,0.9) {$=$};
\node (v1) at (-4.5,0.8) {1};
\node at (-4.5,1.1) {$\su(\tilde 6)^{(2)}$};
\node at (-6.2,0.9) {$\so(8)+2\S+5\F$};
\end{tikzpicture}
\ee

\be \label{SO8w3Sa1Ca4F}
\begin{tikzpicture} [scale=1.9]
\node at (-6.3,0.9) {$\so(8)+3\S+\C+4\F$};
\node at (-5.1,0.9) {$=$};
\node (v1) at (-4.5,0.8) {2};
\node at (-4.5,1.1) {$\fg_2^{(1)}$};
\node (v2) at (-3.5,0.8) {1};
\node at (-3.5,1.1) {$\sp(0)^{(1)}$};
\draw  (v1) edge (v2);
\end{tikzpicture}
\ee

\be \label{SO8w1Sa1Ca5F}
\begin{tikzpicture} [scale=1.9]
\node at (-5.2,0.9) {$=$};
\node (v1) at (-4.5,0.8) {1};
\node at (-4.5,1.1) {$\su(6)^{(2)}$};
\node at (-6.4,0.9) {$\so(8)+\S+\C+5\F$};
\end{tikzpicture}
\ee

\be \label{SO8w3Sa2Ca3F}
\begin{tikzpicture} [scale=1.9]
\node at (-5.2,0.9) {$=$};
\node (v1) at (-4.5,0.8) {1};
\node at (-4.5,1.1) {$\so(7)^{(1)}$};
\node at (-6.5,0.9) {$\so(8)+3\S+2\C+3\F$};
\end{tikzpicture}
\ee

\be \label{SO8w2Sa2Ca4F}
\begin{tikzpicture} [scale=1.9]
\node at (-6.4,0.9) {$\so(8)+2\S+2\C+4\F$};
\node at (-5.1,0.9) {$=$};
\node (v1) at (-4.4,0.8) {2};
\node at (-4.4,1.1) {$\su(4)^{(2)}$};
\node (v2) at (-3.4,0.8) {1};
\node at (-3.4,1.1) {$\sp(0)^{(1)}$};
\draw  (v1) edge (v2);
\end{tikzpicture}
\ee

The $6d$ lifts for some of the cases here were already considered in the literature. Specifically, the $6d$ lift for case \eqref{SO8w2Sa2Ca4F} was conjectured by \cite{Razamat:2018gbu} and for case \eqref{SO8w4Sa4F} by \cite{Sela:2019gbz}. In both cases our findings from geometry support these conjectures. The rest of the cases are, to our knowledge, new. The case \eqref{SO8w2Sa5F} lifts to the outer-automorphism twisted compactification of $6d$ SCFT whose tensor branch is described by the $6d$ gauge theory $\su(6)+\half\L^3+15\F$. The tilde on top of $\su(6)$ differentiates this $6d$ SCFT from the cousin $6d$ SCFT whose tensor branch is described by the $6d$ gauge theory $\su(6)+\L^2+14\F$.

\subsubsection*{\ubf{$\so(9)$}:}

\be \label{SO9w1Sa6F}
\begin{tikzpicture} [scale=1.9]
\node at (-5.2,0.9) {$=$};
\node (v1) at (-4.5,0.8) {1};
\node at (-4.5,1.1) {$\su(7)^{(2)}$};
\node at (-6.2,0.9) {$\so(9)+\S+6\F$};
\end{tikzpicture}
\ee

\be \label{SO9w2Sa5F}
\begin{tikzpicture} [scale=1.9]
\node at (-6.1,0.9) {$\so(9)+2\S+5\F$};
\node at (-5.1,0.9) {$=$};
\node (v1) at (-4.4,0.8) {1};
\node at (-4.4,1.1) {$\sp(0)^{(1)}$};
\node (v2) at (-3.4,0.8) {2};
\node at (-3.4,1.1) {$\su(5)^{(2)}$};
\draw  (v1) edge (v2);
\end{tikzpicture}
\ee

\be \label{SO9w3Sa3F}
\begin{tikzpicture} [scale=1.9]
\node at (-6.2,0.9) {$=$};
\node (v1) at (-5.5,0.8) {1};
\node at (-5.5,1.1) {$\so(8)^{(2)}$};
\node at (-7.2,0.9) {$\so(9)+3\S+3\F$};
\end{tikzpicture}
\ee

\be \label{SO9w4Sa1F}
\begin{tikzpicture} [scale=1.9]
\node at (-6.9,0.9) {$\so(9)+4\S+\F$};
\node at (-5.9,0.9) {$=$};
\node (v1) at (-5.2,0.8) {2};
\node at (-5.2,1.1) {$\su(1)^{(1)}$};
\node (v2) at (-4.2,0.8) {1};
\node at (-4.2,1.1) {$\sp(0)^{(1)}$};
\draw  (v1) edge (v2);
\node (v3) at (-3.2,0.8) {3};
\node at (-3.2,1.1) {$\su(3)^{(2)}$};
\draw  (v2) edge (v3);
\end{tikzpicture}
\ee

The $6d$ lifts for some of the cases here were already considered in the literature, but there are new cases as well. Notably, the $6d$ lift for case \eqref{SO9w2Sa5F} was conjectured by \cite{Razamat:2018gbu}, and for case \eqref{SO9w4Sa1F} by \cite{Sela:2019gbz}. In both cases our findings from geometry support these conjectures. The remaining cases are, to our knowledge, new.

\subsubsection*{\ubf{$\ff_4$}:}

\be
\begin{tikzpicture} [scale=1.9]
\node at (-6.1,0.9) {$\ff_4+\A$};
\node at (-5.4,0.9) {$=$};
\node (v3) at (-2.9,0.8) {2};
\node at (-2.9,1.1) {$\su(1)^{(1)}$};
\node (v4) at (-3.8,0.8) {2};
\node at (-3.8,1.1) {$\su(1)^{(1)}$};
\node (v5) at (-3.35,0.8) {\tiny{2}};
\draw  (v4) edge (v5);
\draw [->] (v5) edge (v3);
\node (v6) at (-4.7,0.8) {2};
\node at (-4.7,1.1) {$\su(1)^{(1)}$};
\draw  (v6) edge (v4);
\node (v2) at (-2,0.8) {2};
\node at (-2,1.1) {$\su(1)^{(1)}$};
\draw  (v3) edge (v2);
\end{tikzpicture}
\ee

Here the only $6d$ lifting case is the maximally supersymmetric one, whose lift was worked out previously \cite{Tachikawa:2011ch}. There appears to be no $6d$ lifting case for $\ff_4$ with fundamental matter, see \cite{Bhardwaj:2019xeg}. There are, however, $5d$ $\ff_4$ gauge theories with fundamental matter with a $5d$ SCFT UV completion. These will be covered in the next subsection.

\subsubsection{Rank 5}

\subsubsection*{\ubf{$\su(6)$}:}

\be \label{SU6wHHTASa13F}
\begin{tikzpicture} [scale=1.9]
\node at (-6.6,0.9) {$=$};
\node (v1) at (-5.9,0.8) {1};
\node at (-5.9,1.1) {$\su(5)^{(1)}$};
\node at (-7.8,0.9) {$\su(6)_0+\half\L^3+13\F$};
\end{tikzpicture}
\ee

\be \label{SU6wHHTASa9F}
\begin{tikzpicture} [scale=1.9]
\node at (-9.7,0.9) {$\su(6)_3+\half\L^3+9\F$};
\node at (-8.6,0.9) {$=$};
\node (v1) at (-8,0.8) {1};
\node at (-8,1.1) {$\sp(0)^{(1)}$};
\node (v2) at (-7.1,0.8) {2};
\node at (-7.1,1.1) {$\su(1)^{(1)}$};
\draw  (v1) edge (v2);
\node (v3) at (-6.2,0.8) {2};
\node at (-6.2,1.1) {$\su(1)^{(1)}$};
\draw  (v2) edge (v3);
\node (v4) at (-5.3,0.8) {2};
\node at (-5.3,1.1) {$\su(1)^{(1)}$};
\draw  (v3) edge (v4);
\node (v5) at (-4.4,0.8) {2};
\node at (-4.4,1.1) {$\su(1)^{(1)}$};
\draw  (v4) edge (v5);
\end{tikzpicture}
\ee


\be \label{SU6wHHTASaASa9F}
\begin{tikzpicture} [scale=1.9]
\node at (-5.1,0.9) {$=$};
\node (v1) at (-4.4,0.8) {1};
\node at (-4.4,1.1) {$\su(3)^{(1)}$};
\node at (-6.5,0.9) {$\su(6)_0+\half\L^3+\L^2+9\F$};
\node (v2) at (-3.4,0.8) {2};
\node at (-3.4,1.1) {$\su(2)^{(1)}$};
\draw  (v1) edge (v2);
\end{tikzpicture}
\ee

\be \label{SU6wHHTASaASa8F}
\begin{tikzpicture} [scale=1.9]
\node at (-5.9,0.9) {$=$};
\node (v1) at (-5.2,0.8) {1};
\node at (-5.2,1.1) {$\sp(1)^{(1)}$};
\node (v2) at (-4.2,0.8) {2};
\node at (-4.2,1.1) {$\su(2)^{(1)}$};
\draw  (v1) edge (v2);
\node (v3) at (-3.2,0.8) {2};
\node at (-3.2,1.1) {$\su(1)^{(1)}$};
\draw  (v2) edge (v3);
\node at (-7.3,0.9) {$\su(6)_\frac32+\half\L^3+\L^2+8\F$};
\end{tikzpicture}
\ee

\be \label{SU6wHHTASa2ASCSH}
\begin{tikzpicture} [scale=1.9]
\node at (-5.2,0.9) {$=$};
\node (v1) at (-4.7,0.8) {1};
\node at (-4.7,1.1) {$\ff_4^{(1)}$};
\node at (-6.6,0.9) {$\su(6)_\half+\half\L^3+2\L^2+2\F$};
\end{tikzpicture}
\ee

\be \label{SU6wHHTASa2ASCS3H}
\begin{tikzpicture} [scale=1.9]
\node at (-5.1,0.9) {$=$};
\node (v1) at (-3.4,0.8) {1};
\node at (-3.4,1.1) {$\sp(0)^{(1)}$};
\node at (-6.5,0.9) {$\su(6)_\frac32+\half\L^3+2\L^2+2\F$};
\node (v2) at (-2.4,0.8) {3};
\node at (-2.4,1.1) {$\so(8)^{(3)}$};
\draw  (v1) edge (v2);
\node (v0) at (-4.4,0.8) {2};
\draw  (v0) edge (v1);
\node at (-4.4,1.1) {$\su(1)^{(1)}$};
\end{tikzpicture}
\ee

\be \label{SU6w3AS1}
\begin{tikzpicture} [scale=1.9]
\node at (-5.4,0.9) {$=$};
\node (v1) at (-4.9,0.8) {2};
\node at (-4.9,1.1) {$\ff_4^{(1)}$};
\node at (-6.3,0.9) {$\su(6)_1+3\L^2$};
\end{tikzpicture}
\ee

\be \label{SU6w3AS3}
\begin{tikzpicture} [scale=1.9]
\node at (-5.1,0.9) {$=$};
\node (v1) at (-3.4,0.8) {1};
\node at (-3.4,1.1) {$\sp(0)^{(1)}$};
\node at (-6,0.9) {$\su(6)_3+3\L^2$};
\node (v2) at (-2.4,0.8) {4};
\node at (-2.4,1.1) {$\so(8)^{(3)}$};
\draw  (v1) edge (v2);
\node (v0) at (-4.4,0.8) {2};
\draw  (v0) edge (v1);
\node at (-4.4,1.1) {$\su(1)^{(1)}$};
\end{tikzpicture}
\ee

\be \label{SU6wTASa10F}
\begin{tikzpicture} [scale=1.9]
\node at (-5.1,0.9) {$=$};
\node (v1) at (-4.4,0.8) {1};
\node at (-4.4,1.1) {$\sp(2)^{(1)}$};
\node at (-6.2,0.9) {$\su(6)_0+\L^3+10\F$};
\node (v2) at (-3.4,0.8) {2};
\node at (-3.4,1.1) {$\su(2)^{(1)}$};
\draw  (v1) edge (v2);
\end{tikzpicture}
\ee

\be \label{SU6wTASa9F}
\begin{tikzpicture} [scale=1.9]
\node (v1) at (-4.2,0.8) {2};
\node at (-4.2,1.1) {$\su(1)^{(1)}$};
\node (v0) at (-5.2,0.8) {1};
\node at (-5.2,1.1) {$\sp(0)^{(1)}$};
\node (v2) at (-3.2,0.8) {2};
\node at (-3.2,1.1) {$\su(2)^{(1)}$};
\draw  (v1) edge (v2);
\node (v3) at (-2.2,0.8) {2};
\node at (-2.2,1.1) {$\su(1)^{(1)}$};
\draw  (v2) edge (v3);
\node at (-7,0.9) {$\su(6)_\frac32+\L^3+9\F$};
\node at (-5.9,0.9) {$=$};
\draw  (v0) edge (v1);
\end{tikzpicture}
\ee

\be \label{SU6wTASaASa4F}
\begin{tikzpicture} [scale=1.9]
\node at (-5.1,0.9) {$=$};
\node (v1) at (-4.4,0.8) {3};
\node at (-4.4,1.1) {$\so(8)^{(2)}$};
\node at (-6.4,0.9) {$\su(6)_0+\L^3+\L^2+4\F$};
\node (v2) at (-3.4,0.8) {1};
\node at (-3.4,1.1) {$\sp(0)^{(1)}$};
\node (v3) at (-3.9,0.8) {\tiny{2}};
\draw  (v1) edge (v3);
\draw[->]  (v3) edge (v2);
\end{tikzpicture}
\ee

\be \label{SU6wTASaASa3F}
\begin{tikzpicture} [scale=1.9]
\node (v1) at (-5.2,0.8) {3};
\node at (-5.2,1.1) {$\su(3)^{(2)}$};
\node (v2) at (-4.2,0.8) {1};
\node at (-4.2,1.1) {$\sp(0)^{(1)}$};
\draw  (v1) edge (v2);
\node (v3) at (-3.2,0.8) {2};
\node at (-3.2,1.1) {$\su(3)^{(2)}$};
\draw  (v2) edge (v3);
\node at (-7.4,0.9) {$\su(6)_\frac32+\L^3+\L^2+3\F$};
\node at (-6.1,0.9) {$=$};
\end{tikzpicture}
\ee

\be \label{SU6w3HTASa5F}
\begin{tikzpicture} [scale=1.9]
\node at (-7.3,0.9) {$\su(6)_0+\frac32\L^3+5\F$};
\node at (-6,0.9) {$=$};
\node (v1) at (-5.2,0.8) {1};
\node at (-5.2,1.1) {$\so(10)^{(2)}$};
\end{tikzpicture}
\ee

\be \label{SU6w3HTASa1F}
\begin{tikzpicture} [scale=1.9]
\node at (-8.0,0.9) {$\su(6)_3+\frac32\L^3+\F$};
\node at (-6.8,0.9) {$=$};
\node (v1) at (-6.2,0.8) {1};
\node at (-6.2,1.1) {$\fe_6^{(2)}$};
\end{tikzpicture}
\ee

\be \label{SU6wHTASa2AS}
\begin{tikzpicture} [scale=1.9]
\node (v1) at (-6.7,0.8) {1};
\node at (-6.7,1.1) {$\fe_6^{(2)}$};
\node at (-8.5,0.9) {$\su(6)_\frac72+\half\L^3+2\L^2$};
\node at (-7.3,0.9) {$=$};
\end{tikzpicture}
\ee

\be \label{SU6w3HTAS}
\begin{tikzpicture} [scale=1.9]
\node at (-5.2,0.9) {$=$};
\node (v1) at (-4.7,0.8) {3};
\node at (-4.7,1.1) {$\fe_6^{(2)}$};
\node at (-6.1,0.9) {$\su(6)_\frac 92+\frac32\L^3$};
\end{tikzpicture}
\ee

\be \label{SU6w2TAS}
\begin{tikzpicture} [scale=1.9]
\node at (-5.8,0.9) {$\su(6)_0+2\L^3$};
\node at (-4.9,0.9) {$=$};
\node (v1) at (-4.2,0.8) {3};
\node at (-4.2,1.1) {$\su(3)^{(2)}$};
\node (v2) at (-3.2,0.8) {1};
\node at (-3.2,1.1) {$\sp(0)^{(1)}$};
\draw  (v1) edge (v2);
\node (v3) at (-2.2,0.8) {3};
\node at (-2.2,1.1) {$\su(3)^{(2)}$};
\draw  (v2) edge (v3);
\end{tikzpicture}
\ee

\be \label{SU6wHTASaSa1F}
\begin{tikzpicture} [scale=1.9]
\node at (-5.1,0.9) {$=$};
\node (v1) at (-4.4,0.8) {2};
\node at (-4.4,1.1) {$\su(2)^{(1)}$};
\node at (-6.4,0.9) {$\su(6)_0+\S^2+\half\L^3+\F$};
\node (v2) at (-3.4,0.8) {2};
\node at (-3.4,1.1) {$\su(3)^{(1)}$};
\draw  (v1) edge (v2);
\draw (v2) .. controls (-3.9,0.2) and (-2.9,0.2) .. (v2);
\end{tikzpicture}
\ee

\be \label{SU6wHTASaS}
\begin{tikzpicture} [scale=1.9]
\node at (-5.1,0.9) {$=$};
\node (v1) at (-3.4,0.8) {2};
\node at (-3.4,1.1) {$\su(2)^{(1)}$};
\node at (-6.3,0.9) {$\su(6)_\frac32+\S^2+\half\L^3$};
\node (v2) at (-2.4,0.8) {2};
\node at (-2.4,1.1) {$\su(2)^{(1)}$};
\draw  (v1) edge (v2);
\draw (v2) .. controls (-2.9,0.2) and (-1.9,0.2) .. (v2);
\node (v0) at (-4.4,0.8) {2};
\draw  (v0) edge (v1);
\node at (-4.4,1.1) {$\su(1)^{(1)}$};
\end{tikzpicture}
\ee

Some of the cases appearing here have been previously studied in the literature, while some are new. Specifically, the $6d$ lifts for cases \eqref{SU6wHHTASa13F}, \eqref{SU6wTASa10F} and \eqref{SU6wHTASaSa1F} were conjectured in \cite{Hayashi:2019yxj}, and our results from geometry are consistent with these conjectures. Additionally, \cite{Hayashi:2019yxj} also presented brane constructions for cases \eqref{SU6wHHTASa9F}, \eqref{SU6wHHTASaASa9F}, \eqref{SU6wHHTASaASa8F}, \eqref{SU6wTASa9F}, \eqref{SU6wTASaASa4F}, \eqref{SU6w2TAS}, and \eqref{SU6wHTASaS}, from which it was inferred that these are $6d$ lifting though the explicit $6d$ lift was not determined. The remaining cases are new, to our knowledge.

Finally, we note several dualities for theories in this list. Case \eqref{SU6wHHTASa13F} is dual to the $m=6$ case of \eqref{SUmwFa1ASnoCS}, case \eqref{SU6wHHTASa9F} is dual to the $m=6$ case of \eqref{SUmwFa1ASaCS} and to the $m=5$ case of \eqref{SpmwFaAS}, case \eqref{SU6wHHTASaASa8F} is dual to the $m=3$ case of \eqref{SUmevenw7Fa2AS}, and cases \eqref{SU6wHTASa2AS} and \eqref{SU6w3HTASa1F} are dual to each other.

\subsubsection*{\ubf{$\so(10)$}:}

\be \label{SO10w4Sa2F}
\begin{tikzpicture} [scale=1.9]
\node at (-5.1,0.9) {$=$};
\node (v1) at (-3.4,0.8) {1};
\node at (-3.4,1.1) {$\sp(0)^{(1)}$};
\node at (-6.2,0.9) {$\so(10)+4\S+2\F$};
\node (v2) at (-2.4,0.8) {2};
\node at (-2.4,1.1) {$\su(2)^{(1)}$};
\draw  (v1) edge (v2);
\node (v0) at (-4.4,0.8) {3};
\draw  (v0) edge (v1);
\node at (-4.4,1.1) {$\su(3)^{(2)}$};
\end{tikzpicture}
\ee

\be \label{SO10w3Sa4F}
\begin{tikzpicture} [scale=1.9]
\node at (-5.6,0.9) {$=$};
\node (v1) at (-4.9,0.8) {1};
\node at (-4.9,1.1) {$\so(9)^{(1)}$};
\node at (-6.7,0.9) {$\so(10)+3\S+4\F$};
\end{tikzpicture}
\ee

\be \label{SO10w2Sa6F}
\begin{tikzpicture} [scale=1.9]
\node at (-5.1,0.9) {$=$};
\node (v1) at (-4.4,0.8) {1};
\node at (-4.4,1.1) {$\sp(0)^{(1)}$};
\node at (-6.2,0.9) {$\so(10)+2\S+6\F$};
\node (v2) at (-3.4,0.8) {2};
\node at (-3.4,1.1) {$\su(6)^{(2)}$};
\draw  (v1) edge (v2);
\end{tikzpicture}
\ee

\be \label{SO10w1Sa7F}
\begin{tikzpicture} [scale=1.9]
\node at (-5.6,0.9) {$=$};
\node (v1) at (-4.9,0.8) {1};
\node at (-4.9,1.1) {$\su(8)^{(2)}$};
\node at (-6.6,0.9) {$\so(10)+\S+7\F$};
\end{tikzpicture}
\ee

The $6d$ lifts for some of the cases here were already considered in the literature. Specifically, the $6d$ lift for case \eqref{SO10w2Sa6F} was conjectured by \cite{Razamat:2018gbu} and for case \eqref{SO10w4Sa2F} by \cite{Sela:2019gbz}. In both cases our findings from geometry support these conjectures. The rest of the cases are, to our knowledge, new.

\subsubsection*{\ubf{$\so(11)$}:}

\be \label{SO11w2Sa3F}
\begin{tikzpicture} [scale=1.9]
\node at (-7,0.9) {$\so(11)+2\S+3\F$};
\node at (-5.9,0.9) {$=$};
\node (v1) at (-5.2,0.8) {3};
\node at (-5.2,1.1) {$\su(3)^{(2)}$};
\node (v2) at (-4.2,0.8) {1};
\node at (-4.2,1.1) {$\sp(0)^{(1)}$};
\draw  (v1) edge (v2);
\node (v3) at (-3.2,0.8) {2};
\node at (-3.2,1.1) {$\su(3)^{(2)}$};
\draw  (v2) edge (v3);
\end{tikzpicture}
\ee

\be \label{SO11w3HSa5F}
\begin{tikzpicture} [scale=1.9]
\node at (-6,0.9) {$=$};
\node (v1) at (-5.2,0.8) {1};
\node at (-5.2,1.1) {$\so(10)^{(2)}$};
\node at (-7.1,0.9) {$\so(11)+\frac32\S+5\F$};
\end{tikzpicture}
\ee

\be \label{SO11w5HS}
\begin{tikzpicture} [scale=1.9]
\node at (-5.8,0.9) {$=$};
\node (v1) at (-5.2,0.8) {1};
\node at (-5.2,1.1) {$\fe_6^{(2)}$};
\node at (-6.7,0.9) {$\so(11)+\frac52\S$};
\end{tikzpicture}
\ee

\be \label{SO11w1Sa7F}
\begin{tikzpicture} [scale=1.9]
\node at (-5.1,0.9) {$=$};
\node (v1) at (-4.4,0.8) {1};
\node at (-4.4,1.1) {$\sp(0)^{(1)}$};
\node at (-6.1,0.9) {$\so(11)+\S+7\F$};
\node (v2) at (-3.4,0.8) {2};
\node at (-3.4,1.1) {$\su(7)^{(2)}$};
\draw  (v1) edge (v2);
\end{tikzpicture}
\ee

\be \label{SO11w1HSa8F}
\begin{tikzpicture} [scale=1.9]
\node at (-5.6,0.9) {$=$};
\node (v1) at (-4.9,0.8) {1};
\node at (-4.9,1.1) {$\su(9)^{(2)}$};
\node at (-6.7,0.9) {$\so(11)+\half\S+8\F$};
\end{tikzpicture}
\ee

The $6d$ lifts for some of the cases here were already considered in the literature, but there are new cases as well. Notably, the $6d$ lift for case \eqref{SO11w1Sa7F} was conjectured by \cite{Razamat:2018gbu}, and for case \eqref{SO11w2Sa3F} by \cite{Sela:2019gbz}. In both cases our findings from geometry support these conjectures. The remaining cases are, to our knowledge, new. See Section \ref{duality} for dualities.

\subsubsection{Rank 6}

\subsubsection*{\ubf{$\su(7)$}:}

\be \label{SU7wTASa6F}
\begin{tikzpicture} [scale=1.9]
\node at (-5.1,0.9) {$=$};
\node (v1) at (-4.4,0.8) {3};
\node at (-4.4,1.1) {$\so(8)^{(2)}$};
\node at (-6.2,0.9) {$\su(7)_0+\L^3+6\F$};
\node (v2) at (-3.4,0.8) {1};
\node at (-3.4,1.1) {$\sp(1)^{(1)}$};
\node (v3) at (-3.9,0.8) {\tiny{2}};
\draw [dashed] (v1) edge (v3);
\draw[->]  (v3) edge (v2);
\end{tikzpicture}
\ee

\be \label{SU7wTASa5F}
\begin{tikzpicture} [scale=1.9]
\node at (-6.9,0.9) {$\su(7)_\frac32+\L^3+5\F$};
\node at (-5.9,0.9) {$=$};
\node (v1) at (-5.2,0.8) {3};
\node at (-5.2,1.1) {$\su(3)^{(2)}$};
\node (v2) at (-4.2,0.8) {1};
\node at (-4.2,1.1) {$\sp(0)^{(1)}$};
\draw  (v1) edge (v2);
\node (v3) at (-3.2,0.8) {2};
\node at (-3.2,1.1) {$\su(5)^{(2)}$};
\draw  (v2) edge (v3);
\end{tikzpicture}
\ee

Both of these cases are new to our knowledge. The $6d$ SCFT corresponding to the case \eqref{SU7wTASa6F} is denoted as
\be\nn
\begin{tikzpicture} [scale=1.9]
\node (v1) at (-5.2,0.8) {1};
\node at (-5.2,1.1) {$\sp(1)$};
\node (v2) at (-4.2,0.8) {3};
\node at (-4.2,1.1) {$\so(8)$};
\draw [dashed] (v1) edge (v2);
\node (v3) at (-3.2,0.8) {1};
\node at (-3.2,1.1) {$\sp(1)$};
\draw  (v2) edge (v3);
\end{tikzpicture}
\ee
in \cite{Bhardwaj:2019fzv} since (upto triality) one of the $\sp(1)$ gauges a hyper in $\F$ of $\so(8)$ and the other $\sp(1)$ gauges a hyper in $\S$ of $\so(8)$, where the former gauging is denoted by a solid edge and the latter gauging is denoted by a dashed edge. While compactifying on a circle, the $6d$ SCFT is twisted by the $\Z_2$ outer automorphism of $\so(8)$ which exchanges $\F$ and $\S$ thus folding the dashed edge onto the solid edge. Consequently, we denote the KK theory with a partially solid and partially dashed edge.

\subsubsection*{\ubf{$\so(12)$}:}

\be \label{SO12w2Sa4F}
\begin{tikzpicture} [scale=1.9]
\node (v1) at (-4.2,0.8) {1};
\node at (-4.2,1.1) {$\sp(0)^{(1)}$};
\node (v0) at (-5.2,0.8) {3};
\node at (-5.2,1.1) {$\su(3)^{(2)}$};
\node (v2) at (-3.2,0.8) {3};
\node at (-3.2,1.1) {$\su(3)^{(2)}$};
\draw  (v1) edge (v2);
\node (v3) at (-2.2,0.8) {1};
\node at (-2.2,1.1) {$\sp(0)^{(1)}$};
\draw  (v2) edge (v3);
\node at (-7,0.9) {$\so(12)+2\S+4\F$};
\node at (-5.9,0.9) {$=$};
\draw  (v0) edge (v1);
\end{tikzpicture}
\ee

\be \label{SO12w3HSa1Ca1F}
\begin{tikzpicture} [scale=1.9]
\node at (-5,0.9) {$=$};
\node (v1) at (-4.4,0.8) {3};
\node at (-4.4,1.1) {$\fe_6^{(2)}$};
\node at (-6.2,0.9) {$\so(12)+\frac32\S+\C+\F$};
\node (v2) at (-3.4,0.8) {1};
\node at (-3.4,1.1) {$\sp(0)^{(1)}$};
\node (v3) at (-3.9,0.8) {\tiny{2}};
\draw (v1) edge (v3);
\draw[->]  (v3) edge (v2);
\end{tikzpicture}
\ee

\be \label{SO12w3HSa1HCa4F}
\begin{tikzpicture} [scale=1.9]
\node at (-5.1,0.9) {$=$};
\node (v1) at (-3.4,0.8) {1};
\node at (-3.4,1.1) {$\sp(0)^{(1)}$};
\node at (-6.4,0.9) {$\so(12)+\frac32\S+\half\C+4\F$};
\node (v2) at (-2.4,0.8) {2};
\node at (-2.4,1.1) {$\fg_2^{(1)}$};
\draw  (v1) edge (v2);
\node (v0) at (-4.4,0.8) {3};
\draw  (v0) edge (v1);
\node at (-4.4,1.1) {$\su(3)^{(2)}$};
\end{tikzpicture}
\ee

\be \label{SO12w1Sa1HCa6F}
\begin{tikzpicture} [scale=1.9]
\node at (-6.6,0.9) {$\so(12)+\S+\half\C+6\F$};
\node at (-5.3,0.9) {$=$};
\node (v2) at (-4.5,0.8) {1};
\node at (-4.5,1.1) {$\so(11)^{(1)}$};
\end{tikzpicture}
\ee

\be \label{SO12w3HSa6F}
\begin{tikzpicture} [scale=1.9]
\node (v2) at (-5.2,0.8) {1};
\node at (-5.2,1.1) {$\so(11)^{(1)}$};
\node at (-7.2,0.9) {$\so(12)+\frac32\S+6\F$};
\node at (-6.1,0.9) {$=$};
\end{tikzpicture}
\ee

\be \label{SO12w1Sa1Ca4F}
\begin{tikzpicture} [scale=1.9]
\node at (-5.1,0.9) {$=$};
\node (v1) at (-3.4,0.8) {1};
\node at (-3.4,1.1) {$\sp(0)^{(1)}$};
\node at (-6.3,0.9) {$\so(12)+\S+\C+4\F$};
\node (v2) at (-2.4,0.8) {2};
\node at (-2.4,1.1) {$\su(4)^{(2)}$};
\draw  (v1) edge (v2);
\node (v0) at (-4.4,0.8) {3};
\draw  (v0) edge (v1);
\node at (-4.4,1.1) {$\su(3)^{(2)}$};
\end{tikzpicture}
\ee

\be \label{SO12w1Sa8F}
\begin{tikzpicture} [scale=1.9]
\node at (-5.1,0.9) {$=$};
\node (v1) at (-4.4,0.8) {1};
\node at (-4.4,1.1) {$\sp(0)_0^{(1)}$};
\node at (-6.1,0.9) {$\so(12)+\S+8\F$};
\node (v2) at (-3.4,0.8) {2};
\node at (-3.4,1.1) {$\su(8)^{(2)}$};
\draw  (v1) edge (v2);
\end{tikzpicture}
\ee

\be \label{SO12w1HSa1HCa8F}
\begin{tikzpicture} [scale=1.9]
\node at (-5.1,0.9) {$=$};
\node (v1) at (-4.4,0.8) {1};
\node at (-4.4,1.1) {$\sp(0)_\pi^{(1)}$};
\node at (-6.4,0.9) {$\so(12)+\half\S+\half\C+8\F$};
\node (v2) at (-3.4,0.8) {2};
\node at (-3.4,1.1) {$\su(8)^{(2)}$};
\draw  (v1) edge (v2);
\end{tikzpicture}
\ee

\be \label{SO12w1HSa9F}
\begin{tikzpicture} [scale=1.9]
\node at (-5.6,0.9) {$=$};
\node (v1) at (-4.8,0.8) {1};
\node at (-4.8,1.1) {$\su(10)^{(2)}$};
\node at (-6.7,0.9) {$\so(12)+\half\S+9\F$};
\end{tikzpicture}
\ee

The $6d$ lifts for some of these cases were already considered in the literature. Specifically, the $6d$ lift for cases \eqref{SO12w1HSa1HCa8F} and \eqref{SO12w1Sa8F} were conjectured by \cite{Razamat:2018gbu}. As explained there, these two cases differ by the embedding of the $\su(8)$ gauge symmetry on the $-2$ curve in the $\fe_8$ global symmetry associated with the empty $-1$ curve. As this difference becomes the theta angle of the $\sp(n)$ gauge group if it is turned on the $-1$ curve \cite{Mekareeya:2017jgc}, we differentiate the two cases by denoting this angle even though $n=0$ here. Additionally, cases \eqref{SO12w2Sa4F}, \eqref{SO12w3HSa1HCa4F} and \eqref{SO12w1Sa1Ca4F} were conjectured by \cite{Sela:2019gbz}. Our results from geometry support these conjectures in all cases. 

The remaining cases are all new to our knowledge. We also note that the two cases \eqref{SO12w1Sa1HCa6F} and \eqref{SO12w3HSa6F} are dual to each other.

\subsubsection*{\ubf{$\so(13)$}:}

\be \label{SO13w1Sa5F}
\begin{tikzpicture} [scale=1.9]
\node at (-6.4,0.9) {$\so(13)+\S+5\F$};
\node at (-5.4,0.9) {$=$};
\node (v1) at (-4.7,0.8) {3};
\node at (-4.7,1.1) {$\su(3)^{(2)}$};
\node (v2) at (-3.7,0.8) {1};
\node at (-3.7,1.1) {$\sp(0)^{(1)}$};
\draw  (v1) edge (v2);
\node (v3) at (-2.7,0.8) {2};
\node at (-2.7,1.1) {$\su(5)^{(2)}$};
\draw  (v2) edge (v3);
\end{tikzpicture}
\ee

\be \label{SO13w1HSa9F}
\begin{tikzpicture} [scale=1.9]
\node at (-5.1,0.9) {$=$};
\node (v1) at (-4.4,0.8) {1};
\node at (-4.4,1.1) {$\sp(0)^{(1)}$};
\node at (-6.2,0.9) {$\so(13)+\half\S+9\F$};
\node (v2) at (-3.4,0.8) {2};
\node at (-3.4,1.1) {$\su(9)^{(2)}$};
\draw  (v1) edge (v2);
\end{tikzpicture}
\ee

The $6d$ lifts for both of these cases were already considered in the literature. Specifically, the $6d$ lift for case \eqref{SO13w1HSa9F} was conjectured by \cite{Razamat:2018gbu}, and for case \eqref{SO13w1Sa5F} by \cite{Sela:2019gbz}. In both cases our findings from geometry support these conjectures.

\subsubsection*{\ubf{$\fe_6$}:}

\be
\begin{tikzpicture} [scale=1.9]
\node at (-5.4,0.9) {$=$};
\node (v1) at (-1.5,0.8) {2};
\node at (-1.5,1.1) {$\su(1)^{(1)}$};
\node (v3) at (-2.9,0.8) {2};
\node (v6) at (-2.9,1.1) {$\su(1)^{(1)}$};
\node (v2) at (-2.2,0.8) {$\cdots$};
\draw  (v2) edge (v3);
\draw  (v2) edge (v1);
\node (v4) at (-3.8,0.8) {2};
\node at (-3.8,1.1) {$\su(1)^{(1)}$};
\begin{scope}[shift={(0,0.45)}]
\node at (-2.2,-0.15) {$3$};
\draw (-3.1,0.15) .. controls (-3.1,0.1) and (-3.1,0.05) .. (-3,0.05);
\draw (-3,0.05) -- (-2.3,0.05);
\draw (-2.2,0) .. controls (-2.2,0.05) and (-2.25,0.05) .. (-2.3,0.05);
\draw (-2.2,0) .. controls (-2.2,0.05) and (-2.15,0.05) .. (-2.1,0.05);
\draw (-2.1,0.05) -- (-1.4,0.05);
\draw (-1.3,0.15) .. controls (-1.3,0.1) and (-1.3,0.05) .. (-1.4,0.05);
\end{scope}
\node at (-6.1,0.9) {$\fe_6+\A$};
\draw  (v4) edge (v3);
\node (v5) at (-2.9,1.8) {2};
\node at (-2.9,2.1) {$\su(1)^{(1)}$};
\draw  (v5) edge (v6);
\node (v7) at (-4.7,0.8) {2};
\node at (-4.7,1.1) {$\su(1)^{(1)}$};
\draw  (v7) edge (v4);
\end{tikzpicture}
\ee

Here the only $6d$ lifting case is the maximally supersymmetric one, whose lift was worked out previously \cite{Tachikawa:2011ch}. There appears to be no $6d$ lifting case for $\fe_6$ with fundamental matter, see \cite{Bhardwaj:2019xeg}. There are, however, $5d$ $\fe_6$ gauge theories with fundamental matter with a $5d$ SCFT UV completion. These will be covered in the next subsection.

\subsubsection{Rank 7}

\subsubsection*{\ubf{$\so(14)$}:}

\be
\begin{tikzpicture} [scale=1.9]
\node at (-5.1,0.9) {$=$};
\node (v1) at (-3.4,0.8) {1};
\node at (-3.4,1.1) {$\sp(0)^{(1)}$};
\node at (-6.1,0.9) {$\so(14)+\S+6\F$};
\node (v2) at (-2.4,0.8) {2};
\node at (-2.4,1.1) {$\su(6)^{(2)}$};
\draw  (v1) edge (v2);
\node (v0) at (-4.4,0.8) {3};
\draw  (v0) edge (v1);
\node at (-4.4,1.1) {$\su(3)^{(2)}$};
\end{tikzpicture}
\ee

The $6d$ lift for this case was conjectured by \cite{Sela:2019gbz}, and our results from geometry match this conjecture.

\subsubsection*{\ubf{$\fe_7$}:}

\be
\begin{tikzpicture} [scale=1.9]
\node at (-5.4,0.9) {$=$};
\node (v1) at (-1.5,0.8) {2};
\node at (-1.5,1.1) {$\su(1)^{(1)}$};
\node (v3) at (-2.9,0.8) {2};
\node (v6) at (-2.9,1.1) {$\su(1)^{(1)}$};
\node (v2) at (-2.2,0.8) {$\cdots$};
\draw  (v2) edge (v3);
\draw  (v2) edge (v1);
\node (v4) at (-3.8,0.8) {2};
\node at (-3.8,1.1) {$\su(1)^{(1)}$};
\begin{scope}[shift={(0,0.45)}]
\node at (-2.2,-0.15) {$4$};
\draw (-3.1,0.15) .. controls (-3.1,0.1) and (-3.1,0.05) .. (-3,0.05);
\draw (-3,0.05) -- (-2.3,0.05);
\draw (-2.2,0) .. controls (-2.2,0.05) and (-2.25,0.05) .. (-2.3,0.05);
\draw (-2.2,0) .. controls (-2.2,0.05) and (-2.15,0.05) .. (-2.1,0.05);
\draw (-2.1,0.05) -- (-1.4,0.05);
\draw (-1.3,0.15) .. controls (-1.3,0.1) and (-1.3,0.05) .. (-1.4,0.05);
\end{scope}
\node at (-6.1,0.9) {$\fe_7+\A$};
\draw  (v4) edge (v3);
\node (v5) at (-2.9,1.8) {2};
\node at (-2.9,2.1) {$\su(1)^{(1)}$};
\draw  (v5) edge (v6);
\node (v7) at (-4.7,0.8) {2};
\node at (-4.7,1.1) {$\su(1)^{(1)}$};
\draw  (v7) edge (v4);
\end{tikzpicture}
\ee

Here the only $6d$ lifting case is the maximally supersymmetric one, whose lift was worked out previously \cite{Tachikawa:2011ch}. There appears to be no $6d$ lifting case for $\fe_7$ with fundamental matter, see \cite{Bhardwaj:2019xeg}. There are, however, $5d$ $\fe_7$ gauge theories with fundamental matter with a $5d$ SCFT UV completion. These will be covered in the next subsection.

\subsubsection{Rank 8}

\subsubsection*{\ubf{$\fe_8$}:}

\be
\begin{tikzpicture} [scale=1.9]
\node at (-5.4,0.9) {$=$};
\node (v1) at (-1.5,0.8) {2};
\node at (-1.5,1.1) {$\su(1)^{(1)}$};
\node (v3) at (-2.9,0.8) {2};
\node (v6) at (-2.9,1.1) {$\su(1)^{(1)}$};
\node (v2) at (-2.2,0.8) {$\cdots$};
\draw  (v2) edge (v3);
\draw  (v2) edge (v1);
\node (v4) at (-3.8,0.8) {2};
\node at (-3.8,1.1) {$\su(1)^{(1)}$};
\begin{scope}[shift={(0,0.45)}]
\node at (-2.2,-0.15) {$5$};
\draw (-3.1,0.15) .. controls (-3.1,0.1) and (-3.1,0.05) .. (-3,0.05);
\draw (-3,0.05) -- (-2.3,0.05);
\draw (-2.2,0) .. controls (-2.2,0.05) and (-2.25,0.05) .. (-2.3,0.05);
\draw (-2.2,0) .. controls (-2.2,0.05) and (-2.15,0.05) .. (-2.1,0.05);
\draw (-2.1,0.05) -- (-1.4,0.05);
\draw (-1.3,0.15) .. controls (-1.3,0.1) and (-1.3,0.05) .. (-1.4,0.05);
\end{scope}
\node at (-6.1,0.9) {$\fe_8+\A$};
\draw  (v4) edge (v3);
\node (v5) at (-2.9,1.8) {2};
\node at (-2.9,2.1) {$\su(1)^{(1)}$};
\draw  (v5) edge (v6);
\node (v7) at (-4.7,0.8) {2};
\node at (-4.7,1.1) {$\su(1)^{(1)}$};
\draw  (v7) edge (v4);
\end{tikzpicture}
\ee

Here the only $6d$ lifting case is the maximally supersymmetric one, whose lift was worked out previously \cite{Tachikawa:2011ch}.

\subsection{SCFTs}\label{scftr}

We next turn to summarizing the cases having a $5d$ SCFT UV completion. Most of these cases can be generated by integrating out matter from the $6d$ lifting cases in the previous list, but there are a handful of cases that can not be generated by integrating out matter from a $6d$ lifting gauge theory. They still can be obtained from a $5d$ KK theory but the transition process requires a (generalized) ungauging along with integrating out matter (see \cite{Bhardwaj:2019xeg} for more details).

\subsubsection{General Rank}

As previously, we first start with the cases existing for generic rank, and later innumerate the finite number of special cases for low rank.  

\subsubsection*{\ubf{$\su(m)$}:}

\be \label{FSUmwFO}
\su(m)_\frac {n-p}2+(2m+4-n-p)\F ,
\ee

\be \label{FSUmw1AS}
\su(m)_\frac {n-p}2+\L^2+(m+6-n-p)\F ,
\ee

\be \label{FSUmw1ASa}
\su(m)_{\frac{m+n}2}+\L^2+(8-n)\F ,
\ee

\be \label{FSUmw2AS}
\su(m)_\frac {n-p}2+2\L^2+(8-n-p)\F ,
\ee

\be \label{FSUmw2ASa}
\su(m)_\frac {3+n}2+2\L^2+(7-n)\F ,
\ee

\be \label{FSUmwS}
\su(m)_\frac{n-p}2+\S^2+(m-2-n-p)\F ,
\ee

All cases here are generated by integrating matter from the cases in \ref{SSGRSUm}. Specifically, case \eqref{FSUmwFO} is generated by integrating $n$ fundamentals with a positive mass and $p$ fundamentals with a negative mass out of the $6d$ lifting case \eqref{SUmwFO}. Likewise, case \eqref{FSUmw1AS} is generated by integrating $n$ fundamentals with a positive mass and $p$ fundamentals with a negative mass out of the $6d$ lifting case \eqref{SUmwFa1ASnoCS}. Case \eqref{FSUmw1ASa} contains the cases generated by integrating matter out of case \eqref{SUmwFa1ASaCS}, where we have restricted only to cases not covered by the previous entry.

In the same vein, case \eqref{FSUmw2AS} covers cases generated by integrating fundamental matter from cases \eqref{SUmevenw8Fa2AS} and \eqref{SUmoddw8Fa2AS}, and case \eqref{FSUmw2ASa} covers cases generated by integrating fundamental matter from cases \eqref{SUmoddw7Fa2AS} and \eqref{SUmevenw7Fa2AS} that where not covered by the previous entry. Finally, case \eqref{FSUmwS} covers cases generated by integrating fundamental matter from case \eqref{SUmwFaS}. We can not get any additional cases by integrating non-fundamental matter or by integrating matter out of the other cases in \ref{SSGRSUm}.

All cases here were known to exist before, and have brane web realizations \cite{Aharony:1997ju,Aharony:1997bh,Bergman:2014kza,Bergman:2015dpa,Hayashi:2015fsa,Jefferson:2017ahm}.

\subsubsection*{\ubf{$\sp(m)$}:}

\be \label{USpwF}
\sp(m)_{0/\pi}+(2m+6-n)\F ,
\ee



\be \label{USpwAS}
\sp(m)_{0/\pi}+\L^2+(8-n)\F ,
\ee



Here also all cases can be generated by integrating out fundamental matter from the $6d$ lifting cases. Specifically, case \eqref{USpwF} can be generated from case \eqref{SpmwF}, and case \eqref{USpwAS} from case \eqref{SpmwFaAS}. Here the theta angle for the $\sp$ group is only physically relevant for the pure case or the case with just a single antisymmetric hyper \cite{Morrison:1996xf}. All the cases here are known to exist. Case \eqref{USpwF} can be realized using brane webs \cite{Brunner:1997ndk,Bergman:2015dpa}. For case \eqref{USpwAS} there is a type I$'$ brane construction \cite{Seiberg:1996bd}, from which one can also get a brane web representation \cite{Bergman:2012rgz}.    

Many of these cases are dual to some of the $\su$ cases discussed previously. Specifically, case \eqref{USpwF} is dual to case \eqref{FSUmwFO} with $p=0$ and $m_{\su} = m_{\sp}+1$, and case \eqref{USpwAS} is dual to case \eqref{FSUmw1ASa} with $m_{\su} = m_{\sp}+1$. Both of these dualities were known previously from other works. Specifically, the duality involving case \eqref{USpwF} was found in \cite{Gaiotto:2017hck} (see also \cite{Hayashi:2015zka} for a brane realization), while the one involving \eqref{USpwAS} was found in \cite{Jefferson:2017ahm}. For both cases, when there is no fundamental flavor, we have two different SCFTs associated with the different theta angles, but only one of each case has an $\su$ dual description. Specifically, for cases where the rank is even, the theta angle with the dual is $\pi$, while for cases where the rank is odd, the theta angle with the dual is $0$.  

\subsubsection*{\ubf{$\so(m)$}:}

\be
\so(m)+(m-2-n)\F ,
\ee

The case here can be conveniently generated by integrating matter out of the $6d$ lifting case \eqref{SOmwV}. This class of theories were known to exist before, notably due to a brane web realization \cite{Brunner:1997ndk,Bergman:2015dpa}.

\subsubsection{Rank 2}

\subsubsection*{\ubf{$\su(3)$}:}

\be \label{SU3CS7Lf}
\su(3)_{4+\frac n2}+(6-n)\F ,
\ee

\be \label{SU3CS6Lf}
\su(3)_6 ,
\ee

\be \label{SU3CS8Lf}
\su(3)_8 ,
\ee

Cases \eqref{SU3CS7Lf} and \eqref{SU3CS6Lf} can be generated by integrating out fundamental matter from case \eqref{SU3CS7L}, while \eqref{SU3CS8Lf} can be generated by integrating out fundamental matter with a positive mass from case \eqref{SU3CS8L}. All three classes of theories were known before, where case \eqref{SU3CS7Lf} was first found from geometry in \cite{Jefferson:2018irk}, case \eqref{SU3CS6Lf} being first noted in \cite{Zafrir:2015rga}, and case \eqref{SU3CS8Lf} was first found, also from geometry, in \cite{Bhardwaj:2019jtr}.

\subsubsection*{\ubf{$\sp(2)$}:}

\be \label{Sp2w2ASf}
\sp(2)_{0/\pi}+2\L^2+(4-n)\F ,
\ee

The case here can be generated by integrating out fundamental matter from the $6d$ lifting case in \eqref{SP2w2AS}. Here the theta angle for the $\sp$ group is only physically relevant for the case with only the two antisymmetric hypermultiplets and no fundamentals (that is $n=4$). This case is dual to \eqref{SU3CS7Lf}, where for $n=4$ the dual $\sp$ case is the one with theta angle $\pi$. Both this class of models and the duality were first found in \cite{Jefferson:2018irk}, with the exception of the $n=4$ case, which is just $\sp(2)+2\L^2 = \so(5)+2F$ and so can also be build from the methods of \cite{Brunner:1997ndk,Bergman:2015dpa,Zafrir:2015ftn,Hayashi:2015vhy} \footnote{This seems to only give the $\theta=0$ case.}.  

\subsubsection*{\ubf{$\fg_2$}:}

\be
\fg_2+(6-n)\F ,
\ee

The case here can be generated by integrating out fundamental matter from the $6d$ lifting case in \eqref{G2wF}. This case is dual to \eqref{SU3CS7Lf} and \eqref{Sp2w2ASf}. Both this class of models and the duality were first found in \cite{Jefferson:2018irk}. Both were also given brane realizations in \cite{Hayashi:2018bkd,Hayashi:2018lyv}.




\subsubsection{Rank 3}

\subsubsection*{\ubf{$\su(4)$}:}

\be \label{SU4CS7Lf}
\su(4)_{4+\frac n2}+(6-n)\F ,
\ee

\be \label{SU4w3ASf}
\su(4)_{k}+3\L^2+(4-n)\F \:;\:\:\: 0\leq k \leq 2+\frac{n}{2} ,
\ee

Case \eqref{SU4CS7Lf} can be generated by integrating fundamental matter with a positive mass from case \eqref{SU4CS7L}, while \eqref{SU4w3ASf} can be generated by integrating fundamental matter from cases \eqref{SU4w3ASaFa0CS}, \eqref{SU4w3ASaFa1CS} and \eqref{SU4w3ASaFa2CS}. Both cases appear new. Case \eqref{SU4w3ASf} can also be regarded as $\so(6)_k+3\F+(4-n)\S$, and so one should be able to build brane webs for these cases, at least for small $k$, using the results of \cite{Zafrir:2015ftn,Hayashi:2017skf}.

\subsubsection*{\ubf{$\sp(3)$}:}

\be \label{SP3wTASf}
\sp(3)+\L^3+(5-n)\F , 
\ee

\be \label{SP3wHTASaFf}
\sp(3)+\half\L^3+\frac{19-2n}{2}\F ,
\ee

\be \label{SP3wHTASaASf}
\sp(3)+\half\L^3+\L^2+\frac{5-2n}2\F ,
\ee

Case \eqref{SP3wTASf} can be generated by integrating fundamental matter from case \eqref{SP3wTASaF}, while \eqref{SP3wHTASaFf} and \eqref{SP3wHTASaASf} can be generated in the same way from cases \eqref{SP3wHTASaF} and \eqref{SP3wHTASaAS}, respectively. Note that as the three index antisymmetric representation of $\sp(3)$ contributes to the anomaly of \cite{Witten:1982fp}, the theta angle in their presence should be physically irrelevant. For the cases \eqref{SP3wHTASaFf} and \eqref{SP3wHTASaASf}, the geometry indicates that the theta angle is physically irrelevant.

Cases \eqref{SP3wTASf} and \eqref{SP3wHTASaFf} have been found previously from brane constructions in \cite{Hayashi:2019yxj}, while \eqref{SP3wHTASaASf} appears new. We also note that case \eqref{SP3wTASf} is dual to case \eqref{SU4CS7Lf}, and case \eqref{SP3wHTASaFf} is dual to the $m=4$ case of \eqref{FSUmw2ASa}. See Section \ref{dualitys} for a list of dualities occurring in gauge theories having a $5d$ SCFT UV completion.

\subsubsection*{\ubf{$\so(7)$}:}

\be \label{SO7wSa1Ff}
\so(7)+(6-n)\S+\F ,
\ee

\be \label{SO7wSa2Ff}
\so(7)+(5-n)\S+2\F ,
\ee

\be \label{SO7wSa3Ff}
\so(7)+(4-n)\S+3\F ,
\ee

\be \label{SO7wSa4Ff}
\so(7)+(2-n)\S+4\F ,
\ee

\be \label{SO7wSa0Ff}
\so(7)+(7-n)\S ,
\ee

Case \eqref{SO7wSa1Ff} can be generated by integrating fundamental matter from case \eqref{SO7w6Sa1V}. Similarly cases \eqref{SO7wSa2Ff}, \eqref{SO7wSa3Ff}, \eqref{SO7wSa4Ff} and \eqref{SO7wSa0Ff} can be generated in the same way from cases \eqref{SO7w5Sa2V}, \eqref{SO7w4Sa3V}, \eqref{SO7w2Sa4V} and \eqref{SO7w7S}, respectively. A brane construction for this class of theories was given in \cite{Zafrir:2015ftn}, which is valid when the number of spinors is four or smaller. It should be possible to use the result of \cite{Zafrir:2016csd}, and lift the class S construction for the $4d$ ${\cal N}=2$ $\so(7)+5\S$ SCFT given in \cite{Chacaltana:2013dsj} to $5d$ to get a brane web description also for the cases with five spinors.    

The cases here are related to the previous cases and to one another by various dualities. See Section \ref{dualitys} for a full account of these dualities.

\subsubsection{Rank 4}\label{r4}

\subsubsection*{\ubf{$\su(5)$}:}

\be \label{SU5w3ASa2Ff}
\su(5)_{\frac{1}{2}}+3\L^2+2\F , 
\ee

\be \label{SU5w3ASa1Ff}
\su(5)_{k}+3\L^2+\F\:;\:\:\:k=0,1,2
\ee

\be \label{SU5w3ASf}
\su(5)_{\frac{2l+1}{2}}+3\L^2\:;\:\:\:l=0,1,2,3
\ee

Most of these cases can be generated by integrating fundamental flavors from cases \eqref{SU5w3ASa3F} and \eqref{SU5w3ASa2F}, with the exception of case \eqref{SU5w3ASf} for $l=3$. This case is one of the few cases of $5d$ gauge theories that have a $5d$ SCFT UV completion, but can not be generated by integrating flavor out of $5d$ gauge theories that lift to $6d$ SCFTs. However, see the end of this subsubsection for a lift of this $5d$ gauge theory to a $5d$ KK theory.

\subsubsection*{\ubf{$\sp(4)$}:}

\be
\sp(4)+\half\L^3+(4-n)\F ,
\ee

This case can be generated by integrating out flavors from the $6d$ lifting case \eqref{SP4wTAS}. We also note that from gometry it appears that the theta angle for the $\sp$ group is physically irrelevant for the $n=4$ case.

\subsubsection*{\ubf{$\so(8)$}:}

\be \label{SO83SnF}
\so(8)+3\S+n\F\:;\:\:\:3\le n\le4
\ee

\be \label{SO82SnF}
\so(8)+2\S+n\F\:;\:\:\:2\le n\le4
\ee

\be
\so(8)+\S+n\F\:;\:\:\:1\le n\le5
\ee

\be \label{SO81C3S3F}
\so(8)+\C+3\S+3\F
\ee

\be
\so(8)+\C+2\S+n\F\:;\:\:\:2\le n\le4
\ee

\be
\so(8)+\C+\S+n\F\:;\:\:\:1\le n\le4
\ee

\be
\so(8)+2\C+2\S+n\F\:;\:\:\:2\le n\le3
\ee

All cases here can be generated by integrating out flavors from the $6d$ lifting cases. To avoid over-counting, we have used the triality outer automorphism of $\so(8)$ to set $n_C \leq n_S \leq n_F$, and hence the lower limitations on $n$. A brane construction for this class of theories was given in \cite{Zafrir:2015ftn}, which can be used to build brane webs for these theories with the exception of cases \eqref{SO83SnF}, \eqref{SO82SnF} and \eqref{SO81C3S3F}. The results in \cite{Hayashi:2017skf} allows the extension of this method also to the case of \eqref{SO82SnF}. It should be possible to use the result of \cite{Zafrir:2016csd}, and lift the class S construction for the $4d$ ${\cal N}=2$ $\so(8)$ SCFTs with spinor matter given in \cite{Chacaltana:2013dsj} to $5d$ to get brane web descriptions also for cases \eqref{SO83SnF} and \eqref{SO81C3S3F}.

\subsubsection*{\ubf{$\so(9)$}:}

\be \label{SO9w3SaFf}
\so(9)+3\S+(3-n)\F ,
\ee

\be \label{SO9w4SaFf}
\so(9)+4\S ,
\ee

\be \label{SO9w2SaFf}
\so(9)+2\S+(5-n) ,
\ee

\be \label{SO9w1SaFf}
\so(9)+\S+(6-n)\F ,
\ee

Case \eqref{SO9w3SaFf} can be generated by integrating fundamental matter from case \eqref{SO9w3Sa3F}. Similarly cases \eqref{SO9w4SaFf}, \eqref{SO9w2SaFf}, and \eqref{SO9w1SaFf} can be generated in the same way from cases \eqref{SO9w4Sa1F}, \eqref{SO9w2Sa5F}, and \eqref{SO9w1Sa6F}, respectively. A brane construction for this class of theories was given in \cite{Zafrir:2015ftn}, which is valid when the number of spinors is two or smaller. It should be possible to use the result of \cite{Zafrir:2016csd}, and lift the class S construction for the $4d$ ${\cal N}=2$ $\so(9)+3\S+\F$ SCFT given in \cite{Chacaltana:2015dat} to $5d$ to get a brane web description also for the cases with three spinors.    

The cases here are related to the previous cases and to one another by various dualities. See Section \ref{dualitys} for a list of these dualities.

\subsubsection*{\ubf{$\ff_4$}:}

\be
\ff_4+(3-n)\F\:;\:\:\:0\le n\le 3
\ee

This case is one of the few cases of $5d$ gauge theories that have a $5d$ SCFT UV completion, but can not be generated by integrating flavor out of $5d$ gauge theories that lift to $6d$ SCFTs, with the exception of the $n=3$ case which can be generated by integrating out the adjoint hyper from the maximally supersymmetric case. We also note that the $n=0$ case is dual to the $l=3$ case of \eqref{SU5w3ASf}, see \cite{Bhardwaj:2019ngx}.

Let us remark here that the $n<3$ cases can be obtained from $5d$ KK theories by performing a generalized ungauging (along with integrating out matter). It can be seen from the geometry for $\ff_4+3\F$ that the $\u(1)$ instanton flavor symmetry of the theory enhances at the conformal point to an $\su(2)$ subgroup of the flavor symmetry of the $5d$ SCFT\footnote{This can also be seen directly from the gauge theory from instanton counting \cite{Zafrir:2015uaa}.}. Gauging this $\su(2)$ symmetry produces the $5d$ KK theory obtained by untwisted compactification of $6d$ SCFT whose tensor branch description is provide by $6d$ $\ff_4+3\F$ gauge theory. Thus, the $5d$ $\ff_4+3\F$ gauge theory can be obtained from this $5d$ KK theory by ungauging the above-mentioned $\su(2)$ symmetry. The theories $ff_4+(3-n)\F$ for $n>0$ can then simply be obtained by integrating out matter from the $\ff_4+3\F$ theory. See \cite{Bhardwaj:2019xeg} for more details.

\subsubsection{Rank 5}

\subsubsection*{\ubf{$\su(6)$}:}

\be \label{SU6wHTASf}
\su(6)_{\frac{n-p}2}+\half\L^3+(13-n-p)\F ,
\ee

\be \label{SU6wHTASaCSf}
\su(6)_{3+\frac n2}+\half\L^3+(9-n)\F ,
\ee

\be \label{SU6wHTASaASf}
\su(6)_{\frac{n-p}2}+\half\L^3+\L^2+(9-n-p)\F ,
\ee

\be \label{SU6wHTASaASaCSf}
\su(6)_\frac{3+n}{2}+\half\L^3+\L^2+(8-n)\F ,
\ee

\be \label{SU6wHTASa2ASf}
\su(6)_\frac{2l+n-p-1}2+\half\L^3+2\L^2+(2-n-p)\F , l=1,2 
\ee

\be \label{SU6wTASaFf}
\su(6)_\frac{n-p}2+\L^3+(10-n-p)\F ,
\ee

\be \label{SU6wTASaFaCSf}
\su(6)_\frac{3+n}2+\L^3+(9-n)\F ,
\ee

\be \label{SU6wTASaASf}
\su(6)_\frac{n-p}2+\L^3+\L^2+(4-n-p)\F ,
\ee

\be \label{SU6wTASaASaCSf}
\su(6)_\frac{3+n}2+\L^3+\L^2+(3-n)\F ,
\ee

\be \label{SU6w3HTASf}
\su(6)_\frac{n-p}2+\frac32\L^3+(5-n-p)\F ,
\ee

\be \label{SU6w3HTASaCSf}
\su(6)_\frac{7}2+\frac32\L^3
\ee

\be \label{SU6wHTASaSf}
\su(6)_\frac{1}2+\S^2+\frac12\L^3
\ee

All cases here can be generated by integrating out fundamental flavors from $6d$ lifting cases. Specifically, case \eqref{SU6wHTASf} can be generated by integrating out $n$ fundamentals with a positive mass and $p$ fundamentals with a negative mass from case \eqref{SU6wHHTASa13F}. Case \eqref{SU6wHTASaCSf} can be generated by integrating out $n$ fundamentals with a positive mass from case \eqref{SU6wHHTASa9F}. Case \eqref{SU6wHTASaASf} can be generated by integrating out $n$ fundamentals with a positive mass and $p$ fundamentals with a negative mass from case \eqref{SU6wHHTASaASa9F}. Case \eqref{SU6wHTASaASaCSf} can be generated by integrating out $n$ fundamentals with a positive mass from case \eqref{SU6wHHTASaASa8F}. Case \eqref{SU6wHTASa2ASf} can be generated by integrating out $n$ fundamentals with a positive mass and $p$ fundamentals with a negative mass from cases \eqref{SU6wHHTASa2ASCSH} and \eqref{SU6wHHTASa2ASCS3H}. Case \eqref{SU6wTASaFf} can be generated by integrating out $n$ fundamentals with a positive mass and $p$ fundamentals with a negative mass from case \eqref{SU6wTASa10F}. Case \eqref{SU6wTASaFaCSf} can be generated by integrating out $n$ fundamentals with a positive mass from case \eqref{SU6wTASa9F}. Case \eqref{SU6wTASaASf} can be generated by integrating out $n$ fundamentals with a positive mass and $p$ fundamentals with a negative mass from case \eqref{SU6wTASaASa4F}. Case \eqref{SU6wTASaASaCSf} can be generated by integrating out $n$ fundamentals with a positive mass from case \eqref{SU6wTASaASa3F}. Case \eqref{SU6w3HTASf} can be generated by integrating out $n$ fundamentals with a positive mass and $p$ fundamentals with a negative mass from case \eqref{SU6w3HTASa5F}. Cases \eqref{SU6w3HTASaCSf} and \eqref{SU6wHTASaSf} can be generated by integrating out the fundamental flavor from cases \eqref{SU6w3HTASa1F} and \eqref{SU6wHTASaSa1F}, respectively.

Many of the cases here were found previously using brane constructions. Specifically, \cite{Hayashi:2019yxj} presented brane web constructions for cases \eqref{SU6wHTASf}, \eqref{SU6wHTASaCSf}, \eqref{SU6wHTASaASf}, \eqref{SU6wHTASaASaCSf}, \eqref{SU6wTASaFf}, \eqref{SU6wTASaFaCSf}, \eqref{SU6wTASaASf}, \eqref{SU6w3HTASf}, and \eqref{SU6wHTASaSf}. The remaining cases are new, to our knowledge.

We also note that case \eqref{SU6wHTASf} with $p=0$ is dual to the $m=6$, $p=0$ case of \eqref{FSUmw1AS}, and that case \eqref{SU6wHTASaCSf} is dual to both the $m=6$ case of \eqref{FSUmw1ASa} and the $m=5$ case of \eqref{USpwAS}, where for $n=8$ the $\sp$ theta angle of the dual theory is $0$. Case \eqref{SU6wHTASaASaCSf} is dual to the $m=6$ case of \eqref{FSUmw2ASa}.

\subsubsection*{\ubf{$\so(11)$}:}

\be \label{SO11w2Sf}
\so(11)+2\S+(3-n)\F ,
\ee

\be \label{SO11w3HSf}
\so(11)+\frac32\S+(5-n)\F ,
\ee

\be \label{SO11w1Sf}
\so(11)+\S+(7-n)\F ,
\ee

\be \label{SO11wHSf}
\so(11)+\half\S+(8-n)\F ,
\ee

All cases here can be generated by integrating fundamental matter from the $6d$ lifting cases. A brane construction for this class of theories was given in \cite{Zafrir:2015ftn}, which can be used to build brane webs for cases \eqref{SO11wHSf} and \eqref{SO11w1Sf}. It should be possible to use the result of \cite{Zafrir:2016csd}, and lift the class S construction for the $4d$ ${\cal N}=2$ $\so(11)$ SCFTs with spinor matter given in \cite{Chacaltana:2015dat} to $5d$ to get brane web descriptions also for cases \eqref{SO11w3HSf} and \eqref{SO11w2Sf}.

We also note that case \eqref{SO11w2Sf} is dual to case \eqref{SU6wTASaASaCSf}, while case \eqref{SO11w3HSf} is dual to the $p=0$ case of \eqref{SU6w3HTASf}.

\subsubsection*{\ubf{$\so(10)$}:}

\be
\so(10)+4\S+(2-n)\F ,
\ee

\be
\so(10)+3\S+(4-n)\F ,
\ee

\be
\so(10)+2\S+(6-n)\F ,
\ee

\be
\so(10)+\S+(7-n)\F ,
\ee

All cases here can be generated by integrating fundamental matter from the $6d$ lifting cases. A brane construction for this class of theories was given in \cite{Zafrir:2015ftn}, which can be used to build brane webs for cases with two or less spinors. It should be possible to use the result of \cite{Zafrir:2016csd}, and lift the class S construction for the $4d$ ${\cal N}=2$ $\so(12)$ SCFTs with spinor matter given in \cite{Chacaltana:2015dat} to $5d$ to get brane web descriptions also for the other cases.

\subsubsection{Rank 6}

\subsubsection*{\ubf{$\su(7)$}:}

\be \label{SU7wTASa6Ff}
\su(7)_\frac{n-p}2+\L^3+(6-n-p)\F ,
\ee

\be \label{SU7wTASa5Ff}
\su(7)_\frac{3+n}2+\L^3+(5-n)\F
\ee

Case \eqref{SU7wTASa6Ff} can be generated by integrating out $n$ fundamentals with a positive mass and $p$ fundamentals with a negative mass from the $6d$ lifting case \eqref{SU7wTASa6F}. Case \eqref{SU7wTASa5Ff} can be generated by integrating out $n$ fundamentals with a positive mass from the $6d$ lifting case \eqref{SU7wTASa5F}.

\subsubsection*{\ubf{$\so(13)$}:}

\be \label{SO13w1Sf}
\so(13)+\S+(5-n)\F ,
\ee

\be
\so(13)+\half\S+(9-n)\F ,
\ee

These cases can be generated by integrating out fundamental flavors from the $6d$ lifting cases of $\so(13)$ with spinor matter. Case \eqref{SO13w1Sf} is dual to case \eqref{SU7wTASa5Ff}.

\subsubsection*{\ubf{$\so(12)$}:}

\be \label{SO12w2Sf}
\so(12)+2\S+(4-n)\F ,
\ee

\be \label{SO12w3HSf}
\so(12)+\frac32\S+(6-n)\F ,
\ee

\be \label{SO12w1Sf}
\so(12)+\S+(8-n)\F ,
\ee

\be \label{SO12wHSf}
\so(12)+\half\S+(9-n)\F ,
\ee

\be
\so(12)+\frac32\S+\C
\ee

\be \label{SO12w3HSaHCf}
\so(12)+\frac32\S+\half\C+(4-n)\F ,
\ee

\be \label{SO12w1Sa1Cf}
\so(12)+\S+\C+(4-n)\F ,
\ee

\be \label{SO12w1SaHCf}
\so(12)+\S+\half\C+(6-n)\F ,
\ee

\be \label{SO12wHSaHCf}
\so(12)+\half\S+\half\C+(8-n)\F ,
\ee

All cases here can be generated by integrating fundamental matter from the $6d$ lifting cases. A brane construction for this class of theories was given in \cite{Zafrir:2015ftn}, which can be used to build brane webs for cases \eqref{SO12wHSf} and \eqref{SO12w1Sf}. The results in \cite{Hayashi:2017skf} allows the extension of this method also to the case of \eqref{SO12wHSaHCf}. It should be possible to use the result of \cite{Zafrir:2016csd}, and lift the class S construction for the $4d$ ${\cal N}=2$ $\so(12)$ SCFTs with spinor matter given in \cite{Chacaltana:2015dat} to $5d$ to get brane web descriptions also for cases \eqref{SO12w3HSf}, \eqref{SO12w2Sf}, \eqref{SO12w1SaHCf}, \eqref{SO12w3HSaHCf} and \eqref{SO12w1Sa1Cf}.

\subsubsection*{\ubf{$\fe_6$}:}

\be
\fe_6+n\F\:;\:\:\:0\le n\le4 ,
\ee

This case is one of the few cases of $5d$ gauge theories that have a $5d$ SCFT UV completion, but can not be generated by integrating flavor out of $5d$ gauge theories that lift to $6d$ SCFTs, with the exception of the $n=0$ case which can be generated by integrating out the adjoint hyper from the maximally supersymmetric case.

The $n>0$ cases can be obtained from $5d$ KK theories in the same fashion as discussed towards the end of subsubsection (\ref{r4}). The $5d$ gauge theory $\fe_6+4\F$ admits an instantonic $\su(2)$ flavor symmetry which can be gauged to produce the $5d$ KK theory obtained from an untwisted compactification of the $6d$ SCFT with tensor branch described by $6d$ gauge theory $\fe_6+4\F$. See \cite{Bhardwaj:2019xeg} for more details.

\subsubsection{Rank 7}

\subsubsection*{\ubf{$\so(14)$}:}

\be
\so(14)+\S+(6-n)\F ,
\ee

This case can be generated by integrating out flavors from the $6d$ lifting case.

\subsubsection*{\ubf{$\fe_7$}:}

\be
\fe_7+\frac{n}{2}\F\:;\:\:\: 0\le n\le 6 ,
\ee

This case is one of the few cases of $5d$ gauge theories that have a $5d$ SCFT UV completion, but can not be generated by integrating flavor out of $5d$ gauge theories that lift to $6d$ SCFTs, with the exception of the $n=0$ case which can be generated by integrating out the adjoint hyper from the maximally supersymmetric case. We also note that as one cannot integrate out an odd number of half-hyper multiplets, the cases with even and odd $n$ sit in distinct flow families.

The $n=6$ and $n=5$ cases can be obtained from $5d$ KK theories by a generalized ungauging. To construct the $n=6$ case, we start with the $5d$ KK theory produced by untwisted comapctification of the $6d$ SCFT whose tensor branch is described by the $6d$ gauge theory $\fe_7+3\F$. This $5d$ KK theory can be obtained by gauging an $\su(2)$ instantonic flavor symmetry of $5d$ $\fe_7+3\F$, the ungauging of which leading to the above $n=6$ case. The $n=5$ case is obtained by applying a generalized ungauging on the $5d$ KK theory obtained by untwisted comapctification of $6d$ SCFT with tensor branch $6d$ gauge theory $\fe_7+\frac52\F$. In this case, the generalized ungauging process cannot be interpreted as ungauging of an instantonic symmetry. See \cite{Bhardwaj:2019xeg} for more details.

\subsubsection{Rank 8}

\subsubsection*{\ubf{$\fe_8$}:}

\be
\fe_8
\ee

This case can be generated by integrating the adjoint hyper out of the maximally supersymmetric case.

\subsection{Inconsistent theories}\label{it}
In this section, we collect the $5d$ gauge theories allowed by \cite{Jefferson:2017ahm}, but disallowed by our analysis. These theories are as follows:
\be
\su(3)_\frac{13}2+3\F\:=\:\sp(2)_\pi+3\L^2
\ee

\be
\su(3)_7+2\F
\ee

\be
\su(4)_3+8\F
\ee

\be
\su(4)_\frac72+7\F
\ee

\be
\su(4)_1+4\L^2
\ee

\be
\su(4)_3+4\L^2
\ee

\be
\sp(3)_\pi+2\L^2
\ee

\be
\su(5)_\frac{11+n}2+(5-n)\F\:;\:\:\:0\le n\le4
\ee

\be
\su(5)_3+3\L^2+\F
\ee

\be
\su(6)_0+3\L^2
\ee

\be
\su(6)_2+3\L^2
\ee

\be
\su(6)_2+\frac32\L^3+3\F
\ee

\be
\su(6)_\frac52+\frac32\L^3+2\F
\ee

\be
\so(12)+2\S+\frac12\C
\ee

We have taken the help of two kinds of arguments to rule these theories out:
\ben
\item In the first argument, a $5d$ gauge theory satisfying the conditions of \cite{Jefferson:2017ahm} is shown to be dual to a $5d$ gauge theory which does not satisfy the conditions of \cite{Jefferson:2017ahm}. Since the latter theory is not supposed to admit an SCFT UV completion, the former theory should not admit an SCFT UV completion either.
\item In the second argument, by deforming a $5d$ gauge theory we land onto another $5d$ gauge theory which is known to admit no SCFT UV completion, either by the conditions of \cite{Jefferson:2017ahm} or by the first argument. Since deforming a theory with pure field-theoretic UV completion should lead to a theory with purely field-theoretic UV completion, we are lead to the conclusion that the gauge theory before the deformation should not admit an SCFT UV completion.
\een
The detailed arguments for each of the above cases can be found in the appropriate subsections of Section \ref{da}.

\subsection{Undetermined theories}\label{ur}
Finally, we collect all the theories which satisfy the criteria of \cite{Jefferson:2017ahm}, but we are neither able to confirm the existence of these theories nor rule them out. That is, we are neither able to put the geometry corresponding to these gauge theories in a form manifesting the structure of a $5d$ KK theory (which is discussed in Section \ref{5dkk}), nor are we able to apply either of the two kinds of arguments discussed at the end of Section \ref{it}. 

These theories are as follows:
\be
\su(4)_7+\L^2
\ee

\be\label{exc}
\su(5)_8
\ee

\be
\su(6)_9
\ee

\be
\su(6)_4+\L^3+\L^2
\ee

\be
\su(7)_5+\L^3
\ee

\be
\so(12)+\frac52\S
\ee

According to the criteria proposed in \cite{Jefferson:2017ahm}, all of the above cases except for the case of \eqref{exc} may either have a UV completion as a $6d$ SCFT or may have no UV completion at all. The case \eqref{exc}, on the other hand, may either have UV completion as a $5d$ SCFT, or as a $6d$ SCFT, or no UV completion at all. The case \eqref{exc} descends from the marginal case $\su(5)_\frac{11}{2}+5\F$ of \cite{Jefferson:2017ahm}. We show later in this paper the following duality
\be\nn
\su(5)_{\frac{11+n}2}+(5-n)\F\:=\:\sp(4)+(4-n)\F+\L^4
\ee
according to which the above marginal theory and its descendants are dual to $\sp(4)$ theories containing $\L^4$, but such $\sp(4)$ theories are ruled out by the criteria of \cite{Jefferson:2017ahm}. This duality is not applicable to the $n=5$ case, and thus this argument is insufficient to decide the fate of \eqref{exc}.

\subsection{Dualities}\label{dr}
In this subsection, we collect the dualities between different $5d$ gauge theories.
\subsubsection{KK theories}\label{duality}
\be
\begin{tikzpicture} [scale=1.9]
\node at (-5.6,0.9) {$\sp(m+1)+(2m+8)\F$};
\node at (-8.3,0.9) {$\su(m+2)_0+(2m+8)\F$};
\node at (-6.9,0.9) {$=$};
\end{tikzpicture}
\ee

\be
\begin{tikzpicture} [scale=1.9]
\node at (-5.7,0.9) {$\sp(m+1)+\L^2+8\F$};
\node at (-8.3,0.9) {$\su(m+2)_{\frac m2+1}+\L^2+8\F$};
\node at (-6.9,0.9) {$=$};
\end{tikzpicture}
\ee

\be
\begin{tikzpicture} [scale=1.9]
\node at (-7.7,0.9) {$\su(2m)_{m}+\S^2$};
\node at (-6.7,0.9) {$=$};
\node at (-5.6,0.9) {$\sp(2m-1)_0+\A$};
\end{tikzpicture}
\ee

\be
\begin{tikzpicture} [scale=1.9]
\node at (-7.5,0.9) {$\su(2m+1)_{m+\half}+\S^2$};
\node at (-6.3,0.9) {$=$};
\node at (-5.4,0.9) {$\sp(2m)_\pi+\A$};
\end{tikzpicture}
\ee

\be
\begin{tikzpicture} [scale=1.9]
\node at (-7.7,0.9) {$\sp(2)+2\L^2+4\F$};
\node at (-9.7,0.9) {$\su(3)_4+6\F$};
\node at (-8.8,0.9) {$=$};
\node at (-6.6,0.9) {$=$};
\node at (-5.9,0.9) {$\fg_2+6\F$};
\end{tikzpicture}
\ee

\be
\begin{tikzpicture} [scale=1.9]
\node at (-6.4,0.9) {$\fg_2+\A$};
\node at (-8,0.9) {$\su(3)_\frac{15}2+\F$};
\node at (-7.1,0.9) {$=$};
\end{tikzpicture}
\ee

\be
\begin{tikzpicture} [scale=1.9]
\node at (-6.1,0.9) {$\sp(3)+\L^3+5\F$};
\node at (-8,0.9) {$\su(4)_4+6\F$};
\node at (-7.1,0.9) {$=$};
\end{tikzpicture}
\ee

\be
\begin{tikzpicture} [scale=1.9]
\node at (-8.2,0.9) {$\su(4)_\frac 32+2\L^2+7\F$};
\node at (-7.1,0.9) {$=$};
\node at (-6,0.9) {$\sp(3)+\frac12\L^3+\frac{19}2\F$};
\end{tikzpicture}
\ee

\be
\begin{tikzpicture} [scale=1.9]
\node at (-6.2,0.9) {$\sp(3)_0+2\L^2$};
\node at (-8,0.9) {$\su(4)_6+2\L^2$};
\node at (-7.1,0.9) {$=$};
\end{tikzpicture}
\ee

\be
\begin{tikzpicture} [scale=1.9]
\node at (-6.3,0.9) {$\so(7)+6\S+\F$};
\node at (-8.4,0.9) {$\su(4)_0+3\L^2+4\F$};
\node at (-7.3,0.9) {$=$};
\end{tikzpicture}
\ee

\be
\begin{tikzpicture} [scale=1.9]
\node at (-7.9,0.9) {$\so(7)+5\S+2\F$};
\node at (-10,0.9) {$\su(4)_1+3\L^2+4\F$};
\node at (-8.9,0.9) {$=$};
\node at (-6.9,0.9) {$=$};
\node at (-6.1,0.9) {$\so(7)+7\S$};
\end{tikzpicture}
\ee

\be
\begin{tikzpicture} [scale=1.9]
\node at (-8,0.9) {$\su(4)_2+3\L^2+4\F$};
\node at (-6.9,0.9) {$=$};
\node at (-5.9,0.9) {$\so(7)+4\S+3\F$};
\end{tikzpicture}
\ee

\be
\begin{tikzpicture} [scale=1.9]
\node at (-8.4,0.9) {$\su(4)_5+3\L^2$};
\node at (-7.5,0.9) {$=$};
\node at (-6.2,0.9) {$\sp(3)+\frac12\L^3+\L^2+\frac{5}2\F$};
\end{tikzpicture}
\ee

\be
\begin{tikzpicture} [scale=1.9]
\node at (-8.3,0.9) {$\su(5)_0+3\L^2+3\F$};
\node at (-7.2,0.9) {$=$};
\node at (-6.2,0.9) {$\so(9)+3\S+3\F$};
\end{tikzpicture}
\ee

\be
\begin{tikzpicture} [scale=1.9]
\node at (-5.9,0.9) {$\so(9)+4\S+\F$};
\node at (-8.1,0.9) {$\su(5)_{\frac{3}2}+3\L^2+2\F$};
\node at (-6.9,0.9) {$=$};
\end{tikzpicture}
\ee

\be
\begin{tikzpicture} [scale=1.9]
\node at (-9.1,0.9) {$\su(6)_0+\L^2+12\F$};
\node at (-6.8,0.9) {$\su(6)_0+\half\L^3+13\F$};
\node at (-8,0.9) {$=$};
\end{tikzpicture}
\ee

\be
\begin{tikzpicture} [scale=1.9]
\node at (-5.5,0.9) {$\sp(5)+\L^2+8\F$};
\node at (-7.6,0.9) {$\su(6)_3+\half\L^3+9\F$};
\node at (-6.5,0.9) {$=$};
\node at (-8.7,0.9) {$=$};
\node at (-9.8,0.9) {$\su(6)_3+\L^2+8\F$};
\end{tikzpicture}
\ee

\be
\begin{tikzpicture} [scale=1.9]
\node at (-8.9,0.9) {$\su(6)_\frac32+2\L^2+7\F$};
\node at (-6.3,0.9) {$\su(6)_\frac32+\half\L^3+\L^2+8\F$};
\node at (-7.7,0.9) {$=$};
\end{tikzpicture}
\ee

\be
\begin{tikzpicture} [scale=1.9]
\node at (-6,0.9) {$\so(11)+2\S+3\F$};
\node at (-8.4,0.9) {$\su(6)_\frac32+\L^3+\L^2+3\F$};
\node at (-7.1,0.9) {$=$};
\end{tikzpicture}
\ee

\be
\begin{tikzpicture} [scale=1.9]
\node at (-8.3,0.9) {$\su(6)_0+\frac32\L^3+5\F$};
\node at (-7.2,0.9) {$=$};
\node at (-6.1,0.9) {$\so(11)+\frac32\S+5\F$};
\end{tikzpicture}
\ee

\be
\begin{tikzpicture} [scale=1.9]
\node at (-7.7,0.9) {$\su(6)_3+\frac32\L^3+\F$};
\node at (-6.6,0.9) {$=$};
\node at (-5.7,0.9) {$\so(11)+\frac52\S$};
\node at (-10,0.9) {$\su(6)_\frac72+\half\L^3+2\L^2$};
\node at (-8.8,0.9) {$=$};
\end{tikzpicture}
\ee

\be
\begin{tikzpicture} [scale=1.9]
\node at (-5.9,0.9) {$\so(13)+\S+5\F$};
\node at (-8,0.9) {$\su(7)_\frac32+\L^3+5\F$};
\node at (-6.9,0.9) {$=$};
\end{tikzpicture}
\ee

\be
\begin{tikzpicture} [scale=1.9]
\node at (-5.6,0.9) {$\so(12)+\S+\half\C+6\F$};
\node at (-8,0.9) {$\so(12)+\frac32\S+6\F$};
\node at (-6.9,0.9) {$=$};
\end{tikzpicture}
\ee

\subsubsection{SCFTs}\label{dualitys}
\be
\su(m+2)_\frac n2+(2m+8-n)\F\:=\:\sp(m+1)+(2m+8-n)\F\:;\:\:\:n\le2m+7
\ee

\be
\su(m+2)_{m+4}\:=\:\sp(m+1)_{m\pi}
\ee

\be
\su(m+2)_{\frac{m+n}2+1}+\L^2+(8-n)\F\:=\:\sp(m+1)+\L^2+(8-n)\F\:;\:\:\:n\le7
\ee

\be
\su(m+2)_{\frac{m}2+5}+\L^2\:=\:\sp(m+1)_{m\pi}+\L^2
\ee

\be
\su(3)_{4+\frac n2}+(6-n)\F\:=\:\sp(2)+2\L^2+(4-n)\F\:=\:\fg_2+(6-n)\F\:;\:\:\:n\le3
\ee

\be
\su(3)_{6}+2\F\:=\:\sp(2)_\pi+2\L^2\:=\:\fg_2+2\F
\ee

\be
\su(3)_{6+\frac n2}+(2-n)\F\:=\:\fg_2+(2-n)\F
\ee

\be
\su(4)_{4+\frac n2}+(6-n)\F\:=\:\sp(3)+\L^3+(5-n)\F
\ee

\be
\su(4)_{\frac{3+n}2}+2\L^2+(7-n)\F\:=\:\sp(3)+\half\L^3+\frac{19-2n}{2}\F\:;\:\:\:n\le7
\ee

\be
\su(4)_{\half}+3\L^2+3\F\:=\:\so(7)+5\S+\F\:=\:\so(7)+6\S
\ee

\be
\su(4)_{\frac{n-1}2}+3\L^2+(3-n)\F\:=\:\so(7)+(6-n)\S\:;\:\:\:n\le3
\ee

\be
\su(4)_{\frac{n+1}2}+3\L^2+(3-n)\F\:=\:\so(7)+(5-n)\S+\F\:;\:\:\:n\le2
\ee

\be
\su(4)_{1+\frac{n}2}+3\L^2+(4-n)\F\:=\:\so(7)+(5-n)\S+2\F\:;\:\:\:n\le2
\ee

\be
\su(4)_{\frac52}+3\L^2+3\F\:=\:\so(7)+3\S+3\F
\ee

\be
\su(5)_\frac n2+3\L^2+(3-n)\F\:=\:\so(9)+3\S+(3-n)\F
\ee

\be
\su(5)_2+3\L^2+\F\:=\:\so(9)+4\S
\ee

\be
\su(5)_\frac72+3\L^2\:=\:\ff_4+3\F
\ee

\be
\su(6)_{\frac n2}+\L^2+(12-n)\F\:=\:\su(6)_{\frac n2}+\half\L^3+(13-n)\F\:;\:\:\:n\le12
\ee

\be
\su(6)_{3+\frac n2}+\L^2+(8-n)\F\:=\:\su(6)_{3+\frac n2}+\half\L^3+(9-n)\F\:=\:\sp(5)+\L^2+(8-n)\F\:;\:\:\:n\le7
\ee

\be
\su(6)_{7}+\L^2\:=\:\su(6)_{7}+\half\L^3+\F\:=\:\sp(5)_0+\L^2
\ee

\be
\su(6)_\frac{3+n}2+2\L^2+(7-n)\F\:=\:\su(6)_\frac{3+n}{2}+\half\L^3+\L^2+(8-n)\F\:;\:\:\:n\le7
\ee

\be
\su(6)_\frac{3+n}2+\L^3+\L^2+(3-n)\F\:=\:\so(11)+2\S+(3-n)\F
\ee

\be
\su(6)_\frac{n}2+\frac32\L^3+(5-n)\F\:=\:\so(11)+\frac32\S+(5-n)\F
\ee

\be
\su(7)_\frac{3+n}2+\L^3+(5-n)\F\:=\:\so(13)+\S+(5-n)\F
\ee

\subsection{Relationship of our work to the classification of $5d$ SCFTs}\label{rr}
As we have seen, in this work, we are able to divide the theories appearing in \cite{Jefferson:2017ahm} into the following four sets:
\ben
\item The theories appearing in Section \ref{kkr} which UV complete into $5d$ KK theories.
\item The theories appearing in Section \ref{scftr} which UV complete into $5d$ SCFTs. Combining the results of this paper with the results of \cite{Bhardwaj:2019xeg}, we conclude that all these $5d$ SCFTs descend from $5d$ KK theories. However, the descent is not as simple as integrating out some BPS particles. In some cases that were discussed in \cite{Bhardwaj:2019xeg}, the descent requires integrating out BPS strings as well.
\item The theories appearing in Section \ref{it} which do not UV complete into $5d$ KK theories or $5d$ SCFTs.
\item The theories appearing in Section \ref{ur} for which it is not clear whether or not they admit an SCFT UV completion.
\een
Thus, our results provide evidence for the conjectures made in \cite{Bhardwaj:2019xeg} regarding the classification of $5d$ SCFTs. We have found that all the theories appearing in \cite{Jefferson:2017ahm} which can be shown to admit UV completions into $5d$ SCFTs indeed descend from $5d$ KK theories. We have also identified a set of theories in Section \ref{ur} for which it is not clear whether or not they admit an SCFT UV completion. Further analysis of these theories should provide another opportunity to test and challenge the conjectures of \cite{Bhardwaj:2019xeg}.

\section{Geometric description of $5d$ theories}\label{gd}
Throughout this paper, we will use a graphical notation to represent a local neighborhood of a collection of Hirzebruch surfaces intersecting each other inside a Calabi-Yau threefold. This notation and relevant background on Hirzebruch surfaces can be found in Section 2 of \cite{Bhardwaj:2019ngx}, and Section 5.2.1 and Appendix A of \cite{Bhardwaj:2019fzv}. A special role will be played by the automorphism $\cS$ exchanging $e$ and $f$ curves inside the Hirzebruch surface $\bF_0$, which is described in Section 2.6 of \cite{Bhardwaj:2019ngx}. Also relevant are isomorphisms $\cI_n$ and $\cI_n^{-1}$ between Hirzebruch sufaces of different degrees which are described in Section 2.1 of \cite{Bhardwaj:2019ngx}.

\subsection{General features}\label{gf}
Each Calabi-Yau threefold $X$ appearing in this paper is described as a local neighborhood of a collection of intersecting compact Kahler surfaces $S_i$. An intersection between the surfaces $S_i$ and $S_j$ is described as a ``gluing'' between the two surfaces, with the intersection locus being described as the identification of two curves
\be\label{id}
C^{(\alpha)}_{ij}\sim C^{(\alpha)}_{ji}
\ee
where $C^{(\alpha)}_{ij}$ is a curve in $S_i$ and $C^{(\alpha)}_{ji}$ is a curve in $S_j$. $\alpha$ parametrizes different intersections between $S_i$ and $S_j$ with the corresponding \emph{gluing curves} being $C^{(\alpha)}_{ij}$ and $C^{(\alpha)}_{ji}$. The curves
\be
C_{ij}:=\sum_\alpha C^{(\alpha)}_{ij}
\ee
in $S_i$ and
\be
C_{ji}:=\sum_\alpha C^{(\alpha)}_{ji}
\ee
in $S_j$ are referred to as \emph{total gluing curves} for the intersections between $S_i$ and $S_j$.

Consistency of (\ref{id}) requires
\be
C^{(\alpha)}_{ij}\cdot S_k=C^{(\alpha)}_{ji}\cdot S_k
\ee
for all $S_k$. We can compute the intersection number of a compact curve $C$ in $X$ with a surface $S_i$ as follows. If $C$ lives in a surface $S_j\neq S_i$, then
\be
C\cdot S_i = C\cdot C_{ji}
\ee
where the right hand side is computed inside $S_j$. If $C$ lives in $S_i$, then
\be
C\cdot S_i=K'_i\cdot C
\ee
where
\be
K'_i=K_i+\sum_\alpha\left(C^{(\alpha)}_{i}+D^{(\alpha)}_i\right)
\ee
where
\be
C^{(\alpha)}_{i}\sim D^{(\alpha)}_{i}
\ee
describe different \emph{self-gluings} of $S_i$, and $K_i$ is the canonical divisor of $S_i$. The genus $g$ of a curve $C$ in $S_i$ is computed by using
\be
2g-2=(K_i+ C)\cdot C+2\sum_\alpha n^{(\alpha)}
\ee
where
\be
n^{(\alpha)}:=\text{min}(n^{(\alpha)}_1,n^{(\alpha)}_2)
\ee
where
\be
n^{(\alpha)}_1:=C\cdot C^{(\alpha)}_i
\ee
and
\be
n^{(\alpha)}_2:=C\cdot D^{(\alpha)}_i
\ee
are the intersections of $C$ with the curves involved in the self-gluings of $S_i$.

Moreover, for the gluing (\ref{id}) to be consistent with Calabi-Yau structure of $X$, we must have
\be\label{CY}
\left(C^{(\alpha)}_{ij}\right)^2+\left(C^{(\alpha)}_{ji}\right)^2=2g-2
\ee
where $g$ is the genus of $C^{(\alpha)}_{ij}$ which must be equal to the genus of $C^{(\alpha)}_{ji}$ for consistency. Let us emphasize that the self-intersection $\left(C^{(\alpha)}_{ij}\right)^2$ of $C^{(\alpha)}_{ij}$ is computed inside surface $S_i$ since the curve $C^{(\alpha)}_{ij}$ lives in $S_i$ by definition. The condition (\ref{CY}) is referred to as the \emph{Calabi-Yau condition}.

Another consistency condition on the gluing curves $C^{(\alpha)}_{ij}$ comes from equating the various ways of computing the triple intersection number $S_i\cdot S_j\cdot S_k$ for three distinct surfaces. This triple intersection number can be computed in three different ways
\be
S_i\cdot S_j\cdot S_k=C_{ij}\cdot C_{ik}=C_{ji}\cdot C_{jk}=C_{ki}\cdot C_{kj}
\ee
There are two different ways of computing intersection numbers of the form $S_i^2\cdot S_j$ for $S_i\neq S_j$ as well
\be
S_i^2\cdot S_j=C^2_{ji}=K'_i\cdot C_{ij}
\ee
which provide another consistency condition on the gluings.

The (normalizable part of the) Kahler class is defined as
\be
J:=\sum_i\phi_i S_i
\ee
where $\phi_i$ are the normalizable Kahler parameters which are identified as the Coulomb branch moduli of the $5d$ theory. We are ignoring the contribution from non-normalizable Kahler parameters which are identified as the (supersymmetry preserving) mass parameters of the $5d$ theory. The contribution of the Coulomb branch moduli to the mass of a BPS particle coming from an M2 brane wrapping a compact curve $C$ in $X$ can be computed by
\be
\text{vol}(C):=-J\cdot C
\ee
The contribution of the Coulomb branch moduli to the tension of a BPS string coming from an M5 brane wrapping $S_i$ can be computed by
\be
\text{vol}(S_i):=\half J^2\cdot S_i
\ee
The contribution of the Coulomb branch moduli to the prepotential of the $5d$ theory can be computed by
\be
\cF=\frac16J^3
\ee

\subsection{Structure of $5d$ gauge theory}\label{5dgauge}
For a geometry $X$ to describe a $5d$ $\cN=1$ gauge theory, a necessary condition is that all of the surfaces have to be presented as Hirzebruch surfaces
\be
S_i=\bF_{n_i}^{b_i}
\ee
where $n_i$ is the degree of the Hirzebruch surface and $b_i$ are the number of blowups on the Hirzebruch surface. Once such a description is chosen\footnote{We note that it might not always be possible to choose a description in terms of Hirzebruch surfaces. For example, one of the $S_i$ might be equal to $\P^2$ which is not isomorphic to any Hirzebruch surface.}, we can associate an \emph{intersection matrix} to the geometry, which is defined as
\be
\cI_{ij}:=-f_i\cdot S_j
\ee
where $f_i$ is the $\P^1$ fiber of the Hirzebruch surface $S_i$. 

One of the requirements for the geometry to describe a $5d$ gauge theory with gauge algebra $\fg$ is that the intersection matrix $\cI_{ij}$ associated to the geometry equals the Cartan matrix of $\fg$ \cite{Bhardwaj:2019ngx}. Another requirement is that if a gluing curve $C_{ij}^{(\alpha)}$ can be written as
\be\label{p1}
C_{ij}^{(\alpha)}=\alpha_i f_i+\sum_{a=1}^{b_i}\beta_{i,a}x_{i,a}
\ee
with $x_{i,a}$ being the blowups on $S_i$ and $\alpha_i,\beta_{i,a}$ being integers\footnote{$\alpha_i$ must be non-negative integer for the curve to be holomorphic.}, then the gluing curve $C_{ji}^{(\alpha)}$ must take a similar form
\be\label{p2}
C_{ji}^{(\alpha)}=\alpha_j f_j+\sum_{a=1}^{b_j}\beta_{j,a}x_{j,a}
\ee
This is because an M2 brane wrapping a curve of the form shown on the right hand side of (\ref{p1}) describes a perturbative BPS particle in the $5d$ gauge theory, while M2 brane wrapping a curve of the more general form
\be
C_{ji}^{(\alpha)}=\gamma_j e_j+\alpha_j f_j+\sum_{a=1}^{b_j}\beta_{j,a}x_{j,a}
\ee
with $\gamma_j>0$ describes an instantonic BPS particle in the $5d$ gauge theory. So, the identification
\be
C_{ij}^{(\alpha)}\sim C_{ji}^{(\alpha)}
\ee
is compatible with the structure of a $5d$ gauge theory only if $\gamma_j$ is zero and $C_{ji}^{(\alpha)}$ takes the form shown on right hand side of (\ref{p2}).

The matter content for the $5d$ gauge theory is encoded in the blowups $x_{i,a}$ and their gluings. Our task now is to describe how one can read the matter content associated to a geometry $X$ giving rise to a $5d$ gauge theory with gauge algebra $\fg$. First of all, notice that the intersection matrix of a geometry remains unchanged if we flop a $-1$ curve of the form\footnote{We refer to such flops as \emph{perturbative flops} in what follows.} $x_{i,a}$ or $f_i-x_{i,a}$. That is, the geometry $X'$ obtained after performing such a flop on $X$ describes a $5d$ gauge theory with the same gauge algebra $\fg$. In fact, $X'$ describes the same $5d$ gauge theory, and the flop transition corresponds to a phase transition on the mass-deformed Coulomb branch\footnote{In this paper, ``mass-deformed Coulomb branch'' refers to the space obtained by adjoining the space of mass parameters with Coulomb branch moduli space for each value of mass parameters. All phase transitions occur on this combined space.} of the $5d$ gauge theory.

Thus, by performing such flops we can simplify $X$ into a geometry from which it is straightforward to read the associated matter content. So, we associate a simple geometry $X_{\fg,R}$ to every $5d$ gauge theory with a gauge algebra $\fg$ and matter transforming in a representation $R$ of $\fg$. If performing perturbative flops on $X$ converts $X$ to $X_{\fg,R}$, then $X$ describes a $5d$ gauge theory with gauge algebra $\fg$ and matter transforming in representation $R$ of $\fg$. 

As long as $R$ contains no half-hypermultiplets, the geometry $X_{\fg,R}$ can be described easily in terms of the geometry $X_\fg$ associated to the pure $5d$ gauge theory with gauge algebra $\fg$ described in detail in Section 2.4 of \cite{Bhardwaj:2019ngx}. Notice that, unlike the gauge theories carrying non-trivial amount of matter, there is a unique geometry describing a pure $5d$ gauge theory. This is because there is a single perturbative phase for a pure $5d$ gauge theory. Consequently, the Hirzebruch surfaces $S_i$ inside $X_\fg$ carry no blowups. 

Now, let us describe the construction of $X_{\fg,R}$ when $R$ contains no half-hypermultiplets. Let
\be
R=\sum_{\mu=1}^{m} R_\mu
\ee
where each\footnote{We allow the possibility of $R_\mu=R_{\mu'}$ for $\mu\neq\mu'$.} $R_\mu$ is an irreducible representation of $\fg$. We build $X_{\fg,R}$ inductively starting from $X_\fg$ at step zero. At each step $\mu$ for $1\le \mu\le m$, we construct a geometry $X_\fg^\mu$ out of the geometry $X_\fg^{\mu-1}$ obtained at step $(\mu-1)$. To do this, let $n_i$ be the Dynkin coefficients of the highest weight of $R_\mu$. For each $i\in[1,r]$ where $r$ is the rank of $\fg$, perform $n_i$ blowups on the surface $S_i$. That is, we have performed a total of $n_\mu:=\sum_{i=1}^rn_i$ blowups. Now, glue each pair of blowups in this set of $n_\mu$ blowups. Let $C_{ij,\mu}$ and $C_{ji,\mu}$ be the total gluing curves describing the gluing between $S_i$ and $S_j$ in $X_\fg^\mu$. Then, we have
\be
C_{ij,\mu}=C_{ij,(\mu-1)}+n_j\sum_{a=1}^{n_i}x_{i,a}
\ee
where $x_{i,a}$ are the $n_i$ blowups performed on $S_i$ at step $\mu$. Similarly, let $K'_{i,\mu}$ be the $K'$ for $S_i$ in $X_\fg^\mu$. Then, we have
\be
K'_{i,\mu}=K'_{i,(\mu-1)}+n_i\sum_{a=1}^{n_i}x_{i,a}
\ee
Finally
\be
X_{\fg,R}:=X_\fg^m
\ee
That is, $X_{\fg,R}$ is defined to be the geometry obtained after completing step $m$.

The story applies to a general semi-simple $\fg$ so far. However, to discuss the inclusion of CS levels and theta angles, it is easier to restrict to the case of a simple $\fg$. This is justified since we only need to consider the case of simple $\fg$ in this paper. The more general semi-simple case was discussed in \cite{Bhardwaj:2019ngx}. For $\fg=\su(n)$, let $k_m$ be the CS level associated to $X_{\fg,R}$ and $k_0$ be the CS level associated to $X_\fg$. Then
\be
k_m=k_0+\half\sum_{\mu=1}^mA_\mu
\ee
where $A_\mu$ is the cubic Dynkin index (also known as the \emph{anomaly coefficient}) associated to the representation $R_\mu$ of $\fg=\su(n)$. For $\fg=\sp(n)$, the theta angle is relevant if none of the irreps $R_\mu$ are pseudo-real with an odd quadratic Dynkin index, in which case the theta angle associated to $X_{\fg,R}$ equals the theta angle associated to $X_\fg$. If some $R_\mu$ contributes to the $4d$ Witten's anomaly, the choice of theta angle for $X_\fg$ is not relevant in the following sense. Let $X^0_{\fg,R}$ be obtained by applying the above procedure to $X_\fg$ with theta angle zero and let $X^\pi_{\fg,R}$ be obtained by applying the above procedure to $X_\fg$ with theta angle $\pi$. Then, $X^0_{\fg,R}$ is related by perturbative flops to $X^\pi_{\fg,R}$.

If $R$ contains half-hypermultiplets, then we write it as
\be
R=\half\tilde R+\sum_{\mu=1}^{m} R_\mu
\ee
where $\tilde R$ denotes the representation of all half-hypermultiplets and $R_\mu$ denote different irreps for full hypermultiplets. The same inductive construction for $X_{\fg,R}$ as above applies for this case as well, but the geometry at step zero is taken instead to be $X_{\fg,\half\tilde R}$ which is a geometry describing the $5d$ gauge theory with gauge algebra $\fg$ and half-hypers transforming in representation $\tilde R$ of $\fg$. In this paper, we do not tackle the problem of describing $X_{\fg,\half\tilde R}$ for arbitrary $\tilde R$ and $\fg$, but only for the cases that will be relevant for $5d$ gauge theories describing $5d$ SCFTs and $5d$ KK theories. These cases are\footnote{We will later see (geometrically) that the $\sp(3)$ and $\sp(4)$ cases do not admit a physically relevant theta angle.}
\be\label{sp3}
\sp(3)+\half\L^3+\half\F ,
\ee
\be\label{sp4}
\sp(4)+\half\L^3,
\ee
\be
\su(6)_k+\half\L^3 ,
\ee
\be
\so(11)+\half\S ,
\ee
\be
\so(12)+\half\S ,
\ee
\be
\so(12)+\half\S+\half\C ,
\ee
\be
\so(13)+\half\S ,
\ee
\be
\fe_7+\half\F .
\ee
Below, we will assign a $X_{\fg,\half\tilde R}$ to each of these cases. It should be noted that, unlike the case of pure gauge theories, these theories admit multiple perturbative phases. Correspondingly, there is no unique or canonical choice for $X_{\fg,\half\tilde R}$. We only present one of the possible choices for each of the above cases. The other choices can be obtained by performing perturbative flops on our presented choices. For more details on understanding why the geometries displayed below describe the matter content we claim they describe, we refer the reader to the general discussion on matter content presented in \cite{Bhardwaj:2019ngx}. We assign:

\begin{center}
\ubf{$\sp(3)+\half\L^3+\half\F$}:
\end{center}
\be\label{sp3G}
\begin{tikzpicture} [scale=1.9]
\node (v1) at (-4.3,-0.5) {$\mathbf{1}_{11}$};
\node (v2) at (-2.3,-0.5) {$\mathbf{2}_{5}$};
\node (v3) at (-0.5,-0.5) {$\mathbf{3}^{1+1}_{0}$};
\draw  (v1) edge (v2);
\draw  (v2) edge (v3);
\node at (-4,-0.4) {\scriptsize{$e$}};
\node at (-2.7,-0.4) {\scriptsize{$h$+$2f$}};
\node at (-2,-0.4) {\scriptsize{$e$}};
\node at (-1.1,-0.4) {\scriptsize{$2e$+$f$-$x$}};
\node (v4) at (-2.3,-1.1) {\scriptsize{2}};
\draw (v1) .. controls (-4.3,-1) and (-3,-1.1) .. (v4);
\draw (v3) .. controls (-0.5,-1) and (-1.6,-1.1) .. (v4);
\node at (-4.5,-0.8) {\scriptsize{$f,f$}};
\node at (-0.1,-0.8) {\scriptsize{$x$-$y,f$-$x$-$y$}};
\end{tikzpicture}
\ee

\medskip

\begin{center}
\ubf{$\sp(4)+\half\L^3$}:
\end{center}
\be\label{sp4G}
\begin{tikzpicture} [scale=1.9]
\node (v1) at (-5.6,-0.5) {$\mathbf{2}_{12}$};
\node (v2) at (-3.6,-0.5) {$\mathbf{3}_{4}$};
\node (v3) at (-1.5,-0.5) {$\mathbf{4}^{2+2}_{0}$};
\node (v0) at (-7,-0.5) {$\mathbf{1}_{14}$};
\draw  (v1) edge (v2);
\draw  (v2) edge (v3);
\node at (-2.2,-0.4) {\scriptsize{$2e$+$f$-$\sum x_i$}};
\node at (-4,-0.4) {\scriptsize{$h$+$3f$}};
\node at (-3.3,-0.4) {\scriptsize{$e$}};
\node at (-5.3,-0.4) {\scriptsize{$e$}};
\node at (-5.1,-0.8) {\scriptsize{$f,f,f$}};
\node at (-2.55,-0.8) {\scriptsize{$f$-$x_1$-$y_1,x_1$-$y_1,x_2$-$y_2$}};
\node (v4) at (-3.8,-1.1) {\scriptsize{3}};
\draw (v1) .. controls (-5.6,-1) and (-4.5,-1.1) .. (v4);
\draw (v3) .. controls (-1.5,-1) and (-3.1,-1.1) .. (v4);
\draw  (v0) edge (v1);
\node at (-6.7,-0.4) {\scriptsize{$e$}};
\node at (-5.9,-0.4) {\scriptsize{$h$}};
\node (v5) at (-4.4,-1.4) {\scriptsize{3}};
\draw (v0) .. controls (-7,-1.3) and (-5.5,-1.4) .. (v5);
\draw (v3) .. controls (-1.3,-1.5) and (-3.3,-1.4) .. (v5);
\node at (-7.3,-0.8) {\scriptsize{$f,f,f$}};
\node at (-0.7,-1) {\scriptsize{$f$-$x_1$-$x_2,x_1$-$x_2,y_1$-$y_2$}};
\end{tikzpicture}
\ee

\medskip

\begin{center}
\ubf{$\su(6)_k+\half\L^3$, $k=\half-l$, $1\le l\le7$}:
\end{center}
\be
\begin{tikzpicture} [scale=1.9]
\node (v1) at (-5.3,-0.5) {$\mathbf{3}_{l}$};
\node (v2) at (-3.3,-0.5) {$\mathbf{4}_{l-4}$};
\node (v3) at (-9.5,-0.5) {$\mathbf{1}_{4+l}$};
\node (v0) at (-7.6,-0.5) {$\mathbf{2}^1_{2+l}$};
\draw  (v1) edge (v2);
\node at (-5,-0.4) {\scriptsize{$e$}};
\node at (-3.8,-0.4) {\scriptsize{$h$+$f$}};
\node at (-8,-0.4) {\scriptsize{$h$}};
\node at (-9.1,-0.4) {\scriptsize{$e$}};
\draw  (v0) edge (v1);
\node at (-7.2,-0.4) {\scriptsize{$e$}};
\node at (-5.6,-0.4) {\scriptsize{$h$}};
\node (v4) at (-3.3,-1.8) {$\mathbf{5}^{1+1}_{l-6}$};
\draw  (v0) edge (v3);
\draw  (v2) edge (v4);
\node at (-3.1,-0.8) {\scriptsize{$e$}};
\node at (-3.1,-1.5) {\scriptsize{$h$}};
\node at (-6.5,-0.7) {\scriptsize{$f$-$x$}};
\node at (-4.7,-0.7) {\scriptsize{$f$}};
\node at (-4.4,-1.7) {\scriptsize{$x$-$y$}};
\node at (-9.1,-0.7) {\scriptsize{$f$}};
\node at (-3.7,-1.3) {\scriptsize{$f$-$x$-$y$}};
\node at (-4.3,-1.4) {\scriptsize{$y$}};
\draw  (v3) edge (v4);
\draw  (v0) edge (v4);
\draw  (v1) edge (v4);
\end{tikzpicture}
\ee

\medskip

\begin{center}
\ubf{$\su(6)_k+\half\L^3$, $k=-\frac{13}2-2m$, $m\ge1$}:
\end{center}
\be
\begin{tikzpicture} [scale=1.9]
\node (v1) at (-5.3,-0.5) {$\mathbf{3}_{7+2m}$};
\node (v2) at (-3.3,-0.5) {$\mathbf{4}_{3+2m}$};
\node (v3) at (-9.5,-0.5) {$\mathbf{1}_{11+2m}$};
\node (v0) at (-7.6,-0.5) {$\mathbf{2}^1_{9+2m}$};
\draw  (v1) edge (v2);
\node at (-4.8,-0.4) {\scriptsize{$e$}};
\node at (-3.9,-0.4) {\scriptsize{$h$+$f$}};
\node at (-8.1,-0.4) {\scriptsize{$h$}};
\node at (-9,-0.4) {\scriptsize{$e$}};
\draw  (v0) edge (v1);
\node at (-7.1,-0.4) {\scriptsize{$e$}};
\node at (-5.8,-0.4) {\scriptsize{$h$}};
\node (v4) at (-3.3,-1.8) {$\mathbf{5}^{1+1}_{1}$};
\draw  (v0) edge (v3);
\draw  (v2) edge (v4);
\node at (-3.1,-0.8) {\scriptsize{$e$}};
\node at (-3,-1.5) {\scriptsize{$h$+$mf$}};
\node at (-6.5,-0.7) {\scriptsize{$f$-$x$}};
\node at (-4.7,-0.7) {\scriptsize{$f$}};
\node at (-4.4,-1.7) {\scriptsize{$x$-$y$}};
\node at (-9.05,-0.75) {\scriptsize{$f$}};
\node at (-3.7,-1.3) {\scriptsize{$f$-$x$-$y$}};
\node at (-4.3,-1.4) {\scriptsize{$y$}};
\draw  (v3) edge (v4);
\draw  (v0) edge (v4);
\draw  (v1) edge (v4);
\end{tikzpicture}
\ee

\medskip

\begin{center}
\ubf{$\su(6)_k+\half\L^3$, $k=-\frac{11}2-2m$, $m\ge1$}:
\end{center}
\be
\begin{tikzpicture} [scale=1.9]
\node (v1) at (-5.3,-0.5) {$\mathbf{3}_{6+2m}$};
\node (v2) at (-3.3,-0.5) {$\mathbf{4}_{2+2m}$};
\node (v3) at (-9.5,-0.5) {$\mathbf{1}_{10+2m}$};
\node (v0) at (-7.6,-0.5) {$\mathbf{2}^1_{8+2m}$};
\draw  (v1) edge (v2);
\node at (-4.8,-0.4) {\scriptsize{$e$}};
\node at (-3.9,-0.4) {\scriptsize{$h$+$f$}};
\node at (-8.1,-0.4) {\scriptsize{$h$}};
\node at (-9,-0.4) {\scriptsize{$e$}};
\draw  (v0) edge (v1);
\node at (-7.1,-0.4) {\scriptsize{$e$}};
\node at (-5.8,-0.4) {\scriptsize{$h$}};
\node (v4) at (-3.3,-1.8) {$\mathbf{5}^{1+1}_{0}$};
\draw  (v0) edge (v3);
\draw  (v2) edge (v4);
\node at (-3.1,-0.8) {\scriptsize{$e$}};
\node at (-3,-1.5) {\scriptsize{$e$+$mf$}};
\node at (-6.5,-0.7) {\scriptsize{$f$-$x$}};
\node at (-4.7,-0.7) {\scriptsize{$f$}};
\node at (-4.4,-1.7) {\scriptsize{$x$-$y$}};
\node at (-9.05,-0.75) {\scriptsize{$f$}};
\node at (-3.7,-1.3) {\scriptsize{$f$-$x$-$y$}};
\node at (-4.3,-1.4) {\scriptsize{$y$}};
\draw  (v3) edge (v4);
\draw  (v0) edge (v4);
\draw  (v1) edge (v4);
\end{tikzpicture}
\ee

\medskip

\begin{center}
\ubf{$\so(11)+\half\S$}:
\end{center}
\be
\begin{tikzpicture} [scale=1.9]
\node (v1) at (-5.3,-0.5) {$\mathbf{3}_{2}$};
\node (v2) at (-3.3,-0.5) {$\mathbf{4}_{0}$};
\node (v3) at (-9.5,-0.5) {$\mathbf{1}_{6}$};
\node (v0) at (-7.6,-0.5) {$\mathbf{2}^1_{4}$};
\draw  (v1) edge (v2);
\node at (-5,-0.4) {\scriptsize{$e$}};
\node at (-3,-0.8) {\scriptsize{$2e$+$f$}};
\node at (-7.9,-0.4) {\scriptsize{$h$}};
\node at (-9.1,-0.4) {\scriptsize{$e$}};
\draw  (v0) edge (v1);
\node at (-7.3,-0.4) {\scriptsize{$e$}};
\node at (-5.6,-0.4) {\scriptsize{$h$}};
\node (v4) at (-3.3,-1.8) {$\mathbf{5}^{1+1}_{6}$};
\draw  (v0) edge (v3);
\draw  (v2) edge (v4);
\node at (-3.6,-0.4) {\scriptsize{$e$}};
\node at (-3.1,-1.5) {\scriptsize{$e$}};
\node at (-6.5,-0.7) {\scriptsize{$f$-$x$}};
\node at (-4.7,-0.7) {\scriptsize{$f$}};
\node at (-4.4,-1.7) {\scriptsize{$x$-$y$}};
\node at (-9.05,-0.75) {\scriptsize{$f$}};
\node at (-3.7,-1.3) {\scriptsize{$f$-$x$-$y$}};
\node at (-4.3,-1.4) {\scriptsize{$y$}};
\draw  (v3) edge (v4);
\draw  (v0) edge (v4);
\draw  (v1) edge (v4);
\end{tikzpicture}
\ee

\medskip

\begin{center}
\ubf{$\so(12)+\half\S$}:
\end{center}
\be
\begin{tikzpicture} [scale=1.9]
\node (v1) at (-5.3,-0.5) {$\mathbf{3}_{2}$};
\node (v2) at (-3.3,-0.5) {$\mathbf{2}_{4}$};
\node (v3) at (-9.5,-0.5) {$\mathbf{5}_{2}$};
\node (v0) at (-7.6,-0.5) {$\mathbf{4}^1_{0}$};
\draw  (v1) edge (v2);
\node at (-5.6,-0.4) {\scriptsize{$e$}};
\node at (-3,-0.8) {\scriptsize{$h$+$f$}};
\node at (-7.9,-0.4) {\scriptsize{$e$}};
\node at (-9.1,-0.4) {\scriptsize{$e$}};
\draw  (v0) edge (v1);
\node at (-7.2,-0.4) {\scriptsize{$e$}};
\node at (-5,-0.4) {\scriptsize{$h$}};
\node (v4) at (-3.3,-1.8) {$\mathbf{1}^{1+1}_{8}$};
\draw  (v0) edge (v3);
\draw  (v2) edge (v4);
\node at (-3.6,-0.4) {\scriptsize{$e$}};
\node at (-3.1,-1.5) {\scriptsize{$e$}};
\node at (-6.5,-0.7) {\scriptsize{$f$-$x$}};
\node at (-4.7,-0.7) {\scriptsize{$f$}};
\node at (-4.4,-1.7) {\scriptsize{$x$-$y$}};
\node at (-9.05,-0.75) {\scriptsize{$f$}};
\node at (-3.7,-1.3) {\scriptsize{$f$-$x$-$y$}};
\node at (-4.3,-1.4) {\scriptsize{$y$}};
\draw  (v3) edge (v4);
\draw  (v0) edge (v4);
\draw  (v1) edge (v4);
\node (v5) at (-7.6,0.8) {$\mathbf{6}_{1}$};
\draw  (v5) edge (v0);
\node at (-7.4,-0.2) {\scriptsize{$e$-$x$}};
\node at (-7.4,0.5) {\scriptsize{$e$}};
\end{tikzpicture}
\ee

\medskip

\begin{center}
\ubf{$\so(12)+\half\S+\half\C$}:
\end{center}
\be
\begin{tikzpicture} [scale=1.9]
\node (v1) at (-5.3,-0.5) {$\mathbf{3}_2$};
\node (v2) at (-3.3,-0.5) {$\mathbf{2}_{4}$};
\node (v3) at (-9.5,-0.5) {$\mathbf{5}_{1}$};
\node (v0) at (-7.6,-0.5) {$\mathbf{4}^2_0$};
\draw  (v1) edge (v2);
\node at (-5.6,-0.4) {\scriptsize{$e$}};
\node at (-3.6,-0.4) {\scriptsize{$e$}};
\node at (-8,-0.4) {\scriptsize{$e$-$x_2$}};
\node at (-9.2,-0.4) {\scriptsize{$e$}};
\draw  (v0) edge (v1);
\node at (-7.3,-0.4) {\scriptsize{$e$}};
\node at (-5,-0.4) {\scriptsize{$h$}};
\node (v4) at (-3.3,-1.8) {$\mathbf{1}^{2+2}_{10}$};
\draw  (v0) edge (v3);
\draw  (v2) edge (v4);
\node at (-3,-0.8) {\scriptsize{$h$+$2f$}};
\node at (-3.1,-1.5) {\scriptsize{$e$}};
\node (v5) at (-5.4,-1.2) {\scriptsize{2}};
\node (v6) at (-4.5,-1) {\scriptsize{2}};
\draw  (v0) edge (v5);
\draw  (v4) edge (v6);
\draw  (v6) edge (v1);
\node at (-6.5,-0.7) {\scriptsize{$f$-$x_i$}};
\node at (-4.8,-0.7) {\scriptsize{$f$}};
\node[rotate=0] at (-3.8,-1.2) {\scriptsize{$f$-$x_i$-$y_i$}};
\node at (-4.9,-1.6) {\scriptsize{$x_1$-$y_1$}};
\draw  (v5) edge (v4);
\node at (-4.3,-1.4) {\scriptsize{$y_i$}};
\node at (-9.2,-0.7) {\scriptsize{$f$}};
\node (v7) at (-7.6,-1.8) {$\mathbf{6}_{1}$};
\draw  (v0) edge (v7);
\node at (-7.8,-0.7) {\scriptsize{$e$-$x_1$}};
\node at (-7.7,-1.6) {\scriptsize{$e$}};
\node at (-3.9,-1.95) {\scriptsize{$x_2$-$y_2$}};
\node at (-7.2,-1.95) {\scriptsize{$f$}};
\draw  (v3) edge (v4);
\draw  (v7) edge (v4);
\end{tikzpicture}
\ee

\medskip

\begin{center}
\ubf{$\so(13)+\half\S$}:
\end{center}
\be
\begin{tikzpicture} [scale=1.9]
\node (v1) at (-4.9,-0.5) {$\mathbf{3}_3$};
\node (v2) at (-3.3,-0.5) {$\mathbf{2}_{5}$};
\node (v3) at (-8.4,-0.5) {$\mathbf{5}_{1}$};
\node (v0) at (-6.3,-0.5) {$\mathbf{4}_1$};
\draw  (v1) edge (v2);
\node at (-6.6,-0.4) {\scriptsize{$e$}};
\node at (-3.6,-0.4) {\scriptsize{$e$}};
\node at (-8.8,-0.4) {\scriptsize{$2h$}};
\node at (-10.6,-0.4) {\scriptsize{$e$}};
\draw  (v0) edge (v1);
\node at (-8,-0.4) {\scriptsize{$e$}};
\node at (-6,-0.4) {\scriptsize{$h$}};
\node (v4) at (-3.3,-2.2) {$\mathbf{1}^{2+2+2+2}_{11}$};
\draw  (v0) edge (v3);
\draw  (v2) edge (v4);
\node at (-3,-0.8) {\scriptsize{$h$+$2f$}};
\node at (-3.1,-1.9) {\scriptsize{$e$}};
\node (v5) at (-5.1,-1.2) {\scriptsize{2}};
\node (v6) at (-4.2,-1.2) {\scriptsize{2}};
\draw  (v0) edge (v5);
\draw  (v4) edge (v6);
\draw  (v6) edge (v1);
\node at (-5.6,-0.7) {\scriptsize{$f$}};
\node at (-4.5,-0.7) {\scriptsize{$f$}};
\node[rotate=0] at (-3.6,-1.5) {\scriptsize{$f$-$x_i$-$y_i$}};
\node at (-5.3,-1.4) {\scriptsize{$x_1$-$y_1,$}};
\draw  (v5) edge (v4);
\node at (-4.25,-1.5) {\scriptsize{$y_i$-$z_i$}};
\node (v6) at (-6.4,-1.2) {\scriptsize{2}};
\draw  (v3) edge (v6);
\draw  (v6) edge (v4);
\node at (-7.4,-0.7) {\scriptsize{$f,f$}};
\node (v7) at (-11,-0.5) {$\mathbf{6}_{6}$};
\draw  (v7) edge (v3);
\node at (-5.2,-0.4) {\scriptsize{$e$}};
\node at (-4.6,-0.4) {\scriptsize{$h$}};
\node at (-4.9,-1.55) {\scriptsize{$z_2$-$w_1$}};
\node (v8) at (-7.4,-1.3) {\scriptsize{4}};
\draw  (v7) edge (v8);
\draw  (v8) edge (v4);
\node at (-5.5,-2) {\scriptsize{$x_2$-$x_1,y_1$-$y_2,z_1$-$z_2,w_1$-$w_2$}};
\node at (-10.6,-0.8) {\scriptsize{$f,f,f,f$}};
\end{tikzpicture}
\ee

\medskip

\begin{center}
\ubf{$\fe_7+\half\F$}:
\end{center}
\be
\begin{tikzpicture} [scale=1.9]
\node (v1) at (-4,0) {$\mathbf{6_{3}}$};
\node (v2) at (-2.8,0) {$\mathbf{5_{1}}$};
\node (v3) at (-1.3,0) {$\mathbf{4_{1}}$};
\node (v4) at (1.7,1.6) {$\mathbf{1_{8}}$};
\draw  (v1) edge (v2);
\draw  (v2) edge (v3);
\node at (-3.7,-0.1) {\scriptsize{$e$ }};
\node (v6) at (0.2,0) {$\mathbf{3_{2}^{2}}$};
\node (v7) at (1.7,0) {$\mathbf{2_{4}}$};
\draw  (v3) edge (v6);
\draw  (v6) edge (v7);
\node at (-3.1,-0.1) {\scriptsize{$h$}};
\node at (-0.2,-0.1) {\scriptsize{$e$-$x_1$}};
\node at (0.6,-0.1) {\scriptsize{$h$}};
\node at (-2.5,-0.1) {\scriptsize{ $e$}};
\draw  (v7) edge (v4);
\node at (-1.6,-0.1) {\scriptsize{$e$}};
\node at (-1.4,0.3) {\scriptsize{$h$}};
\node at (1.4,-0.1) {\scriptsize{$e$}};
\node at (-0.9,-0.1) {\scriptsize{$h$}};
\node (v10) at (-1.3,1.6) {$\mathbf{7_{3}}$};
\node at (2,0.3) {\scriptsize{$h$+$f$}};
\node at (1.9,1.3) {\scriptsize{$e$}};
\node at (-1.4,1.2) {\scriptsize{$e$}};
\draw  (v10) edge (v3);
\node at (-1,1.1) {\scriptsize{$f$}};
\node at (-0.3,0.2) {\scriptsize{$x_1$-$x_2$}};
\node at (1.2,1.3) {\scriptsize{$f$}};
\node at (0.3,0.5) {\scriptsize{$f$-$x_1$-$x_2$}};
\draw  (v10) edge (v6);
\draw  (v6) edge (v4);
\end{tikzpicture}
\ee

\medskip

Let us now use the presented geometries (\ref{sp3G}) and (\ref{sp4G}) to argue that the theta angle is irrelevant for (\ref{sp3}) and (\ref{sp4}). If the theta was physically relevant for these cases, the geometries corresponding to the other theta angle would be
\be\label{sp3G'}
\begin{tikzpicture} [scale=1.9]
\node (v1) at (-4.3,-0.5) {$\mathbf{1}_{11}$};
\node (v2) at (-2.3,-0.5) {$\mathbf{2}_{5}$};
\node (v3) at (-0.5,-0.5) {$\mathbf{3}^{1+1}_{1}$};
\draw  (v1) edge (v2);
\draw  (v2) edge (v3);
\node at (-4,-0.4) {\scriptsize{$e$}};
\node at (-2.7,-0.4) {\scriptsize{$h$+$2f$}};
\node at (-2,-0.4) {\scriptsize{$e$}};
\node at (-1,-0.4) {\scriptsize{$2h$-$x$}};
\node (v4) at (-2.3,-1.1) {\scriptsize{2}};
\draw (v1) .. controls (-4.3,-1) and (-3,-1.1) .. (v4);
\draw (v3) .. controls (-0.5,-1) and (-1.6,-1.1) .. (v4);
\node at (-4.5,-0.8) {\scriptsize{$f,f$}};
\node at (-0.1,-0.8) {\scriptsize{$x$-$y,f$-$x$-$y$}};
\end{tikzpicture}
\ee
and
\be\label{sp4G'}
\begin{tikzpicture} [scale=1.9]
\node (v1) at (-5.6,-0.5) {$\mathbf{2}_{12}$};
\node (v2) at (-3.6,-0.5) {$\mathbf{3}_{4}$};
\node (v3) at (-1.5,-0.5) {$\mathbf{4}^{2+2}_{1}$};
\node (v0) at (-7,-0.5) {$\mathbf{1}_{14}$};
\draw  (v1) edge (v2);
\draw  (v2) edge (v3);
\node at (-2.1,-0.4) {\scriptsize{$2h$-$\sum x_i$}};
\node at (-4,-0.4) {\scriptsize{$h$+$3f$}};
\node at (-3.3,-0.4) {\scriptsize{$e$}};
\node at (-5.3,-0.4) {\scriptsize{$e$}};
\node at (-5.1,-0.8) {\scriptsize{$f,f,f$}};
\node at (-2.55,-0.8) {\scriptsize{$f$-$x_1$-$y_1,x_1$-$y_1,x_2$-$y_2$}};
\node (v4) at (-3.8,-1.1) {\scriptsize{3}};
\draw (v1) .. controls (-5.6,-1) and (-4.5,-1.1) .. (v4);
\draw (v3) .. controls (-1.5,-1) and (-3.1,-1.1) .. (v4);
\draw  (v0) edge (v1);
\node at (-6.7,-0.4) {\scriptsize{$e$}};
\node at (-5.9,-0.4) {\scriptsize{$h$}};
\node (v5) at (-4.4,-1.4) {\scriptsize{3}};
\draw (v0) .. controls (-7,-1.3) and (-5.5,-1.4) .. (v5);
\draw (v3) .. controls (-1.3,-1.5) and (-3.3,-1.4) .. (v5);
\node at (-7.3,-0.8) {\scriptsize{$f,f,f$}};
\node at (-0.7,-1) {\scriptsize{$f$-$x_1$-$x_2,x_1$-$x_2,y_1$-$y_2$}};
\end{tikzpicture}
\ee
respectively. However, these geometries are isomorphic to (\ref{sp3G}) and (\ref{sp4G}) respectively. Applying $\cI_0$ on $S_3$ of (\ref{sp3G}) using the blowup $x$ living in $S_3$ converts (\ref{sp3G}) into (\ref{sp3G'}). Similarly, applying $\cI_0$ on $S_4$ of (\ref{sp4G}) using the blowup $x_1$ living in $S_4$ converts (\ref{sp4G}) into (\ref{sp4G'}).

\subsection{Structure of $6d$ gauge theory compactified on a circle}\label{6dg}
For a geometry $X$ to describe a twisted circle compactification of a $6d$ $\cN=(1,0)$ gauge theory, all the $S_i$ must be Hirzebruch surfaces with their intersection matrix $\cI_{ij}$ being a direct sum of Cartan matrices associated to simple (twisted and untwisted) affine Lie algebras. As for a $5d$ gauge theory, perturbative BPS particles must only be identified with other perturbative BPS particles.

Let us parametrize simple affine algebras as $\fg^{(q_\alpha)}_\alpha$, where $\fg_\alpha$ are the $6d$ gauge algebras and $q_\alpha$ capture the order of the outer-automorphism used for twisting $\fg_\alpha$ along the circle \cite{Bhardwaj:2019fzv}. Let us correspondingly parametrize the surfaces as $S_{a,\alpha}$ with the index $a$ (for a fixed $\alpha$) parametrizing different surfaces whose intersection matrix
\be
-f_{a,\alpha}\cdot S_{b,\alpha}
\ee
gives rise to the Cartan matrix $\cI_{ab,\alpha}$ of $\fg^{(q_\alpha)}_\alpha$. We let $S_{0,\alpha}$ be the surface corresponding to the affine co-root of $\fg^{(q_\alpha)}_\alpha$. We let $d_{a,\alpha}$ and $d^\vee_{a,\alpha}$ be respectively the Coxeter and dual Coxter labels\footnote{These can be found in Tables 14 and 15 of \cite{Bhardwaj:2019fzv}.} associated to $\fg^{(q_\alpha)}_\alpha$. These are minimum positive integers satisfying
\be
\sum_ad_{a,\alpha}\cI_{ab,\alpha}=0
\ee
and
\be
\sum_b\cI_{ab,\alpha}d^\vee_{b,\alpha}=0
\ee
Also let $e_{a,\alpha}$ and $h_{a,\alpha}$ be the $e$ and $h$ curves of the Hirzebruch surface $S_{a,\alpha}$.

The gluing curves between $S_{a,\alpha}$ and $S_{b,\beta}$ for $\alpha\neq\beta$ cannot involve $e_{a,\alpha}$ or $e_{b,\beta}$ since otherwise the intersection matrix would be modified. Thus
\be\label{O}
f_{a,\alpha}\cdot S_{b,\beta}=0
\ee
for all $a,b$ and $\alpha\neq\beta$. Moreover, the curve
\be\label{KK}
f_\alpha:=\sum_ad_{a,\alpha}f_{a,\alpha}
\ee
satisfies
\be
f_\alpha\cdot S_{a,\alpha}=0
\ee
for all $a$. Combining this result with (\ref{O}) we find that
\be\label{OO}
f_\alpha\cdot S_{b,\beta}=0
\ee
for all $b,\beta$. Thus the volume of $f_\alpha$ defines a mass parameter of the $5d$ theory associated to $X$. Since $f_\alpha$ does not involve any blowups, and the matter content is encoded in the blowups, this mass parameter does not arise from holonomies of flavor symmetry groups around the circle. The only other possible mass parameter is given by the radius of the compactification circle, and thus $f_\alpha$ can be identified as the KK mode. Since there is only a single KK mode, the following curves must be equal as classes in $X$
\be
[n_{\alpha,\beta} f_\alpha]=[n_{\beta,\alpha}f_\beta]
\ee
for some positive integers $n_{\alpha,\beta}$ and $n_{\beta,\alpha}$ for all $\alpha,\beta$. This means that if there is any gluing curve between a surface $S_{a,\alpha}$ and a surface $S_{b,\beta}$ for $\alpha\neq\beta$ then the other gluing curves between the family of surfaces $\cup_a S_{a,\alpha}$ and the family of surfaces of $\cup_b S_{b,\beta}$ must be such that a particular linear combination of the gluing curves leads to the following gluing
\be\label{KKg}
n_{\alpha,\beta} f_\alpha\sim n_{\beta,\alpha}f_\beta
\ee
This requirement was used in \cite{Bhardwaj:2019fzv,Bhardwaj:2018vuu} as consistency conditions on the \emph{gluing rules} between $\cup_a S_{a,\alpha}$ and $\cup_b S_{b,\beta}$.

The tensor branch of the $6d$ $\cN=(1,0)$ theory descends to the Coulomb branch of the circle compactified theory with all mass parameters turned off, which in particular implies that the radius of compactification is set to infinity. Along this Coulomb branch, the masses and tensions of all BPS particles and strings must be non-negative. This Coulomb branch is captured by the Kahler cone $\cK(X)$ of $X$ (with all non-normalizable Kahler parameters turned off) along which all the holomorphic curves and surfaces in $X$ have non-negative volume. According to (\ref{OO}), $f_\alpha$ must have zero volume along any direction in $\cK(X)$. The non-negativity of volumes then implies that each $f_{a,\alpha}$ must have zero volume along any direction $\cK(X)$. This fixes $\cK(X)$ to be a sub-cone of the cone $\cT(X)$ formed by
\be
S_\alpha:=\sum_ad^\vee_{a,\alpha}S_{a,\alpha}
\ee
for different values of $\alpha$.

Now let us assume that there is a non-trivial $\cK(X)$ inside $\cT(X)$. Physically, we are assuming that the $6d$ gauge theory describes either a $6d$ SCFT or a little string theory (LST). Let us focus our attention on a fixed $\alpha$. We now decompactify all $S_{b,\beta}$ for $\beta\neq\alpha$ by decompactifying the curves $e_{0,\beta}$ which forces the decompactification of other $e_{b,\beta}$. During this process, it is possible to keep all the fibers and blowups compact \cite{Bhardwaj:2019xeg}. The compact surfaces in the resulting Calabi-Yau threefold $X_\alpha$ are only $S_{a,\alpha}$, while the compact curves in $X_\alpha$ comprise of the compact curves living inside $\cup_a S_{a,\alpha}$ along with the curves comprising solely of blowups and fibers in $S_{b,\beta}$ (which are now non-compact surfaces inside $X_\alpha$) for $\beta\neq\alpha$. Physically, this decmpactification process corresponds to ungauging all $6d$ gauge algebras $\fg_\beta$ for $\beta\neq\alpha$ since the volumes of $e_{0,\beta}$ capture the masses of BPS particles arising by wrapping $6d$ instanton BPS strings on the compactification circle, and these masses are proportional to inverse gauge couplings in the $6d$ gauge theory. If our starting point was a $6d$ SCFT or a LST, we must land on a $6d$ SCFT at the end of this process. Thus $X_\alpha$ must have a non-trivial Kahler cone $\cK(X_\alpha)$ inside $\cT(X_\alpha)$. The latter cone is spanned entirely by $S_\alpha$, and hence $\cK(X_\alpha)$ is spanned by $S_\alpha$. Since
\be
\text{vol}(x)=-\text{vol}\left(f_{b,\beta}-x\right)
\ee
inside $\cK(X_\alpha)$ for a blowup $x$ living in $S_{b,\beta}$ (which is compact for $\beta=\alpha$ and non-compact for $\beta\neq\alpha$), we learn that
\be\label{shift}
x\cdot S_\alpha=0
\ee
for all blowups $x$ in $X_\alpha$. But since none of the blowups were decompactified, we learn that (\ref{shift}) applies to all blowups $x$ in $X$ and to all $\alpha$. Thus, all the perturbative BPS particles have zero mass inside $\cT(X)$. This justifies the ``shifted prepotential'' proposal of \cite{Bhardwaj:2019fzv}.

Now, let us define a matrix
\be\label{M}
M_{\alpha\beta}=-S_\alpha\cdot e_{0,\beta}
\ee
which captures $\text{vol}(e_{0,\beta})$ inside $\cT(X)$. Since the gluing curves between $S_{0,\alpha}$ and $S_{b,\beta}$ for $\beta\neq\alpha$ must correspond to perturbative BPS particles, any off-diagonal entry $M_{\alpha\beta}$ must be non-positive. If $\cup_a S_{a,\alpha}$ does not intersect $\cup_b S_{b,\beta}$, then $M_{\alpha\beta}=M_{\beta\alpha}=0$. If $\cup_a S_{a,\alpha}$ intersects $\cup_b S_{b,\beta}$, then according to (\ref{KKg}), $f_{0,\alpha}$ must participate in some gluing curve between $S_{0,\alpha}$ and $\cup_b S_{b,\beta}$, thus implying that $M_{\beta\alpha}<0$. If that's the case, exchanging the role of $\beta$ and $\alpha$, we must also have $M_{\alpha\beta}<0$. We conclude that the matrix $[M_{\alpha\beta}]$ is a \emph{generalized Cartan matrix}.

According to an important property of generalized Cartan matrices, if $[M_{\alpha\beta}]$ is positive definite, then there is a non-trivial sub-cone inside $\cT(X)$ along which all $e_{0,\alpha}$ have positive volume. This sub-cone can be identified with $\cK(X)$ as we now show. Any compact curve $C$ inside $X$ lives in some surface $S_{a,\alpha}$ and can be written as
\be
C=me_{a,\alpha}+nf_{a,\alpha}+\sum_ip_i x_i
\ee
where $x_i$ are the blowups in $S_{a,\alpha}$ and $m,n,p_i$ are integers with $m,n\ge0$. Thus
\be
\text{vol}(C)=m\text{ vol}(e_{a,\alpha})
\ee
inside $\cT(X)$. Using the intersections between $S_{a,\alpha}$ for different $a$, one can further rewrite the above as
\be\label{volT}
\text{vol}(C)=mb_{a,\alpha}\text{vol}(e_{0,\alpha})
\ee
for some strictly positive integer $b_{a,\alpha}$. The quantity on the right hand side of (\ref{volT}) is manifestly non-negative inside the sub-cone under discussion. Since, inside this sub-cone, the only curve in each surface $S_{a,\alpha}$ that has non-zero volume is $e_{a,\alpha}$ while the fiber $f_{a,\alpha}$ and blowups have zero volume, each surface $S_{a,\alpha}$ has zero volume. Thus all the compact curves and surfaces have non-negative volume in this sub-cone and it can be identified with $\cK(X)$. In this case, $X$ corresponds to a twisted circle compactification of a $6d$ $\cN=(1,0)$ gauge theory which describes the tensor branch of a $6d$ SCFT.

If $[M_{\alpha\beta}]$ is positive semi-definite, then there is a unique ray inside $\cT(X)$ along which all $e_{0,\alpha}$ have non-negative volume. In fact, the volume of each $e_{0,\alpha}$ along this ray is exactly zero. Hence, every compact curve and surface inside $X$ has zero volume along this ray. In this case, $X$ corresponds to a twisted circle compactification of a $6d$ $\cN=(1,0)$ gauge theory which describes the tensor branch of a $6d$ LST. The ray in the Coulomb branch descends from the non-dynamical tensor multiplet associated to the LST. The fact that the BPS particles $e_{0,\alpha}$ originating from $6d$ strings wrapped on the circle have zero mass means that the strings themselves have zero tension. This is due to the fact that we are working with all mass parameters turned off, so we have the little string mass scale $M_s$ turned off. The size of the tensor branch of a $6d$ LST where the strings have positive tension is dictated by the $M_s$, and when $M_s=0$, there is no such tensor branch, which explains our finding.

If $[M_{\alpha\beta}]$ is indefinite, then there is no non-trivial sub-cone inside $\cT(X)$ where all $e_{0,\alpha}$ have non-negative volume. In this case, $X$ corresponds to a twisted circle compactification of a $6d$ $\cN=(1,0)$ gauge theory which describes the tensor branch of neither a $6d$ SCFT nor a $6d$ LST.

\subsection{Structure of a $5d$ KK theory}\label{5dkk}
We define a $5d$ KK theory to be a twisted circle (of finite, non-zero radius) compactification of a $6d$ SCFT. $6d$ SCFTs are built by gluing $6d$ $\cN=(1,0)$ gauge theories with certain non-gauge-theoretic pieces. In this subsection, we let $X$ be the Calabi-Yau threefold corresponding to a $5d$ KK theory. From the previous subsection, we already understand the parts of $X$ descending from the gauge-theoretic sector of the corresponding $6d$ SCFT. So we only need to understand the geometries corresponding to non-gauge-theoretic sectors.

A non-gauge-theoretic sector of a $6d$ SCFT can be thought of as a sector with trivial gauge algebra. Thus, the corresponding piece $\alpha$ in the geometry $X$ contains a single surface $S_{0,\alpha}$ which can be thought of as the affine node for trivial gauge algebra. There are three possibilities for a non-gauge-theoretic sector, corresponding to following $S_{0,\alpha}$
\be\label{sp0}
\begin{tikzpicture} [scale=1.9]
\node (v1) at (-4,0) {$\mathbf{\bF_1^{8}}$};
\end{tikzpicture}
\ee
\be\label{su1}
\begin{tikzpicture} [scale=1.9]
\node (v1) at (-4,0) {$\mathbf{\bF_0^{1+1}}$};
\draw (v1) .. controls (-4.8,0.5) and (-4.8,-0.5) .. (v1);
\node at (-4.3,0.3) {\scriptsize{$x$}};
\node at (-4.3,-0.3) {\scriptsize{$y$}};
\end{tikzpicture}
\ee
\be\label{su1T}
\begin{tikzpicture} [scale=1.9]
\node (v1) at (-4,0) {$\mathbf{\bF_1^{1+1}}$};
\draw (v1) .. controls (-4.8,0.5) and (-4.8,-0.5) .. (v1);
\node at (-4.3,0.3) {\scriptsize{$x$}};
\node at (-4.3,-0.3) {\scriptsize{$y$}};
\end{tikzpicture}
\ee
For the first case, we define $e_{0,\alpha}$ to be one of the eight blowups instead of the $e$ curve. For the second and third cases, we define $e_{0,\alpha}$ to be the corresponding $e$ curve. If one of the blowups $x$ in (\ref{sp0}) is not generic and creates the curve $e-x$ in the Mori cone, we can apply $\cI_1$ using this blowup to write (\ref{sp0}) in the following isomorphic form
\be\label{sp0'}
\begin{tikzpicture} [scale=1.9]
\node (v1) at (-4,0) {$\mathbf{\bF_2^{8}}$};
\end{tikzpicture}
\ee
We will use this isomorphic geometry for this non-gauge-theoretic sector often in this paper. In this isomorphic form, we let $e_{0,\alpha}$ to still be one of the eight blowups. 

The surface $S_\alpha$ is defined to be equal to $S_{0,\alpha}$ for non-gauge-theoretic sectors. The curve $f_\alpha$ for each of the three cases are defined to be $2h+f-\sum x_i$, $e+f-x-y$ and $2h+f-2x-2y$ respectively. In the isomorphism frame (\ref{sp0'}), the $f_\alpha$ for the first case is written as $2h-\sum x_i$ as the reader can check by applying the isomorphism between (\ref{sp0}) and (\ref{sp0'}). The reader can also see that 
\be
f_\alpha\cdot S_{0,\alpha}=0
\ee
is satisfied in each of the three cases. The curve $f_\alpha$ for non-gauge-theoretic sectors is required to satisfy same conditions as $f_\alpha$ for gauge-theoretic sectors. That is, (\ref{OO}) is now viewed as a constraint on the possible ways of gluing non-gauge-theoretic sectors with the rest of the theory. Similarly, these gluings must satisfy (\ref{KKg}) as well.

For a geometry $X$ describing a general $5d$ KK theory including both gauge-theoretic and non-gauge-theoretic sectors, we define a matrix $[M_{\alpha\beta}]$ using (\ref{M}) and the above definitions for $S_\alpha$ and $e_{0,\beta}$. This is again a generalized Cartan matrix, which must be positive definite for $X$ to describe a $5d$ KK theory. 

We now proceed to show how one can represent $5d$ KK theories using the data of $M_{\alpha\beta}$ and $\fg_\alpha$ for all $\alpha,\beta$. We convert this data into a graphical form introduced in \cite{Bhardwaj:2019fzv} to characterize $5d$ KK theories. We will use this graphical notation throughout this paper to represent $5d$ KK theories. Let us first convert the matrix $[M_{\alpha\beta}]$ into another matrix $[\Omega_{\alpha\beta}]$ via
\be
\Omega_{\alpha\beta}=\frac1{u_\beta}M_{\alpha\beta}
\ee
Here, for trivial $\fg_\beta$, we let $u_\beta=1$. For non-trivial $\fg_\beta$, we first unfold the Dynkin diagram for $\fg^{(q_\beta)}_\beta$ until we reach the Dynkin diagram for an untwisted affine Lie algebra $\fh_\beta$. The inverse process $\fh_\beta\to\fg_\beta^{(q_\beta)}$ involves iterated foldings by identifying nodes exchanged under permutation operations. Then $u_\beta$ is defined to be the product of the orders of the permutation operations corresponding to these foldings. Thus, if $q_\beta=1$, then $u_\beta=1$. For the algebras
\be
\su(2n)^{(2)},\so(2n)^{(2)},\fe_6^{(2)}
\ee
we have $u_\beta=2$. For the algebra
\be
\so(8)^{(3)}
\ee
we have $u_\beta=3$. For the algebras
\be
\su(2n+1)^{(2)}
\ee
we have $u_\beta=4$. Notice that if $[M_{\alpha\beta}]$ is positive definite, then so is $[\Omega_{\alpha\beta}]$, and vice-versa.

Now we convert the data of $[\Omega_{\alpha\beta}]$ into a graph. If $\alpha$ is gauge-theoretic, then we assign a node to it according to following rules:
\bit
\item If $\Omega_{\alpha\alpha}>1$, then we assign the node
\be
\begin{tikzpicture} [scale=1.9]
\node (v2) at (-3.5,0.8) {$\Omega_{\alpha\alpha}$};
\node at (-3.5,1.1) {$\fg_\alpha^{(q_\alpha)}$};
\end{tikzpicture}
\ee
to it. 
\item If $\Omega_{\alpha\alpha}=1$ and $\fg^{(q_\alpha)}_\alpha\neq\su(n)^{(1)}$, then we assign the node
\be
\begin{tikzpicture} [scale=1.9]
\node (v2) at (-3.5,0.8) {1};
\node at (-3.5,1.1) {$\fg_\alpha^{(q_\alpha)}$};
\end{tikzpicture}
\ee
to it.
\item If $\Omega_{\alpha\alpha}=1$ and $\fg^{(q_\alpha)}_\alpha=\su(n)^{(1)}$, we consider intersections of all the compact curves composed out of fibers and blowups with surfaces $S_{a,\alpha}$ for $a\neq0$. These intersections imply that perturbative BPS particles associated to these curves are associated to a direct sum $\oplus_\mu R_\mu$ of irreps $R_\mu$ of $\su(n)$. If none of the $R_\mu$ equals 2-index symmetric irrep of $\su(n)$ and $n\ge3$, then we associate the node
\be
\begin{tikzpicture} [scale=1.9]
\node (v2) at (-3.5,0.8) {1};
\node at (-3.5,1.1) {$\su(n)^{(1)}$};
\end{tikzpicture}
\ee
to it. If none of the $R_\mu$ equals 2-index symmetric irrep of $\su(n)$ and $n=2$, then we associate the node
\be\label{L1}
\begin{tikzpicture} [scale=1.9]
\node (v2) at (-3.5,0.8) {1};
\node at (-3.5,1.1) {$\sp(1)^{(1)}$};
\end{tikzpicture}
\ee
to it. If one of $R_\mu$ equals 2-index symmetric irrep of $\su(n)$, then we associate the node
\be
\begin{tikzpicture} [scale=1.9]
\node (v2) at (-3.5,0.8) {2};
\node at (-3.5,1.1) {$\su(n)^{(1)}$};
\draw (v2) .. controls (-4,0.3) and (-3,0.3) .. (v2);
\end{tikzpicture}
\ee
to it.
\eit
If $\alpha$ is non-gauge-theoretic, then we assign a node to it according to following rules:
\bit
\item If $S_{0,\alpha}$ is isomorphic to (\ref{sp0}), then we assign the node
\be
\begin{tikzpicture} [scale=1.9]
\node (v2) at (-3.5,0.8) {1};
\node at (-3.5,1.1) {$\sp(0)^{(1)}$};
\end{tikzpicture}
\ee
to it.
\item If $S_{0,\alpha}$ is isomorphic to (\ref{su1}), then we assign the node
\be\label{L2}
\begin{tikzpicture} [scale=1.9]
\node (v2) at (-3.5,0.8) {2};
\node at (-3.5,1.1) {$\su(1)^{(1)}$};
\end{tikzpicture}
\ee
to it.
\item If $S_{0,\alpha}$ is isomorphic to (\ref{su1T}), then we assign the node
\be
\begin{tikzpicture} [scale=1.9]
\node (v2) at (-3.5,0.8) {2};
\node at (-3.5,1.1) {$\su(1)^{(1)}$};
\draw (v2) .. controls (-4,0.3) and (-3,0.3) .. (v2);
\end{tikzpicture}
\ee
to it.
\eit
Now we move onto the description of edges:
\bit
\item If $\Omega_{\alpha\beta}=\Omega_{\beta\alpha}=-1$ for $\alpha\neq\beta$, then the nodes corresponding to $\alpha$ and $\beta$ are joined by an edge as shown below
\be
\begin{tikzpicture} [scale=1.9]
\node (v1) at (-4.2,0.8) {$\Omega_{\alpha\alpha}$};
\node at (-4.2,1.1) {$\fg_\alpha^{(q_\alpha)}$};
\node (v2) at (-2.9,0.8) {$\Omega_{\beta\beta}$};
\node at (-2.9,1.1) {$\fg_\beta^{(q_\beta)}$};
\draw  (v1) edge (v2);
\end{tikzpicture}
\ee
Here the node corresponding to $\alpha$ or $\beta$ could carry a loop as in (\ref{L1}) and (\ref{L2}). However, we omit the loop throughout our discussion of edges, as it does not influence the discussion.
\item If $\Omega_{\alpha\beta}=\Omega_{\beta\alpha}=-k<-1$ for $\alpha\neq\beta$, then the nodes corresponding to $\alpha$ and $\beta$ are joined by an edge of the following form
\be
\begin{tikzpicture} [scale=1.9]
\node (v1) at (-4.2,0.8) {$\Omega_{\alpha\alpha}$};
\node at (-4.2,1.1) {$\fg_\alpha^{(q_\alpha)}$};
\node (v2) at (-2.8,0.8) {$\Omega_{\beta\beta}$};
\node at (-2.8,1.1) {$\fg_\beta^{(q_\beta)}$};
\node (v3) at (-3.5,0.8) {\tiny{$k$}};
\draw  (v1) edge (v3);
\draw  (v3) edge (v2);
\end{tikzpicture}
\ee
\item Now let us consider the case $\Omega_{\alpha\beta}\neq\Omega_{\beta\alpha}$ for $\beta\neq\alpha$. From the analysis of the structure of $6d$ SCFTs one can deduce that for this to happen \cite{Bhardwaj:2019fzv}, either $\Omega_{\alpha\beta}=-1$ or $\Omega_{\beta\alpha}=-1$. Let us assume without loss of generality that $\Omega_{\beta\alpha}=-1$ and $\Omega_{\alpha\beta}=-k<-1$. We denote this situation by placing the following edge
\be
\begin{tikzpicture} [scale=1.9]
\node (v1) at (-4.2,0.8) {$\Omega_{\alpha\alpha}$};
\node at (-4.2,1.1) {$\fg_\alpha^{(q_\alpha)}$};
\node (v2) at (-2.8,0.8) {$\Omega_{\beta\beta}$};
\node at (-2.8,1.1) {$\fg_\beta^{(q_\beta)}$};
\node (v3) at (-3.5,0.8) {\tiny{$k$}};
\draw  (v1) edge (v3);
\draw [->] (v3) edge (v2);
\end{tikzpicture}
\ee
\eit
Sometimes different KK theories have the same associated graph. In this case, the vertices and edges are decorated to distinguish between different cases. See more details about such decorations in \cite{Bhardwaj:2019fzv}. Finally, unfolding the graph associated to a $5d$ KK theory and removing the subscripts $q_\alpha$ give rise to the graph associated to the $6d$ SCFT (see \cite{Bhardwaj:2019fzv}) whose circle compactification gives rise to the $5d$ KK theory. The $q_\alpha$ capture the outer-automorphism twist performed on $\fg_\alpha$ while going around the circle, and folding captures the permutation of tensor multiplets occurring while going around the circle.

\section{Detailed Analysis}\label{da}
\subsection{General Rank}\label{dagr}
Consider the following geometry describing untwisted compactification of $6d$ SCFT carrying $\sp(m)$ on $-1$ curve
\be

\ee
Applying $\cS$ to $S_1$, we obtain the geometry
\be
%
\ee
which can be rewritten as
\be\label{spm}
%
\ee
which clearly describes the $5d$ gauge theory $\sp(m+1)+(2m+8)\F$. Now, applying $\cS$ on $S_{m+1}$, we obtain the geometry
\be\label{sum}
%
\ee
which describes the $5d$ gauge theory $\su(m+2)_0+(2m+8)\F$. Thus, we obtain
\be
%
\ee
for $m=0$. Removing the blowups sitting on $S_1$ of (\ref{spm}) and (\ref{sum}), the two geometries remain isomorphic, thus implying that the duality between $\su(m+2)$ and $\sp(m+1)$ gauge theories holds true as we integrate out fundamental matter from both sides of the duality. When all blowups are removed, we obtain that the geometry
\be
%
\ee
describing pure $\su(m+2)_{m+4}$ gauge theory is isomorphic to the geometry
\be
%
\ee
describing pure $\sp(m+1)$ gauge theory with theta angle $\theta=m\pi~(\text{mod }2\pi)$ (See Appendix B.3 of \cite{Bhardwaj:2019fzv} for an explanation).

Consider the following geometry describing the untwisted compactification of $6d$ SCFT carrying $\su(2n)$ on $-1$ curve
\be
%
\ee
Applying $\cS$ on $S_1$ and $S_{2n}$, we obtain the geometry
\be
%
\ee
which describes the $5d$ gauge theory $\su(2n+1)_0+\L^2+(2n+7)\F$. Similarly, applying $\cS$ on $S_1$ and $S_{2n+1}$ in
\be
%
\ee
we transition from untwisted compactification of $6d$ SCFT carrying $\su(2n+1)$ on $-1$ curve to $5d$ gauge theory $\su(2n+2)_0+\L^2+(2n+8)\F$. Thus, we obtain
\be
%
\ee

The KK theory
\be
%
\ee
is described by the geometry
\be
%
\ee
which can be rewritten as
\be
%
\ee
which describes the $5d$ gauge theory $\sp(m+1)+\L^2+8\F$. Another flop frame to describe the same $5d$ gauge theory is
\be\label{spmL}
%
\ee
Applying $\cS$ on $S_{m+1}$ of the above geometry implies that the above $5d$ gauge theory is dual to $\su(m+2)_{\frac m2+1}+\L^2+8\F$. Thus we obtain
\be
%
\ee
It is clear from (\ref{spmL}) that the above duality between $\sp(m+1)$ and $\su(m+2)$ gauge theories continues to hold as we integrate out $\F$ from both sides of the duality.

The $5d$ KK theory
\be
%
\ee
can be described by the geometry
\be
%
\ee
Applying $\cS$ on all the surfaces in this geometry, we learn that
\be
%
\ee

Similarly, the KK theory
\be
%
\ee
can be described by the geometry
\be
%
\ee
Applying $\cS$ on all surfaces of the above geometry, we learn that
\be
%
\ee
can be described by the geometry obtained by applying $\cS$ on all surfaces except $S_{2m+2}$ of the following geometry
\be
%
\ee
The above geometry can be seen to be flop equivalent to
\be
%
\ee
which is isomorphic to
\be
%
\ee
In a similar way, we can obtain
\be
%
\ee
we find that
\be
%
\ee

The KK theory
\be
%
\ee
can be described by the geometry
\be
%
\ee
which is isomorphic to
\be
%
\ee
thus describing $\sp(2m-1)_0+\A$. Similarly, the geometry
\be
%
\ee
describing the KK theory
\be
%
\ee
is isomorphic to the geometry
\be
%
\ee
describing $\sp(2m)_\pi+\A$. Moreover, applying $\cS$ on $S_{m+1}$ in the following geometry
\be
%
\ee
we learn that
\be
\sp(m+1)_{m\pi}+\A=\su(m+2)_{1+\frac m2}+\S^2
\ee
Combining all these results, we obtain
\be
%
\ee

Similarly, we also obtain
\be
%
\ee

Applying $\cS$ on all surfaces of
\be
%
\ee

Applying $\cS$ on all surfaces except $S_{m+1}$ of
\be
%
\ee
leads to a geometry describing $\su(2m+4)_0+\S^2+\L^2$. The above geometry is also isomorphic to
\be
%
\ee
Thus we obtain
\be
%
\ee

The fact that the geometry
\be
%
\ee
is isomorphic to the geometry
\be
%
\ee
we see that
\be
%
\ee
is isomorphic to
\be
%
\ee
And since the geometry
\be
%
\ee
is isomorphic to the geometry
\be
%
\ee

\subsection{Rank 2}
Consider the following geometry corresponding to the KK theory obtained by compactifying $6d$ SCFT carrying $\su(3)$ on $-1$ curve with a charge conjugation twist
\be
%
\ee
Applying $\cS$ on $S_2$, we obtain the geometry
\be
%
\ee
which describes the $5d$ gauge theory $\fg_2+6\F$. The above geometry is also isomorphic to the geometry
\be\label{D1s}
%
\ee
Applying $\cS$ on $S_2$ of the above geometry, we obtain the geometry
\be\label{D1e}
%
\ee
which describes the $5d$ gauge theory $\sp(2)+2\L^2+4\F$. The above geometry is also isomorphic to the geometry
\be\label{D2s}
%
\ee
Applying $\cS$ on $S_2$ of the above geometry, we obtain the geometry
\be\label{D2e}
%
\ee
which describes the $5d$ gauge theory $\su(3)_4+6\F$. Thus, we find that
\be\label{Dsu3+6F}
%
\ee
(\ref{D1s}) and (\ref{D1e}) are flop equivalent to
\be
%
\ee
respectively. The above two geometries remain related by $\cS$ if we remove blowups from $S_1$. In other words, the duality between $\sp(2)$ and $\fg_2$ continues to hold as we integrate out $\F$ from both sides of the duality (until a total of four $\F$ have been integrated out). Similarly, (\ref{D2s}) and (\ref{D2e}) imply that the duality between $\sp(2)$ and $\su(3)$ remains as we integrate out (upto four) $\F$ from both sides of the duality in such a way that the CS level for $\su(3)$ increases (in absolute value). Finally, the geometry (\ref{D1s}) is also isomorphic to the geometry
\be
\begin{tikzpicture} [scale=1.9]
\node (v1) at (-4.4,-0.5) {$\mathbf{2}^{6}_0$};
\node (v2) at (-2.3,-0.5) {$\mathbf{1}_2$};
\draw  (v1) edge (v2);
\node at (-3.8,-0.4) {\scriptsize{$3e$+$f$-$\sum x_i$}};
\node at (-2.6,-0.4) {\scriptsize{$e$}};
\end{tikzpicture}
\ee
Applying $\cS$ on $S_2$ of the above geometry we find the following geometry describing $\su(3)_4+6\F$
\be
\begin{tikzpicture} [scale=1.9]
\node (v1) at (-4.4,-0.5) {$\mathbf{2}^{6}_0$};
\node (v2) at (-2.3,-0.5) {$\mathbf{1}_2$};
\draw  (v1) edge (v2);
\node at (-3.8,-0.4) {\scriptsize{$e$+$3f$-$\sum x_i$}};
\node at (-2.6,-0.4) {\scriptsize{$e$}};
\end{tikzpicture}
\ee
These geometries imply that the duality between $\su(3)$ and $\fg_2$ remains preserved if $\F$ are integrated out from both sides of the duality in such a way that the CS level for $\su(3)$ increases. Notice that we can integrate out all the six $\F$ while preserving the duality between $\su(3)$ and $\fg_2$

A geometry describing the marginal theory $\su(3)_\frac{13}{2}+3\F$ is
\be\label{M1}
\begin{tikzpicture} [scale=1.9]
\node (v1) at (-4.2,-0.5) {$\mathbf{2}^3_0$};
\node (v2) at (-2.3,-0.5) {$\mathbf{1}_6$};
\draw  (v1) edge (v2);
\node at (-3.7,-0.4) {\scriptsize{$e$+$2f$}};
\node at (-2.6,-0.4) {\scriptsize{$e$}};
\end{tikzpicture}
\ee
Applying $\cS$ on $S_2$ leads to the geometry
\be
\begin{tikzpicture} [scale=1.9]
\node (v1) at (-4.2,-0.5) {$\mathbf{2}^3_0$};
\node (v2) at (-2.3,-0.5) {$\mathbf{1}_6$};
\draw  (v1) edge (v2);
\node at (-3.7,-0.4) {\scriptsize{$2e$+$f$}};
\node at (-2.6,-0.4) {\scriptsize{$e$}};
\end{tikzpicture}
\ee
thus implying that the above marginal theory is dual to $\sp(2)_\pi+3\L^2$ which is also a marginal theory. The geometry (\ref{M1}) is also isomorphic to the geometry
\be\label{D3s}
\begin{tikzpicture} [scale=1.9]
\node (v1) at (-4.4,-0.5) {$\mathbf{2}^{2+1}_0$};
\node (v2) at (-2.3,-0.5) {$\mathbf{1}_6$};
\draw  (v1) edge (v2);
\node at (-3.7,-0.4) {\scriptsize{$e$+$3f$-$\sum x_i$}};
\node at (-2.6,-0.4) {\scriptsize{$e$}};
\end{tikzpicture}
\ee
Applying $\cS$ on $S_2$ of the above geometry leads to the geometry
\be\label{D3e}
\begin{tikzpicture} [scale=1.9]
\node (v1) at (-4.4,-0.5) {$\mathbf{2}^{2+1}_0$};
\node (v2) at (-2.3,-0.5) {$\mathbf{1}_6$};
\draw  (v1) edge (v2);
\node at (-3.7,-0.4) {\scriptsize{$3e$+$f$-$\sum x_i$}};
\node at (-2.6,-0.4) {\scriptsize{$e$}};
\end{tikzpicture}
\ee
which implies that the above marginal theories are dual to $\fg_2+\A+2\F$. But $\fg_2+\A$ describes the circle compactification of $6d$ $\cN=(2,0)$ SCFT of type $D_4$ twisted along the circle by the order three outer automorphism of $D_4$. Thus, the theory
\be\label{>KK}
\su(3)_\frac{13}{2}+3\F~~=~~\sp(2)_\pi+3\L^2~~=~~\fg_2+\A+2\F
\ee
is obtained by adding matter to a $6d$ SCFT (compactified on a circle), implying that it cannot be a UV complete QFT. Said another way, the geometry corresponding to  (\ref{>KK}) is such that it is not possible to completely shrink all the compact curves and surfaces in the geometry simultaneously to a point. Thus, it is not possible to decouple (\ref{>KK}) from the rest of M-theory.

The isomorphism between (\ref{D3s}) and (\ref{D3e}) implies that
\be
\begin{tikzpicture} [scale=1.9]
\node at (-6.4,0.9) {$\fg_2+\A$};
\node at (-5.7,0.9) {$=$};
\node (v1) at (-5,0.8) {2};
\node at (-5,1.1) {$\su(1)^{(1)}$};
\draw  (-8.8,1.4) rectangle (-3.4,0.5);
\node at (-8,0.9) {$\su(3)_\frac{15}2+\F$};
\node at (-7.1,0.9) {$=$};
\node (v2) at (-4,0.8) {2};
\node at (-4,1.1) {$\su(1)^{(1)}$};
\node (v3) at (-4.5,0.8) {\tiny{3}};
\draw  (v1) edge (v3);
\draw[->]  (v3) edge (v2);
\end{tikzpicture}
\ee

Applying $\cS$ on $S_2$ of
\be
\begin{tikzpicture} [scale=1.9]
\node (v1) at (-4.3,-0.5) {$\mathbf{2}_0$};
\node (v2) at (-2.3,-0.5) {$\mathbf{1}_{10}$};
\draw  (v1) edge (v2);
\node at (-3.8,-0.4) {\scriptsize{$4e$+$f$}};
\node at (-2.7,-0.4) {\scriptsize{$e$}};
\end{tikzpicture}
\ee
we find
\be
\begin{tikzpicture} [scale=1.9]
\node at (-6,0.9) {$\su(3)_9$};
\node at (-5.3,0.9) {$=$};
\node at (-4.6,0.8) {3};
\node at (-4.6,1.1) {$\su(3)^{(2)}$};
\draw  (-6.6,1.4) rectangle (-4,0.5);
\end{tikzpicture}
\ee

Applying $\cS$ on $S_2$ of
\be
\begin{tikzpicture} [scale=1.9]
\node (v1) at (-4.5,-0.5) {$\mathbf{2}^3_0$};
\node (v2) at (-2.3,-0.5) {$\mathbf{1}_{6}$};
\draw  (v1) edge (v2);
\node at (-3.8,-0.4) {\scriptsize{$4e$+$2f$-$2\sum x_i$}};
\node at (-2.6,-0.4) {\scriptsize{$e$}};
\end{tikzpicture}
\ee
we find
\be
\begin{tikzpicture} [scale=1.9]
\node at (-6.2,0.9) {$\sp(2)_0+3\L^2$};
\node at (-5.3,0.9) {$=$};
\node at (-4.6,0.8) {2};
\node at (-4.6,1.1) {$\su(3)^{(2)}$};
\draw  (-7,1.4) rectangle (-4,0.5);
\end{tikzpicture}
\ee

\subsection{Rank 3}
A geometry describing the marginal theory $\su(4)_3+8\F$ is
\be\label{M2}
\begin{tikzpicture} [scale=1.9]
\node (v1) at (-4.7,-0.5) {$\mathbf{3}^{7+1}_0$};
\node (v2) at (-2.8,-0.5) {$\mathbf{2}_{1}$};
\node (v3) at (-1.4,-0.5) {$\mathbf{1}_{1}$};
\draw  (v1) edge (v2);
\draw  (v2) edge (v3);
\node at (-4,-0.4) {\scriptsize{$e$+$2f$-$\sum x_i$}};
\node at (-3.1,-0.4) {\scriptsize{$h$}};
\node at (-2.5,-0.4) {\scriptsize{$e$}};
\node at (-1.7,-0.4) {\scriptsize{$e$}};
\end{tikzpicture}
\ee
Applying $\cS$ on $S_3$ identifies a dual description of this theory as $\sp(3)+\L^3+7\F$. However, 
\be
\sp(3)+\L^3+5\F=\su(4)_4+6\F
\ee
is already a $5d$ KK theory as can be seen by applying $\cS$ on $S_2$ of the following geometry
\be
\begin{tikzpicture} [scale=1.9]
\node (v1) at (-4.9,-0.5) {$\mathbf{3}_1^{5+1+1+1}$};
\node (v2) at (-2.6,-0.5) {$\mathbf{2}_{0}$};
\node (v3) at (-0.5,-0.5) {$\mathbf{1}_{10}$};
\draw  (v1) edge (v2);
\node at (-4,-0.4) {\scriptsize{$2h$-$\sum x_i$-$y$}};
\node at (-2.9,-0.4) {\scriptsize{$f$}};
\node at (-2.2,-0.4) {\scriptsize{$4e$+$f$}};
\node at (-0.8,-0.4) {\scriptsize{$e$}};
\node (v4) at (-2.6,-1.1) {\scriptsize{4}};
\draw (v1) .. controls (-4.9,-1) and (-3.3,-1.1) .. (v4);
\draw (v3) .. controls (-0.5,-1) and (-1.9,-1.1) .. (v4);
\node at (-5.6,-0.8) {\scriptsize{$y$-$w,y$-$w,f$-$y$-$z,f$-$y$-$z$}};
\node at (-0.2,-0.8) {\scriptsize{$f,f,f,f$}};
\draw  (v2) edge (v3);
\end{tikzpicture}
\ee
Thus,
\be
\begin{tikzpicture} [scale=1.9]
\node at (-6.1,0.9) {$\sp(3)+\L^3+5\F$};
\node at (-5.1,0.9) {$=$};
\node (v1) at (-4.4,0.8) {3};
\node at (-4.4,1.1) {$\su(3)^{(2)}$};
\draw  (-8.8,1.4) rectangle (-2.8,0.5);
\node at (-8,0.9) {$\su(4)_4+6\F$};
\node at (-7.1,0.9) {$=$};
\node (v2) at (-3.4,0.8) {1};
\node at (-3.4,1.1) {$\sp(0)^{(1)}$};
\node (v3) at (-3.9,0.8) {\tiny{2}};
\draw  (v1) edge (v3);
\draw[->]  (v3) edge (v2);
\end{tikzpicture}
\ee
(\ref{M2}) implies that the duality between $\sp(3)$ and $\su(4)$ remains preserved if we integrate out $\F$ from both sides of the duality such that the CS level for $\su(4)$ increases.

Applying $\cS$ on $S_1$ of
\be
\begin{tikzpicture} [scale=1.9]
\node (v1) at (-4.6,-0.5) {$\mathbf{1}_0$};
\node (v2) at (-2.8,-0.5) {$\mathbf{2}_{8}$};
\node (v3) at (-1.3,-0.5) {$\mathbf{3}_{10}$};
\draw  (v1) edge (v2);
\draw  (v2) edge (v3);
\node at (-4.1,-0.4) {\scriptsize{$3e$+$f$}};
\node at (-2.5,-0.4) {\scriptsize{$h$}};
\node at (-3.1,-0.4) {\scriptsize{$e$}};
\node at (-1.65,-0.4) {\scriptsize{$e$}};
\end{tikzpicture}
\ee
implies
\be
\begin{tikzpicture} [scale=1.9]
\node at (-5.7,0.9) {$\su(4)_8$};
\node at (-5,0.9) {$=$};
\node (v1) at (-4.3,0.8) {4};
\node at (-4.3,1.1) {$\so(8)^{(3)}$};
\draw  (-6.2,1.4) rectangle (-3.7,0.5);
\end{tikzpicture}
\ee

Consider the following geometry describing $\sp(3)+\half\L^3+\frac{19}2\F$
\be
\begin{tikzpicture} [scale=1.9]
\node (v1) at (-4.3,-0.5) {$\mathbf{1}_9^{7}$};
\node (v2) at (-2.6,-0.5) {$\mathbf{2}_{3}$};
\node (v3) at (-0.5,-0.5) {$\mathbf{3}^{2+2}_{0}$};
\draw  (v1) edge (v2);
\draw  (v2) edge (v3);
\node at (-4,-0.4) {\scriptsize{$e$}};
\node at (-3,-0.4) {\scriptsize{$h$+$2f$}};
\node at (-2.3,-0.4) {\scriptsize{$e$}};
\node at (-1.3,-0.4) {\scriptsize{$2e$+$f$-$x_2$-$\sum y_i$}};
\node (v4) at (-2.3,-1.1) {\scriptsize{2}};
\draw (v1) .. controls (-4.3,-1) and (-3,-1.1) .. (v4);
\draw (v3) .. controls (-0.5,-1) and (-1.6,-1.1) .. (v4);
\node at (-4.5,-0.8) {\scriptsize{$f,f$}};
\node at (0,-0.8) {\scriptsize{$x_2$-$x_1,f$-$x_1$-$x_2$}};
\end{tikzpicture}
\ee
This geometry is isomorphic to the following geometry
\be
\begin{tikzpicture} [scale=1.9]
\node (v1) at (-4.3,-0.5) {$\mathbf{1}_9^{7}$};
\node (v2) at (-2.4,-0.5) {$\mathbf{2}_{3}$};
\node (v3) at (-0.5,-0.5) {$\mathbf{3}^{2+2}_{0}$};
\draw  (v1) edge (v2);
\draw  (v2) edge (v3);
\node at (-4,-0.4) {\scriptsize{$e$}};
\node at (-2.8,-0.4) {\scriptsize{$h$+$2f$}};
\node at (-2.1,-0.4) {\scriptsize{$e$}};
\node at (-1.1,-0.4) {\scriptsize{$e$+$f$-$x_1$}};
\node (v4) at (-2.3,-1.1) {\scriptsize{2}};
\draw (v1) .. controls (-4.3,-1) and (-3,-1.1) .. (v4);
\draw (v3) .. controls (-0.5,-1) and (-1.6,-1.1) .. (v4);
\node at (-4.5,-0.8) {\scriptsize{$f,f$}};
\node at (0,-0.8) {\scriptsize{$x_1$-$x_2,f$-$y_1$-$y_2$}};
\end{tikzpicture}
\ee
which describes $\su(4)_\frac32+2\L^2+7\F$. The latter $5d$ KK theory is known to be a $5d$ KK theory from the results of Section \ref{dagr}. We thus have
\be
\begin{tikzpicture} [scale=1.9]
\node at (-8.2,0.9) {$\su(4)_\frac 32+2\L^2+7\F$};
\node at (-4.9,0.9) {$=$};
\node (v1) at (-4.2,0.8) {1};
\node at (-4.2,1.1) {$\sp(1)^{(1)}$};
\draw  (-9.2,1.4) rectangle (-2.6,0.5);
\node (v2) at (-3.2,0.8) {2};
\node at (-3.2,1.1) {$\su(1)^{(1)}$};
\draw  (v1) edge (v2);
\node at (-7.1,0.9) {$=$};
\node at (-6,0.9) {$\sp(3)+\frac12\L^3+\frac{19}2\F$};
\end{tikzpicture}
\ee

Applying $\cS$ on $S_3$ of
\be
\begin{tikzpicture} [scale=1.9]
\node (v1) at (-4.25,-0.5) {$\mathbf{1}_8$};
\node (v2) at (-3,-0.5) {$\mathbf{2}^2_{6}$};
\node (v3) at (-1.7,-0.5) {$\mathbf{3}_{0}$};
\draw  (v1) edge (v2);
\draw  (v2) edge (v3);
\node at (-4,-0.4) {\scriptsize{$e$}};
\node at (-3.3,-0.4) {\scriptsize{$h$}};
\node at (-2.7,-0.4) {\scriptsize{$e$}};
\node at (-2.1,-0.4) {\scriptsize{$2e$+$f$}};
\end{tikzpicture}
\ee
yields
\be
\sp(3)_0+2\L^2=\su(4)_6+2\L^2
\ee
And applying $\cS$ on $S_3$ of
\be
\begin{tikzpicture} [scale=1.9]
\node (v1) at (-4.3,-0.5) {$\mathbf{1}_{10}$};
\node (v2) at (-2.6,-0.5) {$\mathbf{2}_{4}$};
\node (v3) at (-0.1,-0.5) {$\mathbf{3}^{2+2}_{0}$};
\draw  (v1) edge (v2);
\draw  (v2) edge (v3);
\node at (-4,-0.4) {\scriptsize{$e$}};
\node at (-3,-0.4) {\scriptsize{$h$+$2f$}};
\node at (-2.3,-0.4) {\scriptsize{$e$}};
\node at (-1.1,-0.4) {\scriptsize{$3e$+$2f$-$2\sum x_i$-$\sum y_i$}};
\node (v4) at (-2.1,-1.1) {\scriptsize{2}};
\draw (v1) .. controls (-4.3,-1) and (-2.8,-1.1) .. (v4);
\draw (v3) .. controls (-0.1,-1) and (-1.4,-1.1) .. (v4);
\node at (-4.4,-0.8) {\scriptsize{$f$}};
\node at (0.1,-0.8) {\scriptsize{$x_i$-$y_i$}};
\end{tikzpicture}
\ee
implies
\be
\begin{tikzpicture} [scale=1.9]
\node at (-6.2,0.9) {$\sp(3)_0+2\L^2$};
\node at (-5.3,0.9) {$=$};
\node at (-4.6,0.8) {2};
\node at (-4.6,1.1) {$\so(8)^{(3)}$};
\draw  (-8.8,1.4) rectangle (-4,0.5);
\node at (-8,0.9) {$\su(4)_6+2\L^2$};
\node at (-7.1,0.9) {$=$};
\end{tikzpicture}
\ee

Applying $\cS$ on $S_2$ of
\be
\begin{tikzpicture} [scale=1.9]
\node (v1) at (-4.3,-0.5) {$\mathbf{3}_3^3$};
\node (v2) at (-2.6,-0.5) {$\mathbf{2}^3_{0}$};
\node (v3) at (-0.9,-0.5) {$\mathbf{1}^1_{1}$};
\draw  (v1) edge (v2);
\draw  (v2) edge (v3);
\node at (-4,-0.4) {\scriptsize{$e$}};
\node at (-3.2,-0.4) {\scriptsize{$2e$+$f$-$\sum x_i$}};
\node at (-2,-0.4) {\scriptsize{$e$+$f$-$\sum x_i$}};
\node at (-1.2,-0.4) {\scriptsize{$e$}};
\end{tikzpicture}
\ee
implies that
\be
\so(7)+6\S+\F=\su(4)_0+3\L^2+4\F
\ee
Applying $\cS$ on $S_2$ of
\be
\begin{tikzpicture} [scale=1.9]
\node (v1) at (-5.2,-0.5) {$\mathbf{3}_0$};
\node (v2) at (-2.5,-0.5) {$\mathbf{2}^{3+3+1}_{0}$};
\node (v3) at (-0.2,-0.5) {$\mathbf{1}_{0}$};
\draw  (v1) edge (v2);
\draw  (v2) edge (v3);
\node at (-4.9,-0.4) {\scriptsize{$e$}};
\node at (-3.7,-0.4) {\scriptsize{$2e$+$2f$-$\sum x_i$-$\sum y_i$-$2z$}};
\node at (-1.5,-0.4) {\scriptsize{$2e$+$f$-$\sum x_i$-$\sum y_i$}};
\node at (-0.5,-0.4) {\scriptsize{$e$}};
\end{tikzpicture}
\ee
implies that
\be
\begin{tikzpicture} [scale=1.9]
\node at (-6.3,0.9) {$\so(7)+6\S+\F$};
\node at (-5.3,0.9) {$=$};
\node at (-4.6,0.8) {1};
\node at (-4.6,1.1) {$\su(4)^{(2)}$};
\draw  (-9.4,1.4) rectangle (-4,0.5);
\node at (-8.4,0.9) {$\su(4)_0+3\L^2+4\F$};
\node at (-7.3,0.9) {$=$};
\end{tikzpicture}
\ee

Applying $\cS$ on $S_2$ of
\be
\begin{tikzpicture} [scale=1.9]
\node (v1) at (-4.3,-0.5) {$\mathbf{3}_3^4$};
\node (v2) at (-2.6,-0.5) {$\mathbf{2}^3_{0}$};
\node (v3) at (-0.9,-0.5) {$\mathbf{1}_{1}$};
\draw  (v1) edge (v2);
\draw  (v2) edge (v3);
\node at (-4,-0.4) {\scriptsize{$e$}};
\node at (-3.2,-0.4) {\scriptsize{$2e$+$f$-$\sum x_i$}};
\node at (-2,-0.4) {\scriptsize{$e$+$f$-$\sum x_i$}};
\node at (-1.2,-0.4) {\scriptsize{$e$}};
\end{tikzpicture}
\ee
implies that
\be
\so(7)+7\S=\su(4)_1+3\L^2+4\F
\ee
Similarly, applying $\cS$ on $S_2$ of
\be
\begin{tikzpicture} [scale=1.9]
\node (v1) at (-4.3,-0.5) {$\mathbf{3}_3^2$};
\node (v2) at (-2.6,-0.5) {$\mathbf{2}^3_{0}$};
\node (v3) at (-0.9,-0.5) {$\mathbf{1}^2_{1}$};
\draw  (v1) edge (v2);
\draw  (v2) edge (v3);
\node at (-4,-0.4) {\scriptsize{$e$}};
\node at (-3.2,-0.4) {\scriptsize{$2e$+$f$-$\sum x_i$}};
\node at (-2,-0.4) {\scriptsize{$e$+$f$-$\sum x_i$}};
\node at (-1.2,-0.4) {\scriptsize{$e$}};
\end{tikzpicture}
\ee
implies that
\be
\so(7)+5\S+2\F=\su(4)_1+3\L^2+4\F
\ee
Now, applying $\cS$ on $S_2$ of
\be
\begin{tikzpicture} [scale=1.9]
\node (v1) at (-4.3,-0.5) {$\mathbf{3}_1$};
\node (v2) at (-2.6,-0.5) {$\mathbf{2}^7_{0}$};
\node (v3) at (-0.9,-0.5) {$\mathbf{1}_{1}$};
\draw  (v1) edge (v2);
\draw  (v2) edge (v3);
\node at (-4,-0.4) {\scriptsize{$h$}};
\node at (-3.2,-0.4) {\scriptsize{$e$+$2f$-$\sum x_i$}};
\node at (-2,-0.4) {\scriptsize{$3e$+$f$-$\sum x_i$}};
\node at (-1.2,-0.4) {\scriptsize{$e$}};
\end{tikzpicture}
\ee
implies that
\be
\begin{tikzpicture} [scale=1.9]
\node at (-7.9,0.9) {$\so(7)+5\S+2\F$};
\node at (-5.3,0.9) {$=$};
\node at (-4.7,0.8) {1};
\node at (-4.7,1.1) {$\fg_2^{(1)}$};
\draw  (-11,1.4) rectangle (-4.3,0.5);
\node at (-10,0.9) {$\su(4)_1+3\L^2+4\F$};
\node at (-8.9,0.9) {$=$};
\node at (-6.9,0.9) {$=$};
\node at (-6.1,0.9) {$\so(7)+7\S$};
\end{tikzpicture}
\ee

Similarly, applying $\cS$ on $S_2$ of the following two geometries
\be
\begin{tikzpicture} [scale=1.9]
\node (v1) at (-4.3,-0.5) {$\mathbf{3}_3^1$};
\node (v2) at (-2.6,-0.5) {$\mathbf{2}^3_{0}$};
\node (v3) at (-0.9,-0.5) {$\mathbf{1}^3_{1}$};
\draw  (v1) edge (v2);
\draw  (v2) edge (v3);
\node at (-4,-0.4) {\scriptsize{$e$}};
\node at (-3.2,-0.4) {\scriptsize{$2e$+$f$-$\sum x_i$}};
\node at (-2,-0.4) {\scriptsize{$e$+$f$-$\sum x_i$}};
\node at (-1.2,-0.4) {\scriptsize{$e$}};
\end{tikzpicture}
\ee
\be
\begin{tikzpicture} [scale=1.9]
\node (v1) at (-4.4,-0.5) {$\mathbf{1}_2^{4+4}$};
\node (v2) at (-2.9,-0.5) {$\mathbf{2}^{3}_{0}$};
\node (v3) at (-0.8,-0.5) {$\mathbf{3}_{6}$};
\draw  (v1) edge (v2);
\draw  (v2) edge (v3);
\node at (-4,-0.4) {\scriptsize{$e$}};
\node at (-3.2,-0.4) {\scriptsize{$f$}};
\node at (-2.2,-0.4) {\scriptsize{$4e$+$2f$-$2\sum x_i$}};
\node at (-1.1,-0.4) {\scriptsize{$e$}};
\node at (-4.7,-0.8) {\scriptsize{$f$-$x_i$-$y_i$}};
\node at (-0.7,-0.8) {\scriptsize{$f$}};
\node (v4) at (-2.5,-1.1) {\scriptsize{4}};
\draw (v1) .. controls (-4.4,-1) and (-3.2,-1.1) .. (v4);
\draw (v3) .. controls (-0.8,-1) and (-1.8,-1.1) .. (v4);
\end{tikzpicture}
\ee
we find that
\be
\begin{tikzpicture} [scale=1.9]
\node at (-8,0.9) {$\su(4)_2+3\L^2+4\F$};
\node at (-4.9,0.9) {$=$};
\node (v1) at (-4.2,0.8) {1};
\node at (-4.2,1.1) {$\sp(0)^{(1)}$};
\draw  (-9,1.4) rectangle (-2.6,0.5);
\node (v2) at (-3.2,0.8) {2};
\node at (-3.2,1.1) {$\su(3)^{(2)}$};
\draw  (v1) edge (v2);
\node at (-6.9,0.9) {$=$};
\node at (-5.9,0.9) {$\so(7)+4\S+3\F$};
\end{tikzpicture}
\ee

A geometry describing $\su(4)_5+3\L^2$ is
\be
\begin{tikzpicture} [scale=1.9]
\node (v1) at (-4.8,-0.5) {$\mathbf{3}^{3+3}_1$};
\node (v2) at (-3,-0.5) {$\mathbf{2}_{2}$};
\node (v3) at (-1.6,-0.5) {$\mathbf{1}_{10}$};
\draw  (v1) edge (v2);
\draw  (v2) edge (v3);
\node at (-4.1,-0.4) {\scriptsize{$h$+$f$-$\sum x_i$}};
\node at (-3.3,-0.4) {\scriptsize{$e$}};
\node at (-2.6,-0.4) {\scriptsize{$h$+$3f$}};
\node at (-1.9,-0.4) {\scriptsize{$e$}};
\node at (-5,-0.8) {\scriptsize{$x_i$-$y_i$}};
\node at (-1.5,-0.8) {\scriptsize{$f$}};
\node (v4) at (-3.2,-1.1) {\scriptsize{3}};
\draw (v1) .. controls (-4.8,-1) and (-3.9,-1.1) .. (v4);
\draw (v3) .. controls (-1.6,-1) and (-2.5,-1.1) .. (v4);
\end{tikzpicture}
\ee
which can be rewritten as
\be
\begin{tikzpicture} [scale=1.9]
\node (v1) at (-5.2,-0.5) {$\mathbf{3}^{3+3}_0$};
\node (v2) at (-2.9,-0.5) {$\mathbf{2}_{2}$};
\node (v3) at (-1.5,-0.5) {$\mathbf{1}_{10}$};
\draw  (v1) edge (v2);
\draw  (v2) edge (v3);
\node at (-4.25,-0.4) {\scriptsize{$2e$+$f$-$x_1$-$x_3$-$y_1$-$y_2$}};
\node at (-3.2,-0.4) {\scriptsize{$e$}};
\node at (-2.5,-0.4) {\scriptsize{$h$+$3f$}};
\node at (-1.8,-0.4) {\scriptsize{$e$}};
\node at (-5.9,-0.8) {\scriptsize{$f$-$x_2$-$y_2,y_2$-$x_2,x_3$-$y_3$}};
\node at (-1.3,-0.8) {\scriptsize{$f,f,f$}};
\node (v4) at (-3.2,-1.1) {\scriptsize{3}};
\draw (v1) .. controls (-5.2,-1) and (-3.9,-1.1) .. (v4);
\draw (v3) .. controls (-1.5,-1) and (-2.5,-1.1) .. (v4);
\end{tikzpicture}
\ee
which implies that there is a dual description as $\sp(3)+\half\L^3+\L^2+\frac52\F$. We can further rewrite the above geometry as
\be
\begin{tikzpicture} [scale=1.9]
\node (v1) at (-5.7,-0.5) {$\mathbf{3}^{3+3}_0$};
\node (v2) at (-2.9,-0.5) {$\mathbf{2}_{2}$};
\node (v3) at (-1.5,-0.5) {$\mathbf{1}_{10}$};
\draw  (v1) edge (v2);
\draw  (v2) edge (v3);
\node at (-4.55,-0.4) {\scriptsize{$3e$+$2f$-$x_1$-$2x_2$-$2x_3$-$\sum y_i$}};
\node at (-3.2,-0.4) {\scriptsize{$e$}};
\node at (-2.5,-0.4) {\scriptsize{$h$+$3f$}};
\node at (-1.8,-0.4) {\scriptsize{$e$}};
\node at (-6.4,-0.8) {\scriptsize{$f$-$x_1$-$y_1,x_2$-$y_2,x_3$-$y_3$}};
\node at (-1.3,-0.8) {\scriptsize{$f,f,f$}};
\node (v4) at (-3.5,-1.1) {\scriptsize{3}};
\draw (v1) .. controls (-5.7,-1) and (-4.2,-1.1) .. (v4);
\draw (v3) .. controls (-1.5,-1) and (-2.8,-1.1) .. (v4);
\end{tikzpicture}
\ee
which implies that
\be
\begin{tikzpicture} [scale=1.9]
\node at (-8.4,0.9) {$\su(4)_5+3\L^2$};
\node at (-4.9,0.9) {$=$};
\node (v1) at (-4.2,0.8) {1};
\node at (-4.2,1.1) {$\so(8)^{(3)}$};
\draw  (-9.2,1.4) rectangle (-3.6,0.5);
\node at (-7.5,0.9) {$=$};
\node at (-6.2,0.9) {$\sp(3)+\frac12\L^3+\L^2+\frac{5}2\F$};
\end{tikzpicture}
\ee

Applying $\cS$ on $S_2$ of
\be
\begin{tikzpicture} [scale=1.9]
\node (v1) at (-4.8,-0.5) {$\mathbf{1}_2$};
\node (v2) at (-2.7,-0.5) {$\mathbf{2}^4_{0}$};
\node (v3) at (-0.6,-0.5) {$\mathbf{3}_{2}$};
\draw  (v1) edge (v2);
\draw  (v2) edge (v3);
\node at (-4.5,-0.4) {\scriptsize{$e$}};
\node at (-3.4,-0.4) {\scriptsize{$2e$+$f$-$\sum x_i$}};
\node at (-2,-0.4) {\scriptsize{$2e$+$f$-$\sum x_i$}};
\node at (-0.9,-0.4) {\scriptsize{$e$}};
\end{tikzpicture}
\ee
leads to
\be
\begin{tikzpicture} [scale=1.9]
\node at (-5.8,0.9) {$\su(4)_0+4\L^2$};
\node at (-4.9,0.9) {$=$};
\node (v1) at (-4.3,0.8) {2};
\node at (-4.3,1.1) {$\su(4)^{(2)}$};
\draw  (-6.6,1.4) rectangle (-3.7,0.5);
\end{tikzpicture}
\ee

Applying $\cS$ on $S_2$ of
\be
\begin{tikzpicture} [scale=1.9]
\node (v1) at (-4.5,-0.5) {$\mathbf{1}_1$};
\node (v2) at (-2.7,-0.5) {$\mathbf{2}^{3+1}_{0}$};
\node (v3) at (-1,-0.5) {$\mathbf{3}_{3}$};
\draw  (v1) edge (v2);
\draw  (v2) edge (v3);
\node at (-4.2,-0.4) {\scriptsize{$e$}};
\node at (-3.35,-0.4) {\scriptsize{$e$+$f$-$\sum x_i$}};
\node at (-2,-0.4) {\scriptsize{$2e$+$f$-$\sum x_i$}};
\node at (-1.3,-0.4) {\scriptsize{$e$}};
\end{tikzpicture}
\ee
implies that
\be
\su(4)_1+4\L^2=\so(7)+\A+3\S
\ee
But, since $\so(7)+\A$ is already a $5d$ KK theory, the marginal theory $\su(4)_1+4\L^2$ cannot describe either a $5d$ SCFT or a $5d$ KK theory. Removing matter from the marginal theory leads us to the theory $\su(4)_1+3\L^2$ which is a $5d$ SCFT as it can be obtained by removing matter from $5d$ KK theories discussed above.

Applying $\cS$ on $S_2$ of
\be
\begin{tikzpicture} [scale=1.9]
\node (v1) at (-4.25,-0.5) {$\mathbf{1}_0$};
\node (v2) at (-2.5,-0.5) {$\mathbf{2}^4_{0}$};
\node (v3) at (-0.8,-0.5) {$\mathbf{3}_{4}$};
\draw  (v1) edge (v2);
\draw  (v2) edge (v3);
\node at (-4,-0.4) {\scriptsize{$e$}};
\node at (-3.1,-0.4) {\scriptsize{$e$+$f$-$\sum x_i$}};
\node at (-1.9,-0.4) {\scriptsize{$3e$+$f$-$\sum x_i$}};
\node at (-1.1,-0.4) {\scriptsize{$e$}};
\end{tikzpicture}
\ee
leads to
\be
\begin{tikzpicture} [scale=1.9]
\node at (-5.8,0.9) {$\su(4)_2+4\L^2$};
\node at (-4.9,0.9) {$=$};
\node (v1) at (-4.3,0.8) {2};
\node at (-4.3,1.1) {$\fg_2^{(1)}$};
\draw  (-6.6,1.4) rectangle (-3.9,0.5);
\end{tikzpicture}
\ee

A geometry describing $\su(4)_3+4\L^2$ is obtained by applying $\cS$ on $S_2$ in the following geometry
\be
\begin{tikzpicture} [scale=1.9]
\node (v1) at (-4.2,-0.5) {$\mathbf{3}^{3+3}_2$};
\node (v2) at (-2.8,-0.5) {$\mathbf{2}^{1}_{0}$};
\node (v3) at (-1.3,-0.5) {$\mathbf{1}_{8}$};
\draw  (v1) edge (v2);
\draw  (v2) edge (v3);
\node at (-3.8,-0.4) {\scriptsize{$e$}};
\node at (-3.1,-0.4) {\scriptsize{$f$}};
\node at (-2.4,-0.4) {\scriptsize{$3e$+$f$}};
\node at (-1.6,-0.4) {\scriptsize{$e$}};
\node at (-4.5,-0.8) {\scriptsize{$f$-$x_i$-$y_i$}};
\node at (-1.2,-0.8) {\scriptsize{$f$}};
\node (v4) at (-2.7,-1.1) {\scriptsize{3}};
\draw (v1) .. controls (-4.2,-1) and (-3.4,-1.1) .. (v4);
\draw (v3) .. controls (-1.3,-1) and (-2,-1.1) .. (v4);
\end{tikzpicture}
\ee
The above geometry can be rewritten as
\be
\begin{tikzpicture} [scale=1.9]
\node (v1) at (-4.4,-0.5) {$\mathbf{3}^{3+3}_0$};
\node (v2) at (-2.8,-0.5) {$\mathbf{2}^{1}_{0}$};
\node (v3) at (-1.3,-0.5) {$\mathbf{1}_{8}$};
\draw  (v1) edge (v2);
\draw  (v2) edge (v3);
\node at (-3.9,-0.4) {\scriptsize{$x_1$-$x_2$}};
\node at (-3.1,-0.4) {\scriptsize{$f$}};
\node at (-2.4,-0.4) {\scriptsize{$3e$+$f$}};
\node at (-1.6,-0.4) {\scriptsize{$e$}};
\node at (-5.1,-0.8) {\scriptsize{$f$-$x_1$-$y_1,y_1$-$x_1,x_2$-$y_2$}};
\node at (-1.1,-0.8) {\scriptsize{$f,f,f$}};
\node (v4) at (-2.8,-1.1) {\scriptsize{3}};
\draw (v1) .. controls (-4.4,-1) and (-3.5,-1.1) .. (v4);
\draw (v3) .. controls (-1.3,-1) and (-2.1,-1.1) .. (v4);
\end{tikzpicture}
\ee
which implies that this theory has a dual description as a $\fg_2\oplus\su(2)$ gauge theory with a half-hyper in bifundamental, a hyper in $\A$ of $\fg_2$ and two full hypers in $\F$ of $\su(2)$. Now, turning off the gauge coupling for $\su(2)$ leads to an RG flow producing $\fg_2+\A+\F$ which has more matter than a KK theory. Geometrically this RG flow is implemented by decompactifying the curve $e$ in $S_3$. Thus $\su(4)_3+4\L^2$ can neither be a $5d$ SCFT nor a $5d$ KK theory, since otherwise $\fg_2+\A+\F$ would have to describe a $5d$ SCFT or a $5d$ KK theory, which cannot be the case as $\fg_2+\A$ is already a $5d$ KK theory.

Applying $\cS$ on $S_2$ of
\be\label{G31}

\ee
we find that
\be
%
\ee
which can be rewritten as
\be
%
\ee
implying the duality
\be
\sp(3)_\pi+2\L^2=\sp(3)+\frac32\L^3+\frac32\F
\ee
but the latter theory exceeds the bounds placed on marginal theories in \cite{Jefferson:2017ahm}. Hence, $\sp(3)_\pi+2\L^2$ describes neither a $5d$ SCFT nor a $5d$ KK theory.

A geometry describing $\su(4)_0+\S^2+\L^2$ is obtained by applying $\cS$ on $S_1$ and $S_3$ of
\be
%
\ee
The above geometry can be rewritten as
\be
%
\ee
which implies that
\be
%
\ee
describing $\so(7)+2\S+4\F$ can be rewritten as
\be
%
\ee
which implies that
\be
%
\ee

\subsection{Rank 4}
Consider the geometry
\be
%
\ee
Performing $\cS$ on $S_3$ leads us to an $\so(9)$ description, and performing $\cS$ on $S_3,S_2$ leads us to an $\su(5)$ description. Working out the matter content, we find that
\be
%
\ee

Applying $\cS$ on $S_2$ of the following geometry
\be
%
\ee
leads to
\be
\su(5)_3+3\L^2+\F=\ff_4+4\F
\ee
Since $\ff_4+4\F$ violates the bound for marginal theories, we conclude that $\su(5)_3+3\L^2+\F$ is not a KK theory or an SCFT.

Applying $\cS$ to $S_1$ of
\be
%
\ee
leads to
\be
\su(5)_{\frac{11}2}+5\F=\sp(4)+4\F+\L^4
\ee
Since the latter theory violates the bound for marginal theories, we conclude that the former theory is not a KK theory or an SCFT.

Applying $\cS$ to $S_2$ of
\be\label{ka}
%
\ee
implies the duality
\be
\su(5)_{\frac{3}2}+3\L^2+2\F=\so(9)+4\S+\F
\ee
And applying $\cS$ to $S_2$ of the following geometry
\be
%
\ee
leads to the identification of the $\so(9)$ theory as a KK theory. In full detail, we have
\be
%
\ee

Consider the geometry
\be
%
\ee
which can be rewritten by performing an isomorphism on $S_1$ as
\be
%
\ee
Equating the theories corresponding to these two isomorphic geometries, we obtain
\be
%
\ee

Now, consider the geometry
\be
%
\ee
which is isomorphic to
\be
%
\ee

Similarly, the geometry
\be
%
\ee
is isomorphic to
\be
%
\ee
leading us to the conclusion that
\be
%
\ee
leads us to conclude
\be
%
\ee
leads us to
\be
%
\ee

Applying $\cS$ on $S_2$ of the geometry
\be
%
\ee

\subsection{Rank 5}
Applying $\cS$ on $S_2$ and $S_4$ of
\be
%
\ee
converts the right hand side of the following equality to the left hand side
\be
%
\ee

Now, consider the geometry
\be
%
\ee
Applying $\cS$ to $S_2$, we obtain an $\so(11)$ description, while applying $\cS$ to $S_2$ and $S_4$, we obtain an $\su(6)$ description
\be
%
\ee
implies that
\be
\su(6)_2+\frac32\L^3+3\F=\so(11)+\frac52\S+2\F
\ee
Since the theory on the right hand side of the above equation exceeds the bound for marginal theories, we find that the theory on the left hand side is neither a $5d$ KK theory nor a $5d$ SCFT. Integrating out matter from (\ref{52s}), we find that
\be\label{52s1}
\su(6)_3+\frac32\L^3+\F=\so(11)+\frac52\S
\ee
where the right hand side lifts to a KK theory. See (\ref{52s2}).

Applying $\cS$ on $S_4$ of
\be
%
\ee

Let us consider the following geometry describing $\su(6)_\frac32+\L^3+9\F$
\be
%
\ee
After the flop of $e-x_1$ in $S_5$, the geometry can be written as
\be
%
\ee
Flopping $y$ in $S_5$ leads to the geometry
\be
%
\ee
So, we find that
\be
%
\ee
for $\su(6)_0+\S^2+\half\L^3+\F$ can be flopped to
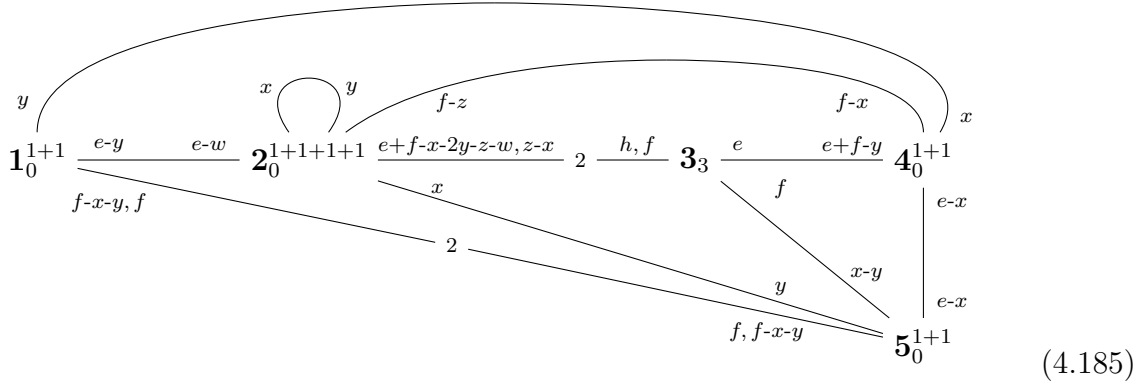
\be
%
\ee
Applying $\cS$ to $S_1$, $S_2$, $S_4$ and $S_5$ we find that
\be
%
\ee

Similarly, we find that
\be
%
\ee

$\su(6)_\frac32+\half\L^3+2\L^2+2\F$ can be described by the geometry
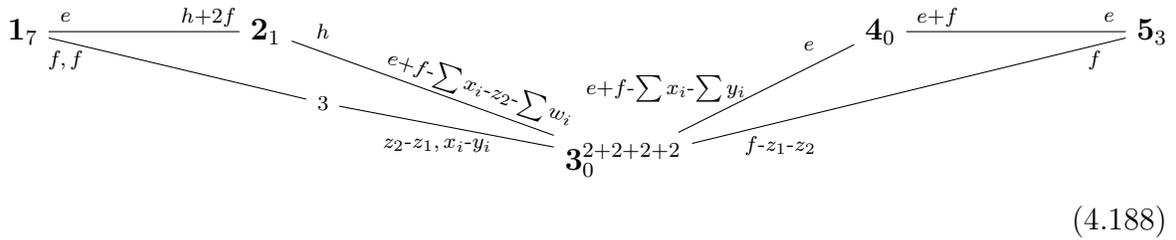
\be
%
\ee
Performing isomorphisms on $S_3$, we can write the above geometry as
\be
%
\ee
Performing some flops associated to $x_i$ and $y_i$, we obtain
\be
%
\ee

Now consider the geometry
\be
%
\ee
which is isomorphic to
\be
%
\ee
(\ref{irrd}) is an irreducible duality, that is, it isn't possible to integrate out any matter while preserving the duality.

Performing $\cS$ on $S_1$, $S_2$, $S_4$ and $S_5$ of
\be
%
\ee
describing $\su(6)_\frac32+\half\L^3+\L^2+8\F$ is isomorphic to
\be
%
\ee
which describes $\su(6)_\frac32+2\L^2+7\F$. The latter theory is known to be KK from Section \ref{dagr}, and we obtain
\be
%
\ee
we learn that
\be
%
\ee
is isomorphic to
\be
%
\ee
We already know from Section \ref{dagr} that the right hand side is obtained by untwisted compactification of rank-5 E-string theory.

Applying $\cS$ to $S_2$ and $S_3$ of the geometry
\be
%
\ee
we see that it gives rise to the $5d$ gauge theory $\su(6)_3+3\L^2$. Flopping $f-w$ in $S_2$, we obtain
\be
%
\ee

The theories $\su(6)_k+3\L^2$ for $k=0,2$ can be reduced to $\sp(3)_\pi+2\L^2$ by Higgsing. As the latter theory is neither $5d$ SCFT nor $5d$ KK theory, the former theories cannot be $5d$ SCFTs or $5d$ KK theories either.

Applying $\cS$ on $S_4$ of
\be
%
\ee
we find that
\be
%
\ee
is isomorphic to the geometry
\be
%
\ee
we find that
\be
%
\ee
we discover that
\be
%
\ee
is isomorphic to the geometry
\be
%
\ee

\subsection{Rank 6}
Applying $\cS$ on $S_2$, $S_3$, $S_4$ and $S_5$ of
\be
%
\ee
leads to
\be
\su(7)_\frac32+\L^3+5\F=\so(13)+\S+5\F
\ee
Now, consider the geometry
\be
%
\ee
Applying $\cS$ on $S_2$ and $S_4$ of the above geometry, we find that
\be
%
\ee
we see that
\be
%
\ee

We claim that
\be\label{so1212}
\so(12)+\frac32\S+6\F=\so(12)+\S+\half\C+6\F
\ee
The proof is slightly involved. Let us start with the following two geometries
\be\label{s1}
%
\ee
describing $\so(12)+\frac32\S+6\F$ and $\so(12)+\S+\half\C+6\F$ respectively. (\ref{s1}) and (\ref{s2}) can be flopped to obtain the following geometries respectively
\be
%
\ee
Applying $\cS$ on $S_2$ of both the above geometries, we obtain the geometries
\be\label{s11}
%
\ee
Performing a few more flops we obtain the following two geometries from (\ref{s11}) and (\ref{s22}) respectively
\be\label{s111}
%
\ee
Relabeling $S_5$ and $S_6$ in (\ref{s222}) we can see that it becomes isomorphic to (\ref{s111}).\\
Now, applying $\cS$ on $S_2$ in (\ref{s1}), we see that
\be
%
\ee
The duality (\ref{so1212}) is an irreducible duality. That is, the duality does not hold if we integrate out matter from both sides of (\ref{so1212}).

Applying $\cS$ on $S_2$ and $S_4$ of the geometry
\be
%
\ee
we learn that
\be
%
\ee
where the theta angle $\theta=0$ for $\sp(0)$ means that the $\su(8)$ is embedded into $\fe_8$ with $\su(2)$ commutant. Similarly, applying $\cS$ on $S_2$ of
\be
%
\ee
where the theta angle $\theta=\pi$ for $\sp(0)$ means that the $\su(8)$ is embedded into $\fe_8$ with $\u(1)$ commutant.

The geometry
\be
%
\ee
is isomorphic to the geometry
\be
%
\ee

Now consider the following geometry
\be
%
\ee
describing $\so(12)+2\S+\half\C$. It is possible to decouple $S_6$ \cite{Bhardwaj:2019xeg} by decompactifying the curves $f-x_1,f-y_2,f-x_3,f-y_4,x_2,y_1,x_4,y_3$ while keeping the curves $x_1,y_2,x_3,y_4,f-x_2,f-y_1,f-x_4,f-y_3,e$ compact. After the decompactification, we obtain the geometry
\be
%
\ee
which describes $\su(6)_\frac92+\half\L^3+2\L^2$ which is neither a $5d$ SCFT nor a $5d$ KK theory. We are thus led to the conclusion that $\so(12)+2\S+\half\C$ is neither a $5d$ SCFT nor a $5d$ KK theory.

The isomorphism between
\be
%
\ee

\subsection{Rank 7}
Applying $\cS$ on $S_4$ of
\be
%
\ee

The isomorphism between
\be
%
\ee

\subsection{Rank 8}
The isomorphism between
\be
%
\ee

\section*{Acknowledgements}
LB thanks IPMU for hospitality during the initial stages of this work.\\
The work of LB is supported by NSF grant PHY-1719924. GZ is supported in part by World Premier International Research Center Initiative (WPI), MEXT, Japan, by the ERC-STG grant 637844-HBQFTNCER and by the INFN.

\bibliographystyle{ytphys}
\let\bbb\bibitem\def\bibitem{\itemsep4pt\bbb}
\bibliography{ref}

\providecommand{\href}[2]{#2}\begingroup\raggedright\begin{thebibliography}{10}

\bibitem{Gaiotto:2009we}
D.~Gaiotto, ``{${\cal N}=2$ dualities},''
\href{http://arxiv.org/abs/arXiv:0904.2715}{{\ttfamily arXiv:arXiv:0904.2715
  [hep-th]}}.

\bibitem{Genolini:2020htk}
P.~B. Genolini, M.~Honda, H.-C. Kim, D.~Tong, and C.~Vafa, ``{Evidence for a
  Non-Supersymmetric 5d CFT from Deformations of 5d $SU(2)$ SYM},''
  \href{http://arxiv.org/abs/2001.00023}{{\ttfamily arXiv:2001.00023
  [hep-th]}}.

\bibitem{Peskin:1980ay}
M.~E. Peskin, ``{Critical Point Behavior Of The Wilson Loop},''
  \href{http://dx.doi.org/10.1016/0370-2693(80)90848-5}{{\em Phys. Lett. B}
  {\bfseries 94} (1980) 161--165}.

\bibitem{Dienes:2004rt}
K.~Dienes, E.~Dudas, and T.~Gherghetta, ``{GUT precursors and fixed points in
  higher-dimensional theories},''
  \href{http://dx.doi.org/10.1007/BF02705084}{{\em Pramana} {\bfseries 62}
  (2004) 219--228}, \href{http://arxiv.org/abs/hep-th/0210294}{{\ttfamily
  arXiv:hep-th/0210294}}.

\bibitem{Morris:2004mg}
T.~R. Morris, ``{Renormalizable extra-dimensional models},''
  \href{http://dx.doi.org/10.1088/1126-6708/2005/01/002}{{\em JHEP} {\bfseries
  01} (2005) 002}, \href{http://arxiv.org/abs/hep-ph/0410142}{{\ttfamily
  arXiv:hep-ph/0410142}}.

\bibitem{Seiberg:1996bd}
N.~Seiberg, ``{Five-dimensional SUSY field theories, nontrivial fixed points
  and string dynamics},''
  \href{http://dx.doi.org/10.1016/S0370-2693(96)01215-4}{{\em Phys. Lett.}
  {\bfseries B388} (1996) 753--760},
\href{http://arxiv.org/abs/hep-th/9608111}{{\ttfamily arXiv:hep-th/9608111
  [hep-th]}}.

\bibitem{Morrison:1996xf}
D.~R. Morrison and N.~Seiberg, ``{Extremal transitions and five-dimensional
  supersymmetric field theories},''
  \href{http://dx.doi.org/10.1016/S0550-3213(96)00592-5}{{\em Nucl. Phys.}
  {\bfseries B483} (1997) 229--247},
\href{http://arxiv.org/abs/hep-th/9609070}{{\ttfamily arXiv:hep-th/9609070
  [hep-th]}}.

\bibitem{Intriligator:1997pq}
K.~A. Intriligator, D.~R. Morrison, and N.~Seiberg, ``{Five-dimensional
  supersymmetric gauge theories and degenerations of Calabi-Yau spaces},''
  \href{http://dx.doi.org/10.1016/S0550-3213(97)00279-4}{{\em Nucl. Phys.}
  {\bfseries B497} (1997) 56--100},
\href{http://arxiv.org/abs/hep-th/9702198}{{\ttfamily arXiv:hep-th/9702198
  [hep-th]}}.

\bibitem{Aharony:1997ju}
O.~Aharony and A.~Hanany, ``{Branes, superpotentials and superconformal fixed
  points},'' \href{http://dx.doi.org/10.1016/S0550-3213(97)00472-0}{{\em Nucl.
  Phys.} {\bfseries B504} (1997) 239--271},
\href{http://arxiv.org/abs/hep-th/9704170}{{\ttfamily arXiv:hep-th/9704170
  [hep-th]}}.

\bibitem{Aharony:1997bh}
O.~Aharony, A.~Hanany, and B.~Kol, ``{Webs of (p,q) five-branes,
  five-dimensional field theories and grid diagrams},''
  \href{http://dx.doi.org/10.1088/1126-6708/1998/01/002}{{\em JHEP} {\bfseries
  01} (1998) 002},
\href{http://arxiv.org/abs/hep-th/9710116}{{\ttfamily arXiv:hep-th/9710116
  [hep-th]}}.

\bibitem{DeWolfe:1999hik}
O.~DeWolfe, A.~Hanany, A.~Iqbal, and E.~Katz, ``{Five-branes, seven-branes and
  five-dimensional E(n) field theories},''
  \href{http://dx.doi.org/10.1088/1126-6708/1999/03/006}{{\em JHEP} {\bfseries
  03} (1999) 006}, \href{http://arxiv.org/abs/9902179}{{\ttfamily arXiv:9902179
  [hep-th]}}.

\bibitem{Douglas:1996xp}
M.~R. Douglas, S.~H. Katz, and C.~Vafa, ``{Small instantons, Del Pezzo surfaces
  and type I-prime theory},''
  \href{http://dx.doi.org/10.1016/S0550-3213(97)00281-2}{{\em Nucl. Phys.}
  {\bfseries B497} (1997) 155--172},
\href{http://arxiv.org/abs/hep-th/9609071}{{\ttfamily arXiv:hep-th/9609071
  [hep-th]}}.

\bibitem{Kim:2012gu}
H.-C. Kim, S.-S. Kim, and K.~Lee, ``{5-dim Superconformal Index with Enhanced
  En Global Symmetry},'' \href{http://dx.doi.org/10.1007/JHEP10(2012)142}{{\em
  JHEP} {\bfseries 10} (2012) 142},
\href{http://arxiv.org/abs/1206.6781}{{\ttfamily arXiv:1206.6781 [hep-th]}}.

\bibitem{Bergman:2013rgz}
O.~Bergman, D.~Rodríguez-Gómez, and G.~Zafrir, ``{Discrete $\theta$ and the
  5d superconformal index},''
  \href{http://dx.doi.org/10.1007/JHEP01(2014)079}{{\em JHEP} {\bfseries 01}
  (2014) 079}, \href{http://arxiv.org/abs/1310.2150}{{\ttfamily arXiv:1310.2150
  [hep-th]}}.

\bibitem{Bergman:2013aca}
O.~Bergman, D.~Rodríguez-Gómez, and G.~Zafrir, ``{5-Brane Webs, Symmetry
  Enhancement, and Duality in 5d Supersymmetric Gauge Theory},''
  \href{http://dx.doi.org/10.1007/JHEP03(2014)112}{{\em JHEP} {\bfseries 03}
  (2014) 112},
\href{http://arxiv.org/abs/1311.4199}{{\ttfamily arXiv:1311.4199 [hep-th]}}.

\bibitem{Zafrir:2014ywa}
G.~Zafrir, ``{Duality and enhancement of symmetry in 5d gauge theories},''
  \href{http://dx.doi.org/10.1007/JHEP12(2014)116}{{\em JHEP} {\bfseries 12}
  (2014) 116},
\href{http://arxiv.org/abs/1408.4040}{{\ttfamily arXiv:1408.4040 [hep-th]}}.

\bibitem{Tachikawa:2015mha}
Y.~Tachikawa, ``{Instanton operators and symmetry enhancement in 5d
  supersymmetric gauge theories},''
  \href{http://dx.doi.org/10.1093/ptep/ptv040}{{\em PTEP} {\bfseries 2015}
  no.~4, (2015) 043B06},
\href{http://arxiv.org/abs/1501.01031}{{\ttfamily arXiv:1501.01031 [hep-th]}}.

\bibitem{Zafrir:2015uaa}
G.~Zafrir, ``{Instanton operators and symmetry enhancement in 5d supersymmetric
  USp, SO and exceptional gauge theories},''
  \href{http://dx.doi.org/10.1007/JHEP07(2015)087}{{\em JHEP} {\bfseries 07}
  (2015) 087},
\href{http://arxiv.org/abs/1503.08136}{{\ttfamily arXiv:1503.08136 [hep-th]}}.

\bibitem{Yonekura:2015ksa}
K.~Yonekura, ``{Instanton operators and symmetry enhancement in 5d
  supersymmetric quiver gauge theories},''
\href{http://arxiv.org/abs/1505.04743}{{\ttfamily arXiv:1505.04743 [hep-th]}}.

\bibitem{Cremonesi:2015lsa}
S.~Cremonesi, G.~Ferlito, A.~Hanany, and N.~Mekareeya, ``{Instanton Operators
  and the Higgs Branch at Infinite Coupling},''
  \href{http://dx.doi.org/10.1007/JHEP04(2017)042}{{\em JHEP} {\bfseries 04}
  (2017) 042},
\href{http://arxiv.org/abs/1505.06302}{{\ttfamily arXiv:1505.06302 [hep-th]}}.

\bibitem{Jefferson:2017ahm}
P.~Jefferson, H.-C. Kim, C.~Vafa, and G.~Zafrir, ``{Towards Classification of
  5d SCFTs: Single Gauge Node},''
\href{http://arxiv.org/abs/1705.05836}{{\ttfamily arXiv:1705.05836 [hep-th]}}.

\bibitem{Ferlito:2017xdq}
G.~Ferlito, A.~Hanany, N.~Mekareeya, and G.~Zafrir, ``{3d Coulomb branch and 5d
  Higgs branch at infinite coupling},''
\href{http://arxiv.org/abs/1712.06604}{{\ttfamily arXiv:1712.06604 [hep-th]}}.

\bibitem{Bao:2011pty}
L.~Bao, E.~Pomoni, M.~Taki, and F.~Yagi, ``{M5-branes, toric diagrams and gauge
  theory duality},'' \href{http://dx.doi.org/10.1007/JHEP04(2012)105}{{\em
  JHEP} {\bfseries 04} (2011) 105},
  \href{http://arxiv.org/abs/1112.5228}{{\ttfamily arXiv:1112.5228 [hep-th]}}.

\bibitem{Bergman:2014kza}
O.~Bergman and G.~Zafrir, ``{Lifting 4d dualities to 5d},''
  \href{http://dx.doi.org/10.1007/JHEP04(2015)141}{{\em JHEP} {\bfseries 04}
  (2015) 141},
\href{http://arxiv.org/abs/1410.2806}{{\ttfamily arXiv:1410.2806 [hep-th]}}.

\bibitem{Kim:2015jba}
S.-S. Kim, M.~Taki, and F.~Yagi, ``{Tao Probing the End of the World},''
  \href{http://dx.doi.org/10.1093/ptep/ptv108}{{\em PTEP} {\bfseries 2015}
  no.~8, (2015) 083B02},
\href{http://arxiv.org/abs/1504.03672}{{\ttfamily arXiv:1504.03672 [hep-th]}}.

\bibitem{Hayashi:2015fsa}
H.~Hayashi, S.-S. Kim, K.~Lee, M.~Taki, and F.~Yagi, ``{A new 5d description of
  6d D-type minimal conformal matter},''
  \href{http://dx.doi.org/10.1007/JHEP08(2015)097}{{\em JHEP} {\bfseries 08}
  (2015) 097},
\href{http://arxiv.org/abs/1505.04439}{{\ttfamily arXiv:1505.04439 [hep-th]}}.

\bibitem{Gaiotto:2017hck}
D.~Gaiotto and H.-C. Kim, ``{Duality walls and defects in 5d ${\cal N}=1$
  theories},'' \href{http://dx.doi.org/10.1007/JHEP01(2017)019}{{\em JHEP}
  {\bfseries 01} (2017) 019},
\href{http://arxiv.org/abs/1506.03871}{{\ttfamily arXiv:1506.03871 [hep-th]}}.

\bibitem{Bergman:2015dpa}
O.~Bergman and G.~Zafrir, ``{5d fixed points from brane webs and O7-planes},''
  \href{http://dx.doi.org/10.1007/JHEP12(2015)163}{{\em JHEP} {\bfseries 12}
  (2015) 163},
\href{http://arxiv.org/abs/1507.03860}{{\ttfamily arXiv:1507.03860 [hep-th]}}.

\bibitem{Zafrir:2015rga}
G.~Zafrir, ``{Brane webs, $5d$ gauge theories and $6d$ $\mathcal{N}$$=(1,0)$
  SCFT's},'' \href{http://dx.doi.org/10.1007/JHEP12(2015)157}{{\em JHEP}
  {\bfseries 12} (2015) 157},
\href{http://arxiv.org/abs/1509.02016}{{\ttfamily arXiv:1509.02016 [hep-th]}}.

\bibitem{Hayashi:2015zka}
H.~Hayashi, S.-S. Kim, K.~Lee, and F.~Yagi, ``{6d SCFTs, 5d Dualities and Tao
  Web Diagrams},'' \href{http://dx.doi.org/10.1007/JHEP05(2019)203}{{\em JHEP}
  {\bfseries 05} (2019) 203},
\href{http://arxiv.org/abs/1509.03300}{{\ttfamily arXiv:1509.03300 [hep-th]}}.

\bibitem{Ohmori:2016shy}
K.~Ohmori and H.~Shimizu, ``{S$^{1}$/T$^{2}$ compactifications of 6d
  $\mathcal{N}=\left(1,\;0\right) $ theories and brane webs},''
  \href{http://dx.doi.org/10.1007/JHEP03(2016)024}{{\em JHEP} {\bfseries 03}
  (2016) 024}, \href{http://arxiv.org/abs/1509.03195}{{\ttfamily
  arXiv:1509.03195 [hep-th]}}.

\bibitem{Zafrir:2015ftn}
G.~Zafrir, ``{Brane webs and $O5$-planes},''
  \href{http://dx.doi.org/10.1007/JHEP03(2016)109}{{\em JHEP} {\bfseries 03}
  (2016) 109},
\href{http://arxiv.org/abs/1512.08114}{{\ttfamily arXiv:1512.08114 [hep-th]}}.

\bibitem{Hayashi:2015vhy}
H.~Hayashi, S.-S. Kim, K.~Lee, M.~Taki, and F.~Yagi, ``{More on 5d descriptions
  of 6d SCFTs},'' \href{http://dx.doi.org/10.1007/JHEP10(2016)126}{{\em JHEP}
  {\bfseries 10} (2016) 126},
\href{http://arxiv.org/abs/1512.08239}{{\ttfamily arXiv:1512.08239 [hep-th]}}.

\bibitem{Zafrir:2016csd}
G.~Zafrir, ``{Brane webs in the presence of an $O5^-$-plane and 4d class S
  theories of type D},'' \href{http://dx.doi.org/10.1007/JHEP07(2016)035}{{\em
  JHEP} {\bfseries 07} (2016) 035},
\href{http://arxiv.org/abs/1602.00130}{{\ttfamily arXiv:1602.00130 [hep-th]}}.

\bibitem{Hayashi:2016abm}
H.~Hayashi, S.-S. Kim, K.~Lee, and F.~Yagi, ``{Equivalence of several
  descriptions for 6d SCFT},''
  \href{http://dx.doi.org/10.1007/JHEP01(2017)093}{{\em JHEP} {\bfseries 01}
  (2017) 093},
\href{http://arxiv.org/abs/1607.07786}{{\ttfamily arXiv:1607.07786 [hep-th]}}.

\bibitem{Hayashi:2017skf}
H.~Hayashi, S.-S. Kim, K.~Lee, and F.~Yagi, ``{Discrete theta angle from an
  O5-plane},'' \href{http://dx.doi.org/10.1007/JHEP11(2017)041}{{\em JHEP}
  {\bfseries 11} (2017) 041},
\href{http://arxiv.org/abs/1707.07181}{{\ttfamily arXiv:1707.07181 [hep-th]}}.

\bibitem{Hayashi:2018bkd}
H.~Hayashi, S.-S. Kim, K.~Lee, and F.~Yagi, ``{5-brane webs for 5d $
  \mathcal{N} $ = 1 G$_{2}$ gauge theories},''
  \href{http://dx.doi.org/10.1007/JHEP03(2018)125}{{\em JHEP} {\bfseries 03}
  (2018) 125},
\href{http://arxiv.org/abs/1801.03916}{{\ttfamily arXiv:1801.03916 [hep-th]}}.

\bibitem{Hayashi:2018lyv}
H.~Hayashi, S.-S. Kim, K.~Lee, and F.~Yagi, ``{Dualities and 5-brane webs for
  5d rank 2 SCFTs},'' \href{http://dx.doi.org/10.1007/JHEP12(2018)016}{{\em
  JHEP} {\bfseries 12} (2018) 016},
\href{http://arxiv.org/abs/1806.10569}{{\ttfamily arXiv:1806.10569 [hep-th]}}.

\bibitem{Cabrera:2019hya}
S.~Cabrera, A.~Hanany, and F.~Yagi, ``{Tropical Geometry and Five Dimensional
  Higgs Branches at Infinite Coupling},''
  \href{http://dx.doi.org/10.1007/JHEP01(2019)068}{{\em JHEP} {\bfseries 01}
  (2019) 068}, \href{http://arxiv.org/abs/1810.01379}{{\ttfamily
  arXiv:1810.01379 [hep-th]}}.

\bibitem{Hayashi:2019yxj}
H.~Hayashi, S.-S. Kim, K.~Lee, and F.~Yagi, ``{Rank-3 antisymmetric matter on
  5-brane webs},'' \href{http://dx.doi.org/10.1007/JHEP05(2019)133}{{\em JHEP}
  {\bfseries 05} (2019) 133},
\href{http://arxiv.org/abs/1902.04754}{{\ttfamily arXiv:1902.04754 [hep-th]}}.

\bibitem{Hayashi:2020sly}
H.~Hayashi, S.-S. Kim, K.~Lee, and F.~Yagi, ``{Complete prepotential for 5d
  ${\cal N} = 1$ superconformal field theories},''
  \href{http://dx.doi.org/10.1007/JHEP02(2020)074}{{\em JHEP} {\bfseries 02}
  (2020) 074}, \href{http://arxiv.org/abs/1912.10301}{{\ttfamily
  arXiv:1912.10301 [hep-th]}}.

\bibitem{DelZotto:2017pti}
M.~Del~Zotto, J.~J. Heckman, and D.~R. Morrison, ``{6D SCFTs and Phases of 5D
  Theories},'' \href{http://dx.doi.org/10.1007/JHEP09(2017)147}{{\em JHEP}
  {\bfseries 09} (2017) 147},
\href{http://arxiv.org/abs/1703.02981}{{\ttfamily arXiv:1703.02981 [hep-th]}}.

\bibitem{Xie:2017pfl}
D.~Xie and S.-T. Yau, ``{Three dimensional canonical singularity and five
  dimensional $ \mathcal{N} $ = 1 SCFT},''
  \href{http://dx.doi.org/10.1007/JHEP06(2017)134}{{\em JHEP} {\bfseries 06}
  (2017) 134},
\href{http://arxiv.org/abs/1704.00799}{{\ttfamily arXiv:1704.00799 [hep-th]}}.

\bibitem{EJeK}
M.~Esole, P.~Jefferson, and M.~J. Kang, ``{The Geometry of $F_4$-Models},''
  \href{http://arxiv.org/abs/1704.08251}{{\ttfamily arXiv:1704.08251
  [hep-th]}}.

\bibitem{EJaK1}
M.~Esole, R.~Jagadeesan, and M.~J. Kang, ``{The Geometry of $G_2$, Spin(7), and
  Spin(8)-models},'' \href{http://arxiv.org/abs/1709.04913}{{\ttfamily
  arXiv:1709.04913 [hep-th]}}.

\bibitem{EKY}
M.~Esole, M.~J. Kang, and S.-T. Yau, ``{Mordell-Weil Torsion, Anomalies, and
  Phase Transitions},'' \href{http://arxiv.org/abs/1712.02337}{{\ttfamily
  arXiv:1712.02337 [hep-th]}}.

\bibitem{Jefferson:2018irk}
P.~Jefferson, S.~Katz, H.-C. Kim, and C.~Vafa, ``{On Geometric Classification
  of 5d SCFTs},'' \href{http://dx.doi.org/10.1007/JHEP04(2018)103}{{\em JHEP}
  {\bfseries 04} (2018) 103},
\href{http://arxiv.org/abs/1801.04036}{{\ttfamily arXiv:1801.04036 [hep-th]}}.

\bibitem{EK1}
M.~Esole and M.~J. Kang, ``{Flopping and Slicing: SO(4) and Spin(4)-models},''
  \href{http://arxiv.org/abs/1802.04802}{{\ttfamily arXiv:1802.04802
  [hep-th]}}.

\bibitem{EK2}
M.~Esole and M.~J. Kang, ``{The Geometry of the SU(2)$\times G_2$-model},''
  \href{http://dx.doi.org/10.1007/JHEP02(2019)091}{{\em JHEP} {\bfseries 02}
  (2019) 091}, \href{http://arxiv.org/abs/1805.03214}{{\ttfamily
  arXiv:1805.03214 [hep-th]}}.

\bibitem{Bhardwaj:2018yhy}
L.~Bhardwaj and P.~Jefferson, ``{Classifying 5d SCFTs via 6d SCFTs: Rank
  one},''
\href{http://arxiv.org/abs/1809.01650}{{\ttfamily arXiv:1809.01650 [hep-th]}}.

\bibitem{Bhardwaj:2018vuu}
L.~Bhardwaj and P.~Jefferson, ``{Classifying 5d SCFTs via 6d SCFTs: Arbitrary
  rank},''
\href{http://arxiv.org/abs/1811.10616}{{\ttfamily arXiv:1811.10616 [hep-th]}}.

\bibitem{Apruzzi:2018nre}
F.~Apruzzi, L.~Lin, and C.~Mayrhofer, ``{Phases of 5d SCFTs from M-/F-theory on
  Non-Flat Fibrations},'' \href{http://dx.doi.org/10.1007/JHEP05(2019)187}{{\em
  JHEP} {\bfseries 05} (2019) 187},
\href{http://arxiv.org/abs/1811.12400}{{\ttfamily arXiv:1811.12400 [hep-th]}}.

\bibitem{Closset:2018bjz}
C.~Closset, M.~Del~Zotto, and V.~Saxena, ``{Five-dimensional SCFTs and gauge
  theory phases: an M-theory/type IIA perspective},''
  \href{http://dx.doi.org/10.21468/SciPostPhys.6.5.052}{{\em SciPost Phys.}
  {\bfseries 6} no.~5, (2019) 052},
\href{http://arxiv.org/abs/1812.10451}{{\ttfamily arXiv:1812.10451 [hep-th]}}.

\bibitem{EJaK2}
M.~Esole, R.~Jagadeesan, and M.~J. Kang, ``{48 Crepant Paths to $SU(2) \times
  SU(3)$},'' \href{http://arxiv.org/abs/1905.05174}{{\ttfamily arXiv:1905.05174
  [hep-th]}}.

\bibitem{EJ}
M.~Esole and P.~Jefferson, ``{The Geometry of SO(3), SO(5), and SO(6)
  models},'' \href{http://arxiv.org/abs/1905.12620}{{\ttfamily arXiv:1905.12620
  [hep-th]}}.

\bibitem{Apruzzi:2019vpe}
F.~Apruzzi, C.~Lawrie, L.~Lin, S.~Schafer-Nameki, and Y.-N. Wang, ``{5d
  Superconformal Field Theories and Graphs},''
  \href{http://dx.doi.org/10.1016/j.physletb.2019.135077}{{\em Phys.Lett.}
  {\bfseries B800} (2019) 135077},
\href{http://arxiv.org/abs/1906.11820}{{\ttfamily arXiv:1906.11820 [hep-th]}}.

\bibitem{Apruzzi:2019opn}
F.~Apruzzi, C.~Lawrie, L.~Lin, S.~Schafer-Nameki, and Y.-N. Wang, ``{Fibers add
  Flavor, Part I: Classification of 5d SCFTs, Flavor Symmetries and BPS
  States},'' \href{http://dx.doi.org/10.1007/JHEP11(2019)068}{{\em JHEP}
  {\bfseries 11} (2019) 068},
\href{http://arxiv.org/abs/1907.05404}{{\ttfamily arXiv:1907.05404 [hep-th]}}.

\bibitem{Apruzzi:2019enx}
F.~Apruzzi, C.~Lawrie, L.~Lin, S.~Schafer-Nameki, and Y.-N. Wang, ``{Fibers add
  Flavor, Part II: 5d SCFTs, Gauge Theories, and Dualities},''
\href{http://arxiv.org/abs/1909.09128}{{\ttfamily arXiv:1909.09128 [hep-th]}}.

\bibitem{Bhardwaj:2019jtr}
L.~Bhardwaj, ``{On the classification of $5d$ SCFTs},''
\href{http://arxiv.org/abs/1909.09635}{{\ttfamily arXiv:1909.09635 [hep-th]}}.

\bibitem{Bhardwaj:2019fzv}
L.~Bhardwaj, P.~Jefferson, H.-C. Kim, H.-C. Tarazi, and C.~Vafa, ``{Twisted
  Circle Compactification of 6d SCFTs},''
\href{http://arxiv.org/abs/1909.11666}{{\ttfamily arXiv:1909.11666 [hep-th]}}.

\bibitem{Bhardwaj:2019ngx}
L.~Bhardwaj, ``{Dualities of $5d$ gauge theories from S-duality},''
\href{http://arxiv.org/abs/1909.05250}{{\ttfamily arXiv:1909.05250 [hep-th]}}.

\bibitem{Saxena:2019wuy}
V.~Saxena, ``{Rank-two 5d SCFTs from M-theory at isolated toric singularities:
  a systematic study},''
\href{http://arxiv.org/abs/1911.09574}{{\ttfamily arXiv:1911.09574 [hep-th]}}.

\bibitem{Bhardwaj:2019xeg}
L.~Bhardwaj, ``{Do all 5d SCFTs descend from 6d SCFTs?},''
\href{http://arxiv.org/abs/1912.00025}{{\ttfamily arXiv:1912.00025 [hep-th]}}.

\bibitem{Apruzzi:2019syw}
F.~Apruzzi, S.~Schafer-Nameki, and Y.-N. Wang, ``{5d SCFTs from Decoupling and
  Gluing},'' \href{http://arxiv.org/abs/1912.04264}{{\ttfamily arXiv:1912.04264
  [hep-th]}}.

\bibitem{Closset:2019mdz}
C.~Closset and M.~Del~Zotto, ``{On 5d SCFTs and their BPS quivers. Part I:
  B-branes and brane tilings},''
  \href{http://arxiv.org/abs/1912.13502}{{\ttfamily arXiv:1912.13502
  [hep-th]}}.

\bibitem{Bhardwaj:2013wy}
L.~Bhardwaj and Y.~Tachikawa, ``{Classification of 4d N=2 gauge theories},''
  \href{http://dx.doi.org/10.1007/JHEP12(2013)100}{{\em JHEP} {\bfseries 12}
  (2013) 100}, \href{http://arxiv.org/abs/1309.5160}{{\ttfamily arXiv:1309.5160
  [hep-th]}}.

\bibitem{Bhardwaj:2015xxa}
L.~Bhardwaj, ``{Classification of 6d $ \mathcal{N}=\left(1,0\right) $ gauge
  theories},'' \href{http://dx.doi.org/10.1007/JHEP11(2015)002}{{\em JHEP}
  {\bfseries 11} (2015) 002},
\href{http://arxiv.org/abs/1502.06594}{{\ttfamily arXiv:1502.06594 [hep-th]}}.

\bibitem{Bergman:2012rgz}
O.~Bergman and D.~Rodríguez-Gómez, ``{5d quivers and their AdS(6) duals},''
  \href{http://dx.doi.org/10.1007/JHEP07(2012)171}{{\em JHEP} {\bfseries 07}
  (2012) 171},
\href{http://arxiv.org/abs/1206.3503}{{\ttfamily arXiv:1206.3503 [hep-th]}}.

\bibitem{Brunner:1997ndk}
I.~Brunner and A.~Karch, ``{Branes and six-dimensional fixed points},''
  \href{http://dx.doi.org/10.1016/S0370-2693(97)00935-0}{{\em Phys.Lett.}
  {\bfseries B409} (1997) 109--116},
  \href{http://arxiv.org/abs/hep-th/9705022}{{\ttfamily arXiv:hep-th/9705022
  [hep-th]}}.

\bibitem{Brandhuber:1999yo}
A.~Brandhuber and Y.~Oz, ``{The D-4 - D-8 brane system and five-dimensional
  fixed points},'' \href{http://dx.doi.org/10.1016/S0370-2693(99)00763-7}{{\em
  Phys. Lett.} {\bfseries B460} (1999) 307--312}.

\bibitem{DHoker:2016wak}
E.~D'Hoker, M.~Gutperle, A.~Karch, and C.~F. Uhlemann, ``{Warped $AdS_6 \times
  S_2$ in Type IIB supergravity I: Local solutions},''
  \href{http://dx.doi.org/10.1007/JHEP08(2016)046}{{\em JHEP} {\bfseries 08}
  (2016) 046}, \href{http://arxiv.org/abs/1606.01254}{{\ttfamily
  arXiv:1606.01254 [hep-th]}}.

\bibitem{DHoker:2017prl}
E.~D'Hoker, M.~Gutperle, and C.~F. Uhlemann, ``{Holographic duals for
  five-dimensional superconformal quantum field theories},''
  \href{http://dx.doi.org/10.1103/PhysRevLett.118.101601}{{\em Phys. Rev.
  Lett.} {\bfseries 118} no.~101601, (2017) 066006},
  \href{http://arxiv.org/abs/1611.09411}{{\ttfamily arXiv:1611.09411
  [hep-th]}}.

\bibitem{DHoker:2017muf}
E.~D'Hoker, M.~Gutperle, and C.~F. Uhlemann, ``{Warped $AdS_6 \times S_2$ in
  Type IIB supergravity II: Global solutions and five-brane webs},''
  \href{http://dx.doi.org/10.1007/JHEP05(2017)131}{{\em JHEP} {\bfseries 05}
  (2017) 131}, \href{http://arxiv.org/abs/1703.08186}{{\ttfamily
  arXiv:1703.08186 [hep-th]}}.

\bibitem{DHoker:2017gcu}
E.~D'Hoker, M.~Gutperle, and C.~F. Uhlemann, ``{Warped $AdS_6 \times S_2$ in
  Type IIB supergravity III: Global solutions with seven-branes},''
  \href{http://dx.doi.org/10.1007/JHEP11(2017)200}{{\em JHEP} {\bfseries 11}
  (2017) 200}, \href{http://arxiv.org/abs/1706.00433}{{\ttfamily
  arXiv:1706.00433 [hep-th]}}.

\bibitem{Uhlemann:2019ors}
C.~F. Uhlemann, ``{$AdS_6$/$CFT_5$ with O7-planes},''
\href{http://arxiv.org/abs/1912.09716}{{\ttfamily arXiv:1912.09716 [hep-th]}}.

\bibitem{Heckman:2013pva}
J.~J. Heckman, D.~R. Morrison, and C.~Vafa, ``{On the Classification of 6D
  SCFTs and Generalized ADE Orbifolds},''
  \href{http://dx.doi.org/10.1007/JHEP06(2015)017,
  10.1007/JHEP05(2014)028}{{\em JHEP} {\bfseries 05} (2014) 028},
  \href{http://arxiv.org/abs/1312.5746}{{\ttfamily arXiv:1312.5746 [hep-th]}}.
[Erratum: JHEP06,017(2015)].

\bibitem{Heckman:2015bfa}
J.~J. Heckman, D.~R. Morrison, T.~Rudelius, and C.~Vafa, ``{Atomic
  Classification of 6D SCFTs},''
  \href{http://dx.doi.org/10.1002/prop.201500024}{{\em Fortsch. Phys.}
  {\bfseries 63} (2015) 468--530},
\href{http://arxiv.org/abs/1502.05405}{{\ttfamily arXiv:1502.05405 [hep-th]}}.

\bibitem{Tachikawa:2011ch}
Y.~Tachikawa, ``{On S-duality of 5d super Yang-Mills on $S^1$},''
  \href{http://dx.doi.org/10.1007/JHEP11(2011)123}{{\em JHEP} {\bfseries 11}
  (2011) 123},
\href{http://arxiv.org/abs/1110.0531}{{\ttfamily arXiv:1110.0531 [hep-th]}}.

\bibitem{Douglas:2010iu}
M.~R. Douglas, ``{On D=5 super Yang-Mills theory and (2,0) theory},''
  \href{http://dx.doi.org/10.1007/JHEP02(2011)011}{{\em JHEP} {\bfseries 1102}
  (2011) 011},
\href{http://arxiv.org/abs/1012.2880}{{\ttfamily arXiv:1012.2880 [hep-th]}}.

\bibitem{Lambert:2010iw}
N.~Lambert, C.~Papageorgakis, and M.~Schmidt-Sommerfeld, ``{M5-Branes,
  D4-Branes and Quantum 5D super-Yang-Mills},''
  \href{http://dx.doi.org/10.1007/JHEP01(2011)083}{{\em JHEP} {\bfseries 1101}
  (2011) 083},
\href{http://arxiv.org/abs/1012.2882}{{\ttfamily arXiv:1012.2882 [hep-th]}}.

\bibitem{Ganor:1996pc}
O.~J. Ganor, D.~R. Morrison, and N.~Seiberg, ``{Branes, Calabi-Yau spaces, and
  toroidal compactification of the N=1 six-dimensional E(8) theory},''
  \href{http://dx.doi.org/10.1016/S0550-3213(96)00690-6}{{\em Nucl.Phys.}
  {\bfseries B487} (1997) 93--127},
\href{http://arxiv.org/abs/hep-th/9610251}{{\ttfamily arXiv:hep-th/9610251
  [hep-th]}}.

\bibitem{Razamat:2018gro}
S.~S. Razamat and G.~Zafrir, ``{Compactification of 6d minimal SCFTs on Riemann
  surfaces},'' \href{http://dx.doi.org/10.1103/PhysRevD.98.066006}{{\em Phys.
  Rev.} {\bfseries D98} no.~6, (2018) 066006},
\href{http://arxiv.org/abs/1806.09196}{{\ttfamily arXiv:1806.09196 [hep-th]}}.

\bibitem{Razamat:2018gbu}
S.~S. Razamat, O.~Sela, and G.~Zafrir, ``{Curious patterns of IR symmetry
  enhancement},'' \href{http://dx.doi.org/10.1007/JHEP10(2018)163}{{\em JHEP}
  {\bfseries 10} (2018) 163},
\href{http://arxiv.org/abs/1809.00541}{{\ttfamily arXiv:1809.00541 [hep-th]}}.

\bibitem{Sela:2019gbz}
O.~Sela and G.~Zafrir, ``{Symmetry enhancement in 4d Spin(n) gauge theories and
  compactification from 6d},''
  \href{http://dx.doi.org/10.1007/JHEP12(2019)052}{{\em JHEP} {\bfseries 12}
  (2019) 052},
\href{http://arxiv.org/abs/1910.03629}{{\ttfamily arXiv:1910.03629 [hep-th]}}.

\bibitem{Mekareeya:2017jgc}
N.~Mekareeya, K.~Ohmori, Y.~Tachikawa, and G.~Zafrir, ``{E$_{8}$ instantons on
  type-A ALE spaces and supersymmetric field theories},''
  \href{http://dx.doi.org/10.1007/JHEP09(2017)144}{{\em JHEP} {\bfseries 09}
  (2017) 144},
\href{http://arxiv.org/abs/1707.04370}{{\ttfamily arXiv:1707.04370 [hep-th]}}.

\bibitem{Witten:1982fp}
E.~Witten, ``{An SU(2) Anomaly},''
  \href{http://dx.doi.org/10.1016/0370-2693(82)90728-6}{{\em Phys. Lett.}
  {\bfseries B117} (1982) 324--328}.
[,230(1982)].

\bibitem{Chacaltana:2013dsj}
O.~Chacaltana and J.~Distler, ``{Tinkertoys for the $D_N$ series},''
  \href{http://dx.doi.org/10.1007/JHEP02(2013)110}{{\em JHEP} {\bfseries 02}
  (2013) 110},
\href{http://arxiv.org/abs/1106.5410}{{\ttfamily arXiv:1106.5410 [hep-th]}}.

\bibitem{Chacaltana:2015dat}
O.~Chacaltana, J.~Distler, and A.~Trimm, ``{Seiberg-Witten for $Spin(n)$ with
  Spinors},'' \href{http://dx.doi.org/10.1007/JHEP08(2015)027}{{\em JHEP}
  {\bfseries 08} (2015) 027},
\href{http://arxiv.org/abs/1404.3736}{{\ttfamily arXiv:1404.3736 [hep-th]}}.

\end{thebibliography}\endgroup

\end{document}